\documentclass[twocolumn]{aastex631}

\usepackage{amsmath}
\usepackage{subfigure}

\begin{document}
\title{A Tetrad-First Approach to Robust Numerical Algorithms in General Relativity}

\collaboration{4}{}

\author{Jonathan Gorard}
\affiliation{Princeton University, Princeton, NJ 08544, USA}
\affiliation{Princeton Plasma Physics Laboratory, Princeton, NJ, 08540, USA}

\author{Ammar Hakim}
\affiliation{Princeton Plasma Physics Laboratory, Princeton, NJ, 08540, USA}

\author{James Juno}
\affiliation{Princeton Plasma Physics Laboratory, Princeton, NJ, 08540, USA}

\author{Jason M. TenBarge}
\affiliation{Princeton University, Princeton, NJ 08544, USA}

\correspondingauthor{Jonathan Gorard}
\email{gorard@princeton.edu}

\begin{abstract}
General relativistic Riemann solvers are typically complex, fragile and unwieldy, at least in comparison to their special relativistic counterparts. In this paper, we present a new high-resolution shock-capturing algorithm on curved spacetimes that employs a local coordinate transformation at each inter-cell boundary, transforming all primitive and conservative variables into a locally flat spacetime coordinate basis (i.e., the \textit{tetrad basis}), generalizing previous approaches developed for relativistic hydrodynamics. This algorithm enables one to employ a purely special relativistic Riemann solver, combined with an appropriate post-hoc flux correction step, irrespective of the geometry of the underlying Lorentzian manifold. We perform a systematic validation of the algorithm using the \textsc{Gkeyll} simulation framework for both general relativistic electromagnetism and general relativistic hydrodynamics, highlighting the algorithm's superior convergence and stability properties in each case when compared against standard analytical solutions for black hole magnetosphere and ultra-relativistic black hole accretion problems. However, as an illustration of the generality and practicality of the algorithm, we also apply it to more astrophysically realistic magnetosphere and fluid accretion problems in the limit of high black hole spin, for which standard general relativistic Riemann solvers are often too unstable to produce useful solutions.
\end{abstract}

\section{Introduction}

A wide range of high-energy astrophysical scenarios are modeled as solutions to systems of local conservation laws in curved spacetime, such as the equations of general relativistic electromagnetism or of general relativistic hydrodynamics. For example, the force-free electrodynamics of black hole magnetospheres and the process of energy extraction from a spinning black hole by an electromagnetic field, as first studied analytically by \citet{wald1974} and later by \citet{blandford_znajek1977}, were subsequently studied numerically by \citet{komissarov2004} using a conservation law form of Maxwell's equations in curved spacetime. Similarly, hydrodynamics equations in curved spacetime have been used by \citet{font_ibanez1998} and later \citet{font_ibanez_papadopoulos1999} to study supersonic matter accretion onto both static and spinning black holes, extending the previous, purely Newtonian, work of \citet{bondi_hoyle1944} on non-spherical matter accretion onto compact objects into the fully general relativistic regime.

The coupling of electromagnetic fields and fluids via the formulation of general relativistic magnetohydrodynamics (GRMHD) as a system of conservation laws by \citet{anton_zanotti_miralles_marti_ibanez_font_pons2006}, which can then be integrated numerically using finite-volume methods, has been the modus operandi in recent decades for a wide variety of applications in high-energy astrophysics, culminating in recent predictive and interpretive simulations in support of major collaborations such as the Event Horizon Telescope (\citet{eht2019a}; \citet{eht2019b}; and \citet{eht2021}). There are now several numerical codes which solve the GRMHD equations in both stationary and dynamic spacetimes (see, e.g., \citet{gammie_mckinney_toth2003}; \citet{mckinney_tchekhovskoy_sadowski_narayan2014}; \citet{sadowski_narayan_mckinney_tchekhovskoy2014}; \citet{white_stone_gammie2016}; \citet{liska_hesp_tchekhovskoy_ingram_vanderklis_markoff2017}; and the references therein), and systematic comparisons of these production solvers have been performed to verify their accuracy (\citet{porth2019}). Many of these codes have been extended to include non-ideal physics such as resistivity (see, e.g., \citet{ripperda_porth_sironi_keppens2019} and \citet{ripperda2019}) and electron thermodynamics (see, e.g., \citet{chael_rowan_narayan_johnson_sironi2018} and \citet{chael_narayan_johnson2019}), and the flexibility of these codes has led to their deployment in the simulation of diverse compact object systems, including jet formation within black hole magnetospheres by \citet{koide_meier_shibata_kudoh2000}, jet formation from magnetically arrested disks by \citet{tchekhovskoy_narayan_mckinney2011}, and outflows from binary neutron star ejecta by \citet{nathanail_gill_porth_fromm_rezzolla2021}.

However, as computational capabilities have increased and modest kinetic simulations of compact objects have become feasible, it has become increasingly apparent that the classical single-fluid models of accretion (\citet{parfrey_philippov_cerutti2019}) are missing physics. Microscale instabilities driven by accelerated beams of particles, parallel heat fluxes, and anisotropic temperatures in the parallel versus perpendicular directions to the local magnetic field, in addition to kinetic modifications to magnetic reconnection, can all alter both the dynamics of the underlying plasma and the signatures of what may be observable from these astrophysical laboratories (\citet{galishnikova_philippov_quataert_bacchini_parfrey_ripperda2023}). Furthermore, these recent kinetic studies suggest that previous works utilizing kinetics-inspired closures, which argued that the impact of these non-ideal effects is small (see, e.g., \citet{chandra_gammie_foucart_quataert2015} and \citet{chandra_foucart_gammie2017}), presented an incomplete picture of the evolution of these plasmas; in other words, a first-principles treatment does indeed modify the global evolution of the accretion onto the compact object. However, since kinetic simulations are likely to remain relatively computationally expensive due to their high dimensionality and the underlying requirements of many kinetic numerical methods to resolve all the microscales of the plasma, alternative models which go beyond single-fluid MHD but which are nonetheless more cost effective than kinetic simulations are of significant interest. It is the purpose of this study to lay the groundwork for such an endeavor.

Inspired by the development of non-relativistic multi-fluid solvers in the space physics community and their favorable performance and physics fidelity compared to first-principles kinetic simulations (\citet{shumlak_loverich2003}; \citet{hakim_loverich_shumlak2006}; \citet{hakim2008}; \citet{shumlak_lilly_reddell_sousa_srinivasan2011}; \citet{wang_hakim_bhattacharjee_germaschewski2015}; \citet{ng_huang_hakim_bhattacharjee_stanier_daughton_wang_germaschewski2015}; \citet{ng_hakim_bhattacharjee_stanier_daughton2017}; \citet{ho_datta_shumlak2018}; \citet{allmannrahn_trost_grauer2018}; \citet{ng_hakim_juno_bhattacharjee2019}; \citet{ng_hakim_wang_bhattacharjee2020}; and \citet{allmannrahn_lautenbach_grauer_sydora2021}), we consider what the extension of these methods to curved spacetimes would require. What distinguishes this approach from the typical GRMHD formulation is that the hydrodynamics equations and Maxwell's equations are solved separately, and then their coupling is handled through source terms on the right-hand side of the momentum, energy, and electric field evolution equations. While these source terms are generically stiff, oscillatory components of the overall system of partial differential equations, they are entirely local in space and hence can be solved implicitly (\citet{wang_hakim_ng_dong_germaschewski2020}). One is therefore able to deploy a numerical method which can reasonably approximate the kinetic microphysics where desired by, e.g., resolving the electron inertial length in reconnecting current sheets, while simultaneously robustly under-resolving the microscales throughout the remainder of the simulation domain and bridging the gap to macroscopic simulations of, e.g., planetary magnetospheres ( \citet{wang_germaschewski_hakim_dong_raeder_bhattacharjee2018} and \citet{dong_wang_hakim_bhattacharjee_slavin_dibraccio_germaschewski2019}). Thus, the most critical component in extending this multi-fluid formalism to curved spacetimes is the need for a robust solver for handling generic conservation laws in general relativity. In particular, one requires a solver that can be used to solve both Maxwell's equations and hydrodynamics equations in curved spacetime, analogous to the successes of these multi-fluid solvers in non-relativistic space physics applications, which have made extensive use of, e.g., the wave propagation method of \citet{leveque1997} for general hyperbolic partial differential equations.

In this regard, a standard approach in computational fluid dynamics is to use a high-resolution shock-capturing method, building upon the seminal work of \citet{godunov1959}, in which a local Riemann problem is solved at each inter-cell interface, at every time-step, to yield an approximate solution to a weak form of the underlying equations that allows for a consistent numerical treatment of discontinuities (i.e., shocks). However, Riemann solvers for fully general relativistic hydrodynamics tend to be somewhat unwieldy, as noted by \citet{ibanez_marti1999}, due to the increased complexity of the underlying equations when compared with the flat spacetime case. For this reason, \citet{pons_font_ibanez_marti_miralles1998} proposed exploiting a basic consequence of the equivalence principle in general relativity, whereby at every point in a curved spacetime there exists a local \textit{geodesic} coordinate system within which the metric appears flat, to transform all fluid variables at the inter-cell interfaces into a locally flat coordinate system. This method allows one to solve hydrodynamics equations within any curved spacetime using only special relativistic Riemann solvers (a considerable simplification over the standard approach), with only the additional expense of a coordinate transformation being incurred at every inter-cell interface. Such an approach was later implemented as an extension to the \textsc{Athena++} GRMHD code by \citet{white_stone_gammie2016}, and exhibited strong performance for the steady-state fluid torus problem of \citet{fishbone_moncrief1976} around a spinning black hole.

In this paper, we begin by noting that the original approach proposed by \citet{pons_font_ibanez_marti_miralles1998} for relativistic hydrodynamics may, with only modest generalization, be extended to arbitrary systems of local conservation laws in curved spacetime, including general relativistic electromagnetism (allowing, for instance, force-free electrodynamics simulations around compact objects to be performed using the same algorithmic framework). Within our generalized formulation, all tensorial quantities are transformed into an orthonormal tetrad basis at each inter-cell interface, the special relativistic Riemann problem is then solved in a co-moving frame across that interface, and finally the resulting flux vector is multiplied by a geometric correction term to obtain the correct form of the general relativistic inter-cell flux. Due to this use of the formalism of tetrads (or frame fields) throughout our formulation of the algorithm, we refer to this as a \textit{tetrad-first} approach. We argue that the tetrad-first approach confers two principal advantages over the use of standard general relativistic Riemann solvers. First, the relative simplicity of the flux Jacobian for special relativistic fluxes over general relativistic ones means that, at least for Riemann solvers\footnote{For instance, the linearized solver of \citet{roe1981}, later extended to both special and general relativistic hydrodynamics by \citet{eulderink_mellema1995}} that rely upon details of the eigensystem of the flux Jacobian, the special relativistic Riemann solvers are typically more numerically robust than their general relativistic counterparts. Second, in scenarios involving high spacetime curvatures, the wave-speed estimates that form part of many Riemann solvers can be made more uniform across the computational domain; for instance, in a Riemann solver for the curved spacetime Maxwell equations, the characteristic wave-speeds depend upon the spacetime gauge variables (and, in particular, may become arbitrarily small in regions of high spacetime curvature), whilst for the flat spacetime Maxwell equations, the wave-speeds are always exactly equal to the speed of light. We find, across a variety of test cases, that this simplification often leads to faster and more uniform convergence of the tetrad-first approach than the approach of using standard general relativistic Riemann solvers.

We have implemented the tetrad-first approach into the existing finite-volume solver of the \textsc{Gkeyll} simulation framework, which already applies a local spatial coordinate transformation at each inter-cell interface to facilitate the computation of inter-cell fluxes in arbitrary curvilinear geometries. As a consequence, the only modifications that are required in order to extend any special relativistic equation system supported by \textsc{Gkeyll} into a fully curved spacetime are the addition of a local space\textit{time} coordinate transformation (i.e., one that also transforms the gauge variables, in addition to the spatial metric), and the addition of a post-hoc flux correction step. In Section \ref{sec:general_algorithm}, we present the general algorithm for arbitrary conservation laws. We have validated this approach for two rather different equation systems, namely the equations of general relativistic electromagnetism in Section \ref{sec:gr_electromagnetism} and the equations of general relativistic hydrodynamics in Section \ref{sec:gr_hydrodynamics}. In both cases, we first validate the underlying special relativistic Riemann solver against standard one-dimensional Riemann problems before validating the tetrad-first algorithm against certain test problems in black hole spacetimes that admit analytical solutions: for the curved spacetime Maxwell equations, the solution of \citet{blandford_znajek1977} for a slowly-rotating black hole magnetosphere and the solution of \citet{wald1974} for a spinning black hole immersed in a uniform magnetic field, and for the curved spacetime hydrodynamics equations, the solution of \citet{petrich_shapiro_teukolsky1988} for an ultra-relativistic fluid accreting subsonically onto a spinning black hole. For many of these test problems, faster convergence to the analytical solution is exhibited for the special relativistic Riemann solvers than for the general relativistic ones. We also find, for many general relativistic problems involving very high black hole spin (which do not necessarily admit analytical solutions), such as an extension of the \citet{wald1974} magnetosphere problem for a rapidly-spinning black hole or the supersonic accretion of an ideal gas onto a rapidly-spinning black hole, the general relativistic Riemann solvers tend to become numerically unstable at lower values of the black hole spin than the corresponding special relativistic ones, indicating that the tetrad-first approach may also be inherently more numerically robust when dealing with highly distorted spacetime coordinate systems. Finally, we conclude in Section \ref{sec:concluding_remarks} that the tetrad-first approach therefore represents a promising robust foundation for future coupled multi-fluid simulations in strongly general relativistic regimes and discuss potential avenues for future extension, including extensions to fully dynamic spacetimes and the incorporation of more sophisticated coordinate transformations for the elimination of geometric source terms.

\subsection{Nomenclature}

Throughout this paper, Greek indices ${\mu, \nu, \rho, \sigma}$, etc. are taken to range over the spacetime coordinate basis ${\left\lbrace t, x, y, z \right\rbrace}$ or ${\left\lbrace t, r, \theta, \phi \right\rbrace}$, while Latin indices ${i, j, k, l}$, etc. are taken to range over the spatial coordinate basis ${\left\lbrace x, y, z \right\rbrace}$ or ${\left\lbrace r, \theta, \phi \right\rbrace}$. ${g_{\mu \nu}}$ is taken to denote a spacetime metric tensor, while ${\gamma_{i j}}$ is taken to denote an induced \textit{spatial} metric tensor. In cases where there is the potential for ambiguity, we use a bracketed ``${\left( 4 \right)}$'' to designate geometrical quantities defined over spacetime and ``${\left( 3 \right)}$'' to designate their purely spatial counterparts. In particular, we use the Christoffel symbols:

\begin{equation}
\begin{split}
{}^{\left( 4 \right)} \Gamma_{\mu \nu}^{\rho} &= \frac{1}{2} g^{\rho \sigma} \left( \partial_{\mu} g_{\sigma \nu} + \partial_\nu g_{\mu \sigma} - \partial_{\sigma} g_{\mu \nu} \right),\\
{}^{\left( 3 \right)} \Gamma_{i j}^{k} &= \frac{1}{2} \gamma^{k l} \left( \partial_i \gamma_{l j} + \partial_j \gamma_{i l} - \partial_l \gamma_{i j} \right),
\end{split}
\end{equation}
the Riemann tensors,

\begin{equation}
\begin{split}
{}^{\left( 4 \right)} R_{\sigma \mu \nu}^{\rho} &= \partial_{\mu} \left( {}^{\left( 4 \right)} \Gamma_{\sigma \nu}^{\rho} \right) - \partial_{\nu} \left( {}^{\left( 4 \right)} \Gamma_{\mu \sigma}^{\rho} \right) + {}^{\left( 4 \right)} \Gamma_{\mu \lambda}^{\rho} {}^{\left( 4 \right)} \Gamma_{\sigma \nu}^{\lambda}\\
&- {}^{\left( 4 \right)} \Gamma_{\lambda \nu}^{\rho} {}^{\left( 4 \right)} \Gamma_{\mu \sigma}^{\lambda},\\
{}^{\left( 3 \right)} R_{l i j}^{k} &= \partial_i \left( {}^{\left( 3 \right)} \Gamma_{l j}^{k} \right) - \partial_j \left( {}^{\left( 3 \right)} \Gamma_{i l}^{k} \right) + {}^{\left( 3 \right)} \Gamma_{i m}^{k} {}^{\left( 3 \right)} \Gamma_{l j}^{m}\\
&- {}^{\left( 3 \right)} \Gamma_{m j}^{k} {}^{\left( 3 \right)} \Gamma_{i l}^{m},
\end{split}
\end{equation}
the Ricci tensors,

\begin{equation}
{}^{\left( 4 \right)} R_{\mu \nu} = {}^{\left( 4 \right)} R_{\mu \sigma \nu}^{\sigma}, \qquad {}^{\left( 3 \right)} R_{i j} = {}^{\left( 3 \right)} R_{i k j}^{k},
\end{equation}
the Ricci scalars,

\begin{equation}
{}^{\left( 4 \right)} R = {}^{\left( 4 \right)} R_{\mu}^{\mu}, \qquad {}^{\left( 3 \right)} R = {}^{\left( 3 \right)} R_i^i,
\end{equation}
and the Einstein tensors,

\begin{equation}
\begin{split}
{}^{\left( 4 \right)} G_{\mu \nu} &= {}^{\left( 4 \right)} R_{\mu \nu} - \frac{1}{2} \left( {}^{\left( 4 \right)} R g_{\mu \nu} \right),\\
{}^{\left( 3 \right)} G_{i j} &= {}^{\left( 3 \right)} R_{i j} - \frac{1}{2} 
\left( {}^{\left( 3 \right)} R \gamma_{i j} \right).
\end{split}
\end{equation}
Unless otherwise specified, the covariant derivative operator ${\nabla}$ is assumed to be over spacetime (and therefore to use the ${{}^{\left( 4 \right)} \Gamma_{\mu \nu}^{\rho}}$ connection coefficients, rather than ${{}^{\left( 3 \right)} \Gamma_{i j}^{k}}$). We use $g$ to denote the spacetime metric determinant ${g = \mathrm{det} \left( g_{\mu \nu} \right)}$ and ${\gamma}$ to denote the spatial metric determinant ${\gamma = \mathrm{det} \left( \gamma_{i j} \right)}$. Einstein summation convention, along with a ${\left( -, +, +, + \right)}$ metric signature, is assumed throughout. We adopt geometrized units in which ${G = c = 1}$.

\section{The Algorithm for Generic Conservation Laws}
\label{sec:general_algorithm}

We begin by assuming a smooth, four-dimensional Lorentzian manifold ${\left( \mathcal{M}, g \right)}$ as our underlying spacetime structure. We proceed to decompose ${\mathcal{M}}$ via the ADM formalism of \citet{arnowitt_deser_misner1959}, using the notation of \citet{gorard2024}, into a time-ordered sequence of three-dimensional (Riemannian) spacelike hypersurfaces ${\Sigma_{t_0}}$ for ${t_0 \in \mathbb{R}}$, each equipped with an induced metric tensor ${\gamma_{i j}}$:

\begin{equation}
\begin{split}
d s^2 &= g_{\mu \nu} \, d x^{\mu} \, d x^{\nu}\\
&= \left( -\alpha^2 + \beta_i \, \beta^i \right) d t^2 + 2 \beta_i \, dt \, d x^i + \gamma_{i j} \, d x^i \, d x^j,
\end{split}
\end{equation}
where the scalar field ${\alpha}$ and three-dimensional vector field ${\beta^i}$ (known as the \textit{lapse function} and \textit{shift vector}, respectively) represent the Lagrange multipliers of the ADM formalism. The lapse function ${\alpha}$ designates the proper time distance ${d \tau}$ between corresponding points on the spacelike hypersurfaces ${\Sigma_{t_0}}$ and ${\Sigma_{t_0 + dt}}$:

\begin{equation}
d \tau \left( t_0, t_0 + dt \right) = \alpha \, dt,
\end{equation}
and the shift vector ${\beta^i}$ designates the relabeling of the spatial coordinate basis ${x^i}$ as one moves from hypersurface ${\Sigma_{t_0}}$ to hypersurface ${\Sigma_{t_0 + dt}}$:

\begin{equation}
x^i \left( t_0 + dt \right) = x^i \left( t_0 \right) - \beta^i \, dt.
\end{equation}
The four-dimensional unit normal vector ${\mathbf{n}}$ to each spacelike hypersurface ${\Sigma_{t_0}}$ is given by the contravariant derivative ${\nabla^{\mu}}$ of the time coordinate $t$:

\begin{equation}
n^{\mu} = -\alpha \, \nabla^{\mu} t = - \alpha \, g^{\mu \nu} \, \nabla_{\nu} t = -\alpha \, g^{\mu \nu} \, \partial_{\nu} t,
\end{equation}
while the three-dimensional unit \textit{time vector} ${\mathbf{t}}$ that determines how points on hypersurface ${\Sigma_{t_0}}$ map to corresponding points on hypersurface ${\Sigma_{t_0 + dt}}$ is represented in terms of a spatial projection of the normal vector ${\mathbf{n}}$ as:

\begin{equation}
t^i = \alpha \, n^i + \beta^i = -\alpha^2 \, g^{i \mu} \, \nabla_{\mu} t + \beta^i.
\end{equation}
Henceforth, we shall refer to an observer that is at rest with respect to the hypersurface ${\Sigma_{t_0}}$, and whose four-velocity ${\mathbf{u}}$ is therefore given by the normal vector ${\mathbf{n}}$, as an \textit{Eulerian observer}.

Following the approach of \citet{pons_font_ibanez_marti_miralles1998} (which we recast here into the language of orthonormal tetrads), we now consider a system of \textit{local conservation laws}, i.e., a system of hyperbolic partial differential equations, each of which takes the form of a simple divergence equation:

\begin{equation}
\nabla \cdot \mathbf{A} = S,
\end{equation}
in the local frame of an Eulerian observer. In the above, ${\mathbf{A}}$ denotes an arbitrary four-dimensional vector field, and $S$ denotes an arbitrary scalar source term. If we integrate such a local conservation law over a four-dimensional spacetime control volume ${\Omega \subseteq \mathcal{M}}$, with a three-dimensional boundary ${\partial \Omega}$, then we obtain (via Gauss's theorem) the following balance law:

\begin{equation}
\int_{\partial \Omega} \mathbf{A} \cdot d^3 \mathbf{\Sigma} = \int_{\Omega} S \, d \Omega.
\end{equation}
We define our computational grid in such a way that the spacetime control volumes ${\Omega}$ are bounded by the coordinate hypersurfaces ${\left\lbrace \Sigma_{t_0}, \Sigma_{x_0}, \Sigma_{y_0}, \Sigma_{z_0} \right\rbrace}$ and ${\left\lbrace \Sigma_{t_0 + \Delta t}, \Sigma_{x_0 + \Delta x}, \Sigma_{y_0 + \Delta y}, \Sigma_{z_0 + \Delta z} \right\rbrace}$.
Thus, if we take the integral average of the timelike component ${A^t}$ of the vector field ${\mathbf{A}}$ over a single spatial cell ${\left[ x_0, x_0 + \Delta x \right] \times \left[ y_0, y_0 + \Delta y \right] \times \left[ z_0, z_0 + \Delta z \right]}$ of volume ${\Delta \mathcal{V}}$, where:

\begin{equation}
\Delta \mathcal{V} = \int_{x_0}^{x_0 + \Delta x} \int_{y_0}^{y_0 + \Delta y} \int_{z_0}^{z_0 + \Delta z} \sqrt{-g} \, dz \, dy \, dx,
\end{equation}
then we obtain the volume-averaged quantity ${\overline{A^t}}$:

\begin{equation}
\overline{A^t} = \frac{1}{\sqrt{\Delta \mathcal{V}}} \, \int_{x_0}^{x_0 + \Delta x} \int_{y_0}^{y_0 + \Delta y} \int_{z_0}^{z_0 + \Delta z} A^t \sqrt{-g} \, dz \, dy \, dx.
\end{equation}
The volume-average ${\left( \overline{A^t} \right)_{t_0}}$ evaluated on the ${\Sigma_{t_0}}$ bounding hypersurface may now be related to the volume-average ${\left( \overline{A^t} \right)_{t_0 + \Delta t}}$ evaluated on the ${\Sigma_{t_0 + \Delta t}}$ bounding hypersurface as follows:

\begin{equation}
\begin{split}
\left( \overline{A^t} \Delta \mathcal{V} \right)_{t_0 + \Delta t} &= \left( \overline{A^t} \Delta \mathcal{V} \right)_{t_0} + \int_{\Omega} S \, d \Omega\\
&- \left( \int_{\Sigma_{x_0}} \mathbf{A} \cdot d^3 \mathbf{\Sigma} + \int_{\Sigma_{x_0 + \Delta x}} \mathbf{A} \cdot d^3 \mathbf{\Sigma} \right.\\
&+ \int_{\Sigma_{y_0}} \mathbf{A} \cdot d^3 \mathbf{\Sigma} + \int_{\Sigma_{y_0 + \Delta y}} \mathbf{A} \cdot d^3 \mathbf{\Sigma}\\
&+ \left. \int_{\Sigma_{z_0}} \mathbf{A} \cdot d^3 \mathbf{\Sigma} + \int_{\Sigma_{z_0 + \Delta z}} \mathbf{A} \cdot d^3 \mathbf{\Sigma} \right),
\end{split}
\end{equation}
which represents the explicit time evolution step of a finite-volume numerical scheme. The volume integral of the source term $S$ over the spacetime control volume ${\Omega}$ may be evaluated by means of an explicit ordinary differential equation solver, as we shall outline later. On the other hand, the hypersurface integrals:

\begin{equation}
\begin{split}
& \int_{\Sigma_{x_0}} \mathbf{A} \cdot d^3 \mathbf{\Sigma} + \int_{\Sigma_{x_0 + \Delta x}} \mathbf{A} \cdot d^3 \mathbf{\Sigma} + \int_{\Sigma_{y_0}} \mathbf{A} \cdot d^3 \mathbf{\Sigma}\\
&+ \int_{\Sigma_{y_0 + \Delta y}} \mathbf{A} \cdot d^3 \mathbf{\Sigma} + \int_{\Sigma_{z_0}} \mathbf{A} \cdot d^3 \mathbf{\Sigma} + \int_{\Sigma_{z_0 + \Delta z}} \mathbf{A} \cdot d^3 \mathbf{\Sigma}\\
&= \int_{\Sigma_{x_0}} A^x \, \sqrt{-g} \, dt \, dy \, dz + \int_{\Sigma_{x_0 + \Delta x}} A^x \, \sqrt{-g} \, dt \, dy \, dz\\
&+ \int_{\Sigma_{y_0}} A^y \, \sqrt{-g} \, dt \, dx \, dz + \int_{\Sigma_{y_0 + \Delta y}} A^y \, \sqrt{-g} \, dt \, dx \, dz\\
&+ \int_{\Sigma_{z_0}} A^z \, \sqrt{-g} \, dt \, dx \, dy + \int_{\Sigma_{z_0 + \Delta z}} A^z \, \sqrt{-g} \, dt \, dx \, dy,
\end{split}
\end{equation}
are typically evaluated by solving local Riemann problems at the boundaries of the spatial cells to determine ${A^x}$, ${A^y}$ and ${A^z}$, which designate the spacelike components of the vector field ${\mathbf{A}}$.

We elect now to simplify these local Riemann problem computations by exploiting an elementary consequence of the equivalence principle: namely that, surrounding any point ${x \in \mathcal{M}}$ in a smooth Lorentzian manifold ${\mathcal{M}}$, there must exist a local \textit{geodesic} coordinate system, with respect to which ${\mathcal{M}}$ appears locally flat (i.e., within which the Christoffel symbols ${\Gamma_{\mu \nu}^{\rho}}$, though not necessarily their derivatives, vanish). More precisely, we select a \textit{frame field} ${e_{a}^{\mu}}$ (otherwise known as an \textit{orthonormal tetrad}), with ${\mathbf{e}_t}$ being a four-dimensional timelike unit vector field and ${\mathbf{e}_x}$, ${\mathbf{e}_y}$ and ${\mathbf{e}_z}$ being orthonormal four-dimensional spacelike vector fields. The significance of such a frame field is that it effectively diagonalizes the (inverse) metric tensor:
\begin{equation}
g^{\mu \nu} = e_{a}^{\mu} \, e_{b}^{\nu} \, \eta^{a b},
\end{equation}
where ${\eta_{\mu \nu}}$ designates the standard Minkowksi metric on flat spacetime. The underlying intuition is that the ${\mu, \nu}$ indices index the original spacetime coordinate basis, while the ${a, b}$ indices index the new local \textit{tetrad basis} within which the metric now appears locally flat (i.e., Minkowski); such frame fields play an analogous role in the theory of spinor bundles to that played by the metric tensor in the theory of vector bundles. Indeed, at every point ${x \in \mathcal{M}}$, there exists an infinite equivalence class of possible orthonormal tetrads, every pair of which are related by a Lorentz transformation. In general, a particular orthonormal tetrad may be computed by means of the Gram-Schmidt process, but in the particular case under discussion here, there exists a very natural choice: we select the unit normal vector ${\mathbf{n}}$ to the spacelike hypersurface ${\Sigma_{t_0}}$ as our timelike vector ${\mathbf{e}_t}$, and the unit normal vectors to the hypersurfaces ${\Sigma_{x_0}}$, ${\Sigma_{y_0}}$ and ${\Sigma_{z_0}}$ as our spacelike vectors ${\mathbf{e}_x}$, ${\mathbf{e}_y}$ and ${\mathbf{e}_z}$. This particular choice of local tetrad basis induces a change-of-basis matrix ${M_{\mu}^{a}}$ of the general form:

\begin{equation}
M_{\mu}^{a} = \frac{\partial_{\mu}}{\mathbf{e}_{a}};
\end{equation}
for instance, in the $x$-direction, this will take the explicit form:

\begin{equation}
M_{\mu}^{a} = \begin{bmatrix}
\frac{1}{\sqrt{\gamma^{x x}}} & \frac{-\gamma^{x y} \gamma_{y y} - \gamma^{x z} \gamma_{y z}}{\gamma^{x x} \sqrt{\gamma_{y y}}} & \frac{-\gamma^{x z} \sqrt{\gamma_{y y} \gamma_{z z} - \left( \gamma_{y z} \right)^2}}{\gamma^{x x} \sqrt{\gamma_{y y}}}\\
0 & \sqrt{\gamma_{y y}} & 0\\
0 & \frac{\gamma_{y z}}{\sqrt{\gamma_{y y}}} & \frac{\sqrt{\gamma_{y y} \gamma_{z z} - \left( \gamma_{y z} \right)^2}}{\sqrt{\gamma_{y y}}}
\end{bmatrix},
\end{equation}
and likewise in the other coordinate directions\footnote{Since the finite-volume solver in \textsc{Gkeyll} adopts an operator splitting approach to the handling of higher-dimensional problems, we shall assume a dimensionally split algorithm within all that follows. The extension of the tetrad-first approach to the case of fully unsplit numerical schemes remains a topic for future investigation.}.

Suppose that ${x_{0}^{\mu}}$ denote the spacetime coordinates of the center of the inter-cell boundary hypersurface ${\Sigma_{x_0}}$ (with respect to the \textit{original} spacetime coordinate basis ${x^{\mu}}$), and suppose that we select an orthonormal tetrad at the point ${x_{0}^{\mu}}$, hence yielding the new (locally flat) spacetime coordinate basis ${\widetilde{x^a}}$, given by:

\begin{equation}
\widetilde{x^a} = M_{\mu}^{a} \left( x^{\mu} - x_{0}^{\mu} \right),
\end{equation}
where the matrix ${M_{\mu}^{a}}$ has been evaluated at ${x_{0}^{\mu}}$. Henceforth, we shall use the ${\sim}$ notation to refer to quantities that have been transformed into the ${\widetilde{x^a}}$ coordinate basis. In these coordinates, the hypersurface ${\Sigma_{x_0}}$ is defined by:

\begin{equation}
\widetilde{x} - \frac{\widetilde{\beta^{x}}}{\alpha} \widetilde{t} = 0, \qquad \text{ where } \widetilde{\beta^{a}} = M_{i}^{a} \beta^i,
\end{equation}
thus allowing us to rewrite the hypersurface integral over ${\Sigma_{x_0}}$ as:

\begin{equation}
\int_{\Sigma_{x_0}} A^x \, \sqrt{g} \, dt \, dy \, dz = \int_{\Sigma_{x_0}} \left( \widetilde{A^x} - \frac{\widetilde{\beta^x}}{\alpha} \widetilde{A^t} \right) \sqrt{-\widetilde{g}} \, d \widetilde{t} \, d \widetilde{y} \, d \widetilde{z},
\end{equation}
where the transformed four-vector ${\widetilde{A^a}}$ is given by:

\begin{equation}
\widetilde{A^a} = M_{\mu}^{a} A^{\mu},
\end{equation}
and the transformed spacetime volume factor ${\sqrt{-\widetilde{g}}}$ is of order unity (more precisely, ${\sqrt{-\widetilde{g}} = 1 + \mathcal{O} \left( \widetilde{x^a} \right)}$). The great advantage of performing this transformation is that the transformed ${\widetilde{A^a}}$ terms appearing within the hypersurface integral are evaluated with respect to the Minokwski metric ${\eta_{\mu \nu}}$ (as opposed to the original ${A^{\mu}}$ terms, which were evaluated with respect to the general metric ${g_{\mu \nu}}$), meaning that only the \textit{special relativistic} Riemann problem needs to be solved at the ${\Sigma_{x_0}}$ boundary, rather than the full general relativistic Riemann problem, which is typically much more complicated. The trade-off is that the inter-cell boundary ${\Sigma_{x_0}}$ is no longer at rest with respect to this new coordinate system, as it is in the case of a standard Riemann problem, but rather it is moving in the $x$-direction with speed ${\frac{\widetilde{\beta^x}}{\alpha}}$. This requires that we solve the Riemann problem in a comoving frame, which in turn necessitates implementing a slight generalization of a traditional Riemann solver. Fortunately, a general approach to solving Riemann problems in co-moving frames was developed by \citet{harten_hyman1983}, which we follow here.

Solving the Riemann problem across the inter-cell boundary hypersurface ${\Sigma_{x_0}}$ yields a constant value for the inter-cell flux:

\begin{equation}
\left( \widetilde{A^x} - \frac{\widetilde{\beta^x}}{\alpha} \widetilde{A^t} \right)^{*},
\end{equation}
over the entirety of ${\Sigma_{x_0}}$, simplifying our hypersurface integral down to:

\begin{equation}
\begin{split}
&\int_{\Sigma_{x_0}} \left( \widetilde{A^x} - \frac{\widetilde{\beta^x}}{\alpha} \widetilde{A^t} \right) \sqrt{-\widetilde{g}} \, d \widetilde{t} \, d \widetilde{y} \, d \widetilde{z}\\
&= \left( \widetilde{A^x} - \frac{\widetilde{\beta^x}}{\alpha} \widetilde{A^t} \right)^{*} \int_{\Sigma_{x_0}} \sqrt{-\widetilde{g}} \, d \widetilde{t} \, d \widetilde{y} \, d \widetilde{z},
\end{split}
\end{equation}
and where the resulting surface integral can now be rewritten trivially in terms of the original spacetime coordinate basis ${x^{\mu}}$ as:

\begin{equation}
\int_{\Sigma_{x_0}} \sqrt{-\widetilde{g}} \, d \widetilde{t} \, d \widetilde{y} \, d \widetilde{z} = \int_{\Sigma_{x_0}} \sqrt{\gamma^{x x}} \, \sqrt{-g} \, dt \, dy \, dz,
\end{equation}
by which we can simply multiply the inter-cell flux, as part of a post-hoc correction step. Thus, the overall algorithm (for updates in the $x$-direction) becomes:

\begin{enumerate}
\item
Select a local orthonormal tetrad ${e_{a}^{\mu}}$ at the center ${x_{0}^{\mu}}$ of the inter-cell boundary hypersurface ${\Sigma_{x_0}}$. Specifically, select ${\mathbf{e}_t = \mathbf{n}}$, and choose ${\mathbf{e}_x}$, ${\mathbf{e}_y}$ and ${\mathbf{e}_z}$ to be unit normals to the hypersurfaces ${\Sigma_{x_0}}$, ${\Sigma_{y_0}}$ and ${\Sigma_{z_0}}$, respectively.

\item
Transform all tensorial quantities into the new coordinate basis:

\begin{equation}
\widetilde{x^a} = M_{\mu}^{a} \left( x^{\mu} - x_{0}^{\mu} \right),
\end{equation}
with change-of-basis matrix:

\begin{equation}
M_{\mu}^{a} = \frac{\partial_{\mu}}{\mathbf{e}_a}.
\end{equation}

\item
Solve the \textit{special relativistic} Riemann problem across the inter-cell boundary ${\Sigma_{x_0}}$, in the co-moving frame with $x$-velocity ${{\widetilde{\beta^x}}/{\alpha}}$, to obtain the constant inter-cell \textit{special relativistic} flux:

\begin{equation}
\left( \widetilde{A^x} - \frac{\widetilde{\beta^x}}{\alpha} \widetilde{A^t} \right)^{*}.
\end{equation}

\item
Multiply this special relativistic flux in the co-moving frame by the (purely geometrical) surface integral:

\begin{equation}
\int_{\Sigma_{x_0}} \sqrt{\gamma^{x x}} \, \sqrt{-g} \, dt \, dy \, dz,
\end{equation}
to obtain the full \textit{general relativistic} flux in the original coordinate basis ${x^{\mu}}$:

\begin{equation}
\begin{split}
&\int_{\Sigma_{x^0}} \mathbf{A} \cdot d^3 \mathbf{\Sigma}\\
&= \left( \widetilde{A^x} - \frac{\widetilde{\beta^x}}{\alpha} \widetilde{A^t} \right)^{*} \int_{\Sigma_{x_0}} \sqrt{\gamma^{x x}} \, \sqrt{-g} \, dt \, dy \, dz.
\end{split}
\end{equation}
\end{enumerate}

We have implemented this general algorithm into \textsc{Gkeyll}, which already employs what might be regarded as a non-relativistic \textit{triad} formalism as part of the flux computation step of its finite-volume solver. More specifically, when solving the Riemann problem across an inter-cell boundary, \textsc{Gkeyll}'s finite-volume solver will transform all tensorial quantities into a local spatial coordinate basis ${\mathbf{e}_x}$, ${\mathbf{e}_y}$, ${\mathbf{e}_z}$, in which ${\mathbf{e}_x}$ is the unit normal to the inter-cell boundary, and ${\mathbf{e}_y}$ and ${\mathbf{e}_z}$ are orthonormal tangent vectors to the boundary. This approach ensures that inter-cell fluxes need only ever be calculated in the (local) $x$-direction, irrespective of which global coordinate direction is actually being updated, and irrespective of the global curvilinear coordinate system being used. As a result, implementing the tetrad-first algorithm into \textsc{Gkeyll} required only a very modest generalization of the existing finite-volume infrastructure to accommodate space\textit{time} coordinate transformations as well as spatial ones (thus allowing one to transform gauge quantities as well as spatial metric quantities), plus one additional step to apply the post-hoc geometrical update to the special relativistic flux (thus transforming it from a special relativistic flux in the co-moving tetrad basis to a general relativistic flux in the original coordinate basis): no other modifications were necessary, in contrast to previous attempts to extend the wave propagation method to curved manifolds, such as the approach of \citet{rossmanith_bale_leveque2004}.

\section{Case I: General Relativistic Electromagnetism}
\label{sec:gr_electromagnetism}

To set up our first illustrative example of the tetrad-first approach, we briefly summarize how Maxwell's equations in curved spacetime may be expressed in the general relativistic conservation law form described in Section \ref{sec:general_algorithm}. Following \citet{gorard2023}, let ${\mathbf{A}}$ denote the electromagnetic four-potential (i.e., a four-vector whose timelike component ${A^t}$ constitutes the electric scalar potential ${\phi}$ and whose spacelike components ${A^i}$ constitute the magnetic vector potential), with corresponding 1-form components ${A_{\mu} = g_{\mu \nu} A^{\nu}}$. Then, the electromagnetic field tensor ${F_{\mu \nu}}$ may be expressed in manifestly covariant form as:

\begin{equation}
F_{\mu \nu} = \nabla_{\mu} A_{\nu} - \nabla_{\nu} A_{\mu},
\end{equation}
although due to the anti-symmetry of ${F_{\mu \nu}}$, the Christoffel symbols in the above cancel out, allowing one to replace the covariant derivatives ${\nabla_{\mu}}$ with partial derivatives ${\partial_{\mu}}$ without breaking covariance. The homogeneous and inhomogeneous Maxwell equations in this notation become:

\begin{subequations}
\begin{align}
\nabla_{\nu} \, {}^{\star} F^{\mu \nu} &= 0,\\
\nabla_{\nu} F^{\mu \nu} &= I^{\mu},
\end{align}
\end{subequations}
respectively, with the four-current density ${\mathbf{I}}$ being given by:

\begin{equation}
I^{\mu} = \frac{1}{\mu_0} \partial_{\nu} \left( F^{\mu \nu} \sqrt{-g} \right),
\end{equation}
with vacuum permeability constant ${\mu_0}$. Assuming vanishing electric and magnetic susceptibilities, the dual electromagnetic field tensor ${{}^{\star} F^{\mu \nu}}$ is simply given by the Hodge dual of ${F_{\mu \nu}}$:

\begin{subequations}
\begin{align}
{}^{\star} F^{\mu \nu} &= \frac{\sqrt{-g}}{2} \varepsilon^{\mu \nu \alpha \beta} F_{\alpha \beta},\\
F^{\mu \nu} &= - \frac{\sqrt{-g}}{2} \varepsilon^{\mu \nu \alpha \beta} \, {}^{\star} F_{\alpha \beta},
\end{align}
\end{subequations}
with ${\varepsilon_{\mu \nu \alpha \beta}}$ being the four-dimensional Levi-Civita symbol.

Taking projections of the homogeneous Maxwell equations in the timelike and spacelike directions, we obtain:

\begin{equation}
\frac{1}{\sqrt{\gamma}} \partial_i \left( \alpha \sqrt{\gamma} \, {}^{\star} F^{t i} \right) = 0,
\end{equation}
and

\begin{equation}
\frac{1}{\sqrt{\gamma}} \partial_t \left( \alpha \sqrt{\gamma} \, {}^{\star} F^{j t} \right) + \frac{1}{\sqrt{\gamma}} \partial_i \left( \alpha \sqrt{\gamma} \, {}^{\star} F^{j i} \right) = 0,
\end{equation}
respectively. Introducing the three-dimensional vector fields ${\mathbf{B}}$ and ${\mathbf{E}}$,

\begin{equation}
B^i = \alpha \, {}^{\star} F^{i t},
\end{equation}
and

\begin{equation}
E_i = \frac{1}{2} \alpha \sqrt{\gamma} \varepsilon_{i j k} \, {}^{\star} F^{j k},
\end{equation}
where ${E^i = \gamma^{i j} E_j}$, respectively, with ${\varepsilon_{i j k}}$ now being the three-dimensional Levi-Civita symbol, these equations now take on their familiar form:

\begin{subequations}
\begin{align}
\nabla \cdot \mathbf{B} &= 0, \label{eq:elliptic_constraint1}\\
\partial_t \mathbf{B} + \nabla \times \mathbf{E} &= 0,
\end{align}
\end{subequations}
where ${\nabla}$ denotes the covariant derivative operator on spacelike hypersurfaces ${\Sigma_t}$. Likewise, taking projections of the inhomogeneous Maxwell equations in the timelike and spacelike directions yields:

\begin{equation}
\frac{1}{\sqrt{\gamma}} \partial_i \left( \alpha \sqrt{\gamma} F^{t i} \right) = \alpha I^t,
\end{equation}
and:

\begin{equation}
\frac{1}{\sqrt{\gamma}} \partial_t \left( \alpha \sqrt{\gamma} F^{j t} \right) + \frac{1}{\sqrt{\gamma}} \partial_i \left( \alpha \sqrt{\gamma} F^{j i} \right) = \alpha I^j,
\end{equation}
respectively. Introducing the three-dimensional vector fields ${\mathbf{D}}$ and ${\mathbf{H}}$,

\begin{equation}
D^i = \alpha F^{t i},
\end{equation}
and

\begin{equation}
H_i = \frac{1}{2} \alpha \sqrt{\gamma} \varepsilon_{i j k} F^{j k},
\end{equation}
where ${H^i = \gamma^{i j} H_j}$, respectively, as well as the scalar field ${\rho}$ and three-dimensional vector field ${\mathbf{J}}$ as the timelike and spacelike projections of the four-current ${\mathbf{I}}$,

\begin{subequations}
\begin{align}
\rho &= \alpha I^t,\\
J^i &= \alpha I^i,
\end{align}
\end{subequations}
respectively, these equations also take on the familiar form:

\begin{subequations}
\begin{align}
\nabla \cdot \mathbf{D} &= \rho, \label{eq:elliptic_constraint2}\\
-\partial_t \mathbf{D} + \nabla \times \mathbf{H} &= \mathbf{J}.
\end{align}
\end{subequations}

\begin{figure}[ht]
\centering
\includegraphics[width=0.45\textwidth]{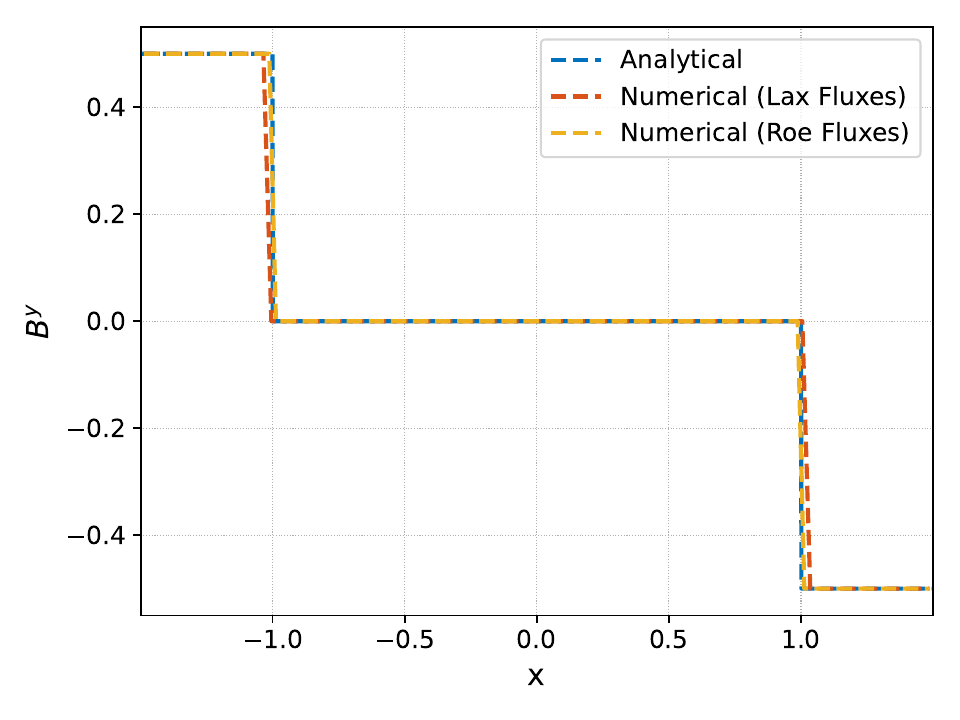}
\caption{The $y$-component of the magnetic field at time ${t = 1}$ for the current sheet Riemann problem, validating the approximate Riemann solver against the analytical solution.}
\label{fig:current_sheet_By}
\end{figure}

\begin{figure}[ht]
\centering
\includegraphics[width=0.45\textwidth]{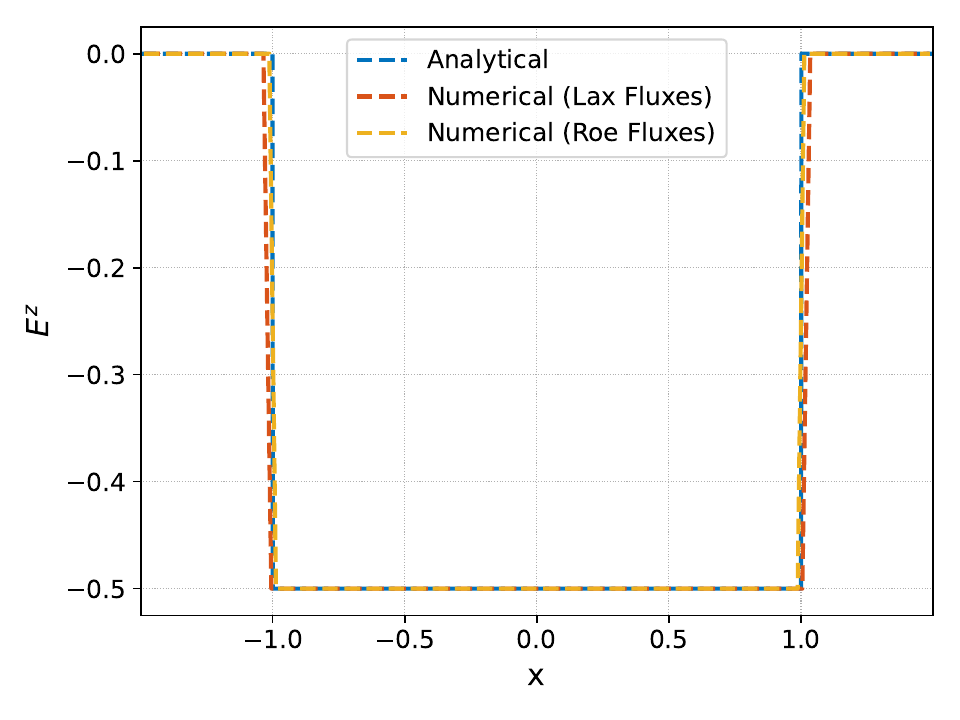}
\caption{The $z$-component of the electric field at time ${t = 1}$ for the current sheet Riemann problem, validating the approximate Riemann solver against the analytical solution.}
\label{fig:current_sheet_Ez}
\end{figure}

The ${\mathbf{D}}$ and ${\mathbf{B}}$ fields represent the electric and magnetic fields perceived by Eulerian observers; ${\mathbf{E}}$ and ${\mathbf{H}}$ are simply auxiliary fields with the vacuum constitutive relations:

\begin{subequations}
\begin{align}
\mathbf{E} &= \alpha \mathbf{D} + \boldsymbol\beta \times \mathbf{B},\\
\mathbf{H} &= \alpha \mathbf{B} - \boldsymbol\beta \times \mathbf{D},
\end{align}
\end{subequations}
respectively. Neglecting the elliptic constraint equations \ref{eq:elliptic_constraint1} and \ref{eq:elliptic_constraint2}, since these are not evolved in time\footnote{The flat spacetime Maxwell solver in \textsc{Gkeyll} uses a hyperbolic divergence cleaning approach for rectifying constraint errors; many other codes adopt a constrained transport approach instead. For the test cases in curved spacetime considered within this paper, the divergence errors remain sufficiently small that for simplicity we do not attempt to correct them.}, and neglecting the charge/current source terms ${\rho}$ and ${\mathbf{J}}$, since these only become relevant when coupling Maxwell's equations to matter fields, we obtain the following conservation law form of the general relativistic Maxwell equations:

\begin{equation}
\partial_t \begin{bmatrix}
D^x\\
D^y\\
D^z\\
B^x\\
B^y\\
B^z
\end{bmatrix} + \partial_x \begin{bmatrix}
0\\
H^z\\
-H^y\\
0\\
-E^z\\
E^y
\end{bmatrix} + \partial_y \begin{bmatrix}
-H^z\\
0\\
H^x\\
E^z\\
0\\
-E^x\\
\end{bmatrix} + \partial_z \begin{bmatrix}
H^y\\
-H^x\\
0\\
-E^y\\
E^x\\
0
\end{bmatrix} = \mathbf{0}.
\end{equation}
The first step in validating a tetrad-first algorithm for solving these equations is to validate the underlying special relativistic Riemann solver, to ensure that we can robustly handle the flat spacetime case where ${\alpha = 1}$, ${\boldsymbol\beta = \mathbf{0}}$, and therefore ${\mathbf{B} = \mathbf{H}}$ and ${\mathbf{E} = \mathbf{D}}$. Fortunately, the existing finite-volume algorithms in \textsc{Gkeyll} for solving the flat spacetime Maxwell equations have already been extensively validated against standard Riemann problems. As an illustrative example, we consider the current sheet Riemann problem of \citet{komissarov2004}, with the following initial conditions:

\begin{equation}
\mathbf{E} = \mathbf{0}, \qquad B^z = 0, \qquad B^x = 1,
\end{equation}
and

\begin{equation}
B^y = \begin{cases}
B_0, \qquad &\text{ for } \qquad x < 0,\\
-B_0, \qquad &\text{ for } \qquad x > 0.
\end{cases}
\end{equation}
This Riemann problem admits a simple analytical solution consisting of two fast electromagnetic waves, one propagating in each direction from the interface. Figures \ref{fig:current_sheet_By} and \ref{fig:current_sheet_Ez} show the field components ${B^y}$ and ${E^z}$ at time ${t = 1}$, assuming magnetic field strength ${B_0 = 0.5}$, and a symmetric domain ${x \in \left[ -1.5, 1.5 \right]}$. The numerical solutions were produced using both Lax-Friedrichs and Roe approximations to the inter-cell flux function ${F_{i + \frac{1}{2}}}$ (i.e., the flux across the boundary between cells $i$ and ${i + 1}$). The two approximate Riemann solvers were run with a CFL coefficient of ${1.0}$, using a spatial discretization of 100 cells, and both show excellent agreement with the analytical solution, with marginally higher levels of numerical diffusion observed in the case of the Lax-Friedrichs fluxes.

The Lax-Friedrichs approximate Riemann solver approximates the inter-cell flux function ${\mathbf{F}_{i + \frac{1}{2}}}$ as:

\begin{equation}
\mathbf{F}_{i + \frac{1}{2}} = \frac{1}{2} \left[ \mathbf{F} \left( \mathbf{U}_i \right) + \mathbf{F} \left( \mathbf{U}_{i + 1} \right) \right] - \frac{\Delta x}{2 \Delta t} \left( \mathbf{U}_{i + 1} - \mathbf{U}_i \right),
\end{equation}
where ${\mathbf{U}_i}$ denotes the value of the conserved variable vector at the center of cell $i$. On the other hand, the approximate Riemann solver of \citet{roe1981} approximates ${\mathbf{F}_{i + \frac{1}{2}}}$ as:

\begin{equation}
\mathbf{F}_{i + \frac{1}{2}} = \frac{1}{2} \left[ \mathbf{F} \left( \mathbf{U}_i \right) + \mathbf{F} \left( \mathbf{U}_{i + 1} \right) \right] - \frac{1}{2} \sum_p \left\lvert \lambda_p \right\rvert \alpha_p \mathbf{r}_p,
\end{equation}
where ${\lambda_p}$ and ${\mathbf{r}_p}$ denote the eigenvalues and (right) eigenvectors of the inter-cell \textit{Roe matrix} ${A_{i + \frac{1}{2}}}$, obtained as a linearized form of the flux Jacobian $A$ in the following linear approximation,

\begin{equation}
\partial_t \mathbf{U} + A \partial_x \mathbf{U} = \mathbf{0},
\end{equation}
to the full non-linear Riemann problem,

\begin{equation}
\partial_t \mathbf{U} + \partial_x \mathbf{F} \left( \mathbf{U} \right) = \mathbf{0},
\end{equation}
and ${\alpha_p}$ are the coefficients of the decomposition,

\begin{equation}
\mathbf{U}_{i + 1} - \mathbf{U}_i = \sum_p \alpha_p \mathbf{r}_p.
\end{equation}
Therefore, the only two ingredients required for the implementation of a Roe solver for an arbitrary system of hyperbolic partial differential equations in conservation law form are knowledge of the eigensystem of the flux Jacobian and a self-consistent procedure for averaging the conserved variables between cells $i$ and ${i + 1}$ to obtain a constant inter-cell Roe matrix ${A_{i + \frac{1}{2}}}$. For the case of the flat spacetime Maxwell equations, the flux Jacobian,

\begin{equation}
\mathcal{J}^{flat} = \frac{\partial \begin{bmatrix}
0 & B^z & -B^y & 0 & -E^z & E^y
\end{bmatrix}^{\intercal}}{\partial \begin{bmatrix}
E^x & E^y & E^z & B^x & B^y & B^z
\end{bmatrix}^{\intercal}},
\end{equation}
has eigenvalues ${\lambda_{-} = -1}$, ${\lambda_{0} = 0}$ and ${\lambda_{+} = 1}$ (each with algebraic multiplicity 2), with corresponding right eigenvectors,

\begin{equation}
\begin{split}
\mathbf{r}_{-}^{1} &= \begin{bmatrix}
0\\
-1\\
0\\
0\\
0\\
1
\end{bmatrix}, \qquad \mathbf{r}_{-}^{2} = \begin{bmatrix}
0\\
0\\
1\\
0\\
1\\
0
\end{bmatrix}, \qquad \mathbf{r}_{0}^{1} = \begin{bmatrix}
0\\
0\\
0\\
1\\
0\\
0
\end{bmatrix},\\
\mathbf{r}_{0}^{2} &= \begin{bmatrix}
1\\
0\\
0\\
0\\
0\\
0
\end{bmatrix}, \qquad \mathbf{r}_{+}^{1} = \begin{bmatrix}
0\\
1\\
0\\
0\\
0\\
1
\end{bmatrix}, \qquad \mathbf{r}_{+}^{2} = \begin{bmatrix}
0\\
0\\
-1\\
0\\
1\\
0
\end{bmatrix}.
\end{split}
\end{equation}
On the other hand, for the case of the curved spacetime Maxwell equations, the flux Jacobian,

\begin{equation}
\mathcal{J}^{curved} = \frac{\partial \begin{bmatrix}
0 & H^z & -H^y & 0 & -E^z & E^y
\end{bmatrix}^{\intercal}}{\partial \begin{bmatrix}
D^x & D^y & D^z & B^x & B^y & B^z
\end{bmatrix}^{\intercal}},
\end{equation}
has eigenvalues ${\lambda_{-} = -\alpha - \beta^x}$, ${\lambda_{0} = 0}$ and ${\lambda_{+} = \alpha - \beta^x}$ (again, each with algebraic multiplicity 2), with the right eigenvectors ${\mathbf{r}_{-}^{1}}$, ${\mathbf{r}_{-}^{2}}$, ${\mathbf{r}_{+}^{1}}$ and ${\mathbf{r}_{+}^{2}}$ being identical to their flat spacetime counterparts, and with ${\mathbf{r}_{0}^{1}}$ and ${\mathbf{r}_{0}^{2}}$ now being given by:

\begin{equation}
\mathbf{r}_{0}^{1} = \begin{bmatrix}
- \frac{\left( \alpha^2 - \left( \beta^x \right)^2 \right) \beta^y}{\alpha \left( \left( \beta^y \right)^2 + \left( \beta^z \right)^2 \right)}\\[1em]
- \frac{- \left( \beta^x \right)^2 \left( \beta^y \right)^2 - \alpha^2 \left( \beta^z \right)^2}{\alpha \beta^x \left( \left( \beta^y \right)^2 + \left( \beta^z \right)^2 \right)}\\[1em]
- \frac{\beta^y \left( \alpha^2 \beta^z - \left( \beta^x \right)^2 \beta^z \right)}{\alpha \beta^x \left( \left( \beta^y \right)^2 + \left( \beta^z \right)^2 \right)}\\[1em]
- \frac{\alpha^2 \beta^z - \left( \beta^x \right)^2 \beta^z}{\beta^x \left( \left( \beta^y \right)^2 + \left( \beta^z \right)^2 \right)}\\
0\\
1
\end{bmatrix}, \quad \mathbf{r}_{0}^{2} = \begin{bmatrix}
\frac{\left( \alpha^2 - \left( \beta^x \right)^2 \right) \beta^z}{\alpha \left( \left( \beta^y \right)^2 + \left( \beta^z \right)^2 \right)}\\[1em]
\frac{\left( \alpha^2 \beta^y - \left( \beta^x \right)^2 \beta^y \right) \beta^z}{\alpha \beta^x \left( \left( \beta^y \right)^2 + \left( \beta^z \right)^2 \right)}\\[1em]
- \frac{\alpha^2 \left( \beta^y \right)^2 + \left( \beta^x \right)^2 \left( \beta^z \right)^2}{\alpha \beta^x \left( \left( \beta^y \right)^2 + \left( \beta^z \right)^2 \right)}\\[1em]
- \frac{\alpha^2 \beta^y - \left( \beta^x \right)^2 \beta^y}{\beta^x \left( \left( \beta^y \right)^2 + \left( \beta^z \right)^2 \right)}\\
1\\
0
\end{bmatrix}.
\end{equation}
For Maxwell's equations (in both flat and curved spacetimes), no inter-cell averaging process is required, and the Riemann problem between cells $i$ and ${i + 1}$ can instead be solved exactly. All simulations described within this section were run using \textit{both} Lax-Friedrichs and Roe fluxes, in order to verify that the post-hoc correction to the special relativistic fluxes remains robust \textit{irrespective} of the choice of underlying special relativistic Riemann solver. However, only the results using the Roe fluxes are shown, due to their favorable convergence properties and lower levels of numerical diffusion compared to the Lax-Friedrichs approximation. Both Riemann solvers have been implemented in both their special relativistic and general relativistic forms, since this enables more direct comparison of our tetrad-first approach against standard curved spacetime Riemann solver approaches. Although both the Lax-Friedrichs and Roe Riemann solvers are nominally first-order, \textsc{Gkeyll} nevertheless achieves overall second-order convergence using the MUSCL reconstruction approach of \citet{vanleer1979}.

\subsection{The Blandford-Znajek Black Hole Magnetosphere}

\citet{blandford_znajek1977} applied methods of perturbation theory to derive an analytical solution to the equations of force-free electrodynamics for the magnetosphere of a slowly-rotating black hole. They start from the Kerr metric for an uncharged, spinning black hole with mass $M$ and dimensionless spin ${a = J / M}$ in the spherical Kerr-Schild coordinate system ${\left\lbrace t, r, \theta, \phi \right\rbrace}$,

\begin{equation}
\begin{split}
ds^2 &= g_{\mu \nu} \, d x^{\mu} \, d x^{\nu}\\
&= - \left( 1 - \frac{2 M r}{\Sigma} \right) d t^2 + \left( \frac{4 M r}{\Sigma} \right) dr \, dt\\
&+ \left( 1 + \frac{2 M r}{\Sigma} \right) d r^2 + \Sigma \, d \theta^2 - \frac{4 a M r \sin^2 \left( \theta \right)}{\Sigma} \, d \phi \, dt\\
&-2a \left( 1 + \frac{2 M r}{\Sigma} \right) \sin^2 \left( \theta \right) d \phi \, dr\\
&+ \left( \Delta + \frac{2 M r \left( r^2 + a^2 \right)}{\Sigma} \right) \sin^2 \left( \theta \right) d \phi^2,
\end{split}
\end{equation}
where we have defined

\begin{subequations}
\begin{align}
\Sigma &= r^2 + a^2 \cos^2 \left( \theta \right),\\
\Delta &= \left( r - r_{+} \right) \left( r - r_{-} \right),
\end{align}
\end{subequations}
where ${r_{-}}$ and ${r_{+}}$ denote the interior and exterior black hole horizons, respectively,

\begin{equation}
r_{\pm} = M \pm \sqrt{M^2 - a^2},
\end{equation}
and where $J$ is the (dimensional) angular momentum of the black hole. They then demand stationarity and axisymmetry for the electromagnetic four-potential ${\mathbf{A}}$ (i.e., ${\partial_t A_{\mu} = 0}$ and ${\partial_{\phi} A_{\mu} = 0}$). They introduce a function ${\psi \left( r, \theta \right)}$ designating the magnetic flux through a circular loop surrounding the black hole spin axis and intersecting the point ${\left( r, \theta \right)}$,

\begin{equation}
\psi \left( r, \theta \right) = A_{\phi} \left( r, \theta \right),
\end{equation}
as well as a function ${\Omega \left( r, \theta \right)}$ designating the angular velocity of the magnetic field lines,

\begin{subequations}
\begin{align}
\partial_r A_t &= -\Omega \, \partial_r \psi,\\
\partial_{\theta} A_t &= - \Omega \, \partial_{\theta} \psi.
\end{align}
\end{subequations}
Finally, they also introduce functions $I$ and ${B^{\phi}}$ designating the total electric current flowing through the loop, and the toroidal magnetic field, respectively,

\begin{subequations}
\begin{align}
I &= \sqrt{-g} F^{\theta r},\\
B^{\phi} &= \frac{1}{\sqrt{-g}} F_{r \theta};
\end{align}
\end{subequations}
collectively, these functions obey the integrability conditions,

\begin{subequations}
\begin{align}
\partial_r \Omega \, \partial_{\theta} \psi &= \partial_{\theta} \Omega \, \partial_r \psi,\\
\partial_r I \, \partial_{\theta} \psi &= \partial_r \psi \, \partial_{\theta} I,
\end{align}
\end{subequations}
as well as the stream equation,

\begin{equation}
-\Omega \partial_{\mu} \left( \sqrt{-g} \, F^{t \mu} \right) + \partial_{\mu} \left( \sqrt{-g} \, F^{\phi \mu} \right) + F_{r \theta} \frac{d I}{d \psi} = 0.
\end{equation}

\citet{blandford_znajek1977} require that the magnetic flux ${\psi}$ vanish at ${\theta = 0}$ (the black hole rotation axis), and that both ${\psi}$ and the toroidal field,

\begin{equation}
B^{\phi} = - \frac{I \Sigma + \left( 2 M \Omega r - a \right) \sin \left( \theta \right) \partial_{\theta} \psi}{\Delta \Sigma \sin^2 \left( \theta \right)},
\end{equation}
be regular at the outer horizon ${r = r_{+}}$. They then proceed to expand these functions perturbatively in powers of the spin parameter ${a / M}$:

\begin{equation}
\begin{split}
\psi &= \psi^{ \left( 0 \right)} + \left( \frac{a}{M} \right)^2 \psi^{\left( 2 \right)} + \mathcal{O} \left( \left( \frac{a}{M} \right)^4 \right),\\
2 M \Omega &= \left( \frac{a}{M} \right) \Omega^{\left( 1 \right)} + \left( \frac{a}{M} \right)^3 \Omega^{\left( 3 \right)} + \mathcal{O} \left( \left( \frac{a}{M} \right)^5 \right),\\
2 M I &= \left( \frac{a}{M} \right) I^{\left( 1 \right)} + \left( \frac{a}{M} \right)^3 I^{\left( 3\right)} + \mathcal{O} \left( \left( \frac{a}{M} \right)^5 \right),\\
B^{\phi} &= \left( \frac{a}{M} \right) B^{\phi \left( 1 \right)} + \left( \frac{a}{M} \right)^3 B^{\phi \left( 3 \right)} + \mathcal{O} \left( \left( \frac{a}{M} \right)^5 \right).
\end{split}
\end{equation}
To zeroth-order, this yields a static monopole solution:

\begin{equation}
\psi^{\left( 0 \right)} = 1 - \cos \left( \theta \right).
\end{equation}
To first-order, this yields a monopole solution that is co-rotating with the event horizon, first derived in the flat spacetime case by \citet{michel1973}:

\begin{equation}
\begin{split}
I^{\left( 1 \right)} &= \frac{1 - 2 \Omega^{\left( 1 \right)}}{2} \sin^2 \left( \theta \right),\\
B^{\phi \left( 1 \right)} &= - \frac{1 - 2 \Omega^{\left( 1 \right)}}{4 M r^2} - \frac{1}{2 r^3},
\end{split}
\end{equation}
and so on. This existence of a perturbative analytical solution for small spin ${a \ll 1}$ makes the \citet{blandford_znajek1977} magnetosphere problem an excellent test case for validating a general relativistic Maxwell solver.

The curved spacetime Maxwell solver in \textsc{Gkeyll} represents Kerr black holes using the Cartesian Kerr-Schild coordinate system ${\left\lbrace t, x, y, z \right\rbrace}$, with the spacetime metric tensor ${g_{\mu \nu}}$ computed as a perturbation of the Minkowski metric ${\eta_{\mu \nu}}$:

\begin{equation}
g_{\mu \nu} = \eta_{\mu \nu} - V l_{\mu} l_{\nu}.
\end{equation}
The scalar $V$ and null (co)vector ${l_{\mu} = l^{\mu}}$ are given by,

\begin{equation}
V = -\frac{M R^3}{R^4 + M^2 a^2 z^2},
\end{equation}
and

\begin{subequations}
\begin{align}
l_0 &= l^0 = -1,\\
l_1 &= l^1 = - \frac{R x + a M y}{R^2 + M^2 a^2},\\
l_2 &= l^2 = - \frac{R y - a M x}{R^2 + M^2 a^2},\\
l_3 &= l^3 = - \frac{z}{R},
\end{align}
\end{subequations}
respectively, with the generalized radial quantity $R$ defined implicitly via the algebraic equation,

\begin{equation}
\frac{x^2 + y^2}{R^2 + M^2 a^2} + \frac{z^2}{R^2} = 1,
\end{equation}
which has the explicit (positive) solution:

\begin{equation}
\begin{split}
R &= \left( \frac{1}{\sqrt{2}} \right) \sqrt{Q},\\
Q &= x^2 + y^2 + z^2 - M^2 a^2\\
&+ \sqrt{\left( x^2 + y^2 + z^2 - M^2 a^2 \right)^2 + 4 M^2 a^2 z^2}.
\end{split}
\end{equation}
We foliate the Kerr spacetime using the lapse and shift conditions:

\begin{subequations}
\begin{align}
\alpha &= \frac{1}{\sqrt{1 - 2V}},\\
\beta^i &= \frac{2V}{1 - 2V} l^i,
\end{align}
\end{subequations}
respectively, yielding a sequence of spacelike hypersurfaces, each with the induced metric tensor ${\gamma_{i j}}$ given by perturbation of the flat (Euclidean) metric ${\delta_{i j}}$,

\begin{equation}
\gamma_{i j} = \delta_{i j} - 2 V l_i l_j.
\end{equation}
These Cartesian Kerr-Schild coordinates are related to the spherical Kerr-Schild coordinates used within the \citet{blandford_znajek1977} setup by the coordinate transformation:

\begin{equation}
\begin{split}
x &= \left[ r \cos \left( \phi \right) - a \sin \left( \phi \right) \right] \sin \left( \theta \right)\\
&= \sqrt{r^2 + a^2} \sin \left( \theta \right) \cos \left( \phi - \arctan \left( \frac{a}{r} \right) \right),\\
y &= \left[ r \sin \left( \phi \right) + a \cos \left( \phi \right) \right] \sin \left( \theta \right)\\
&= \sqrt{r^2 + a^2} \sin \left( \theta \right) \sin \left( \phi - \arctan \left( \frac{a}{r} \right) \right),\\
z &= r \cos \left( \theta \right).
\end{split}
\end{equation}
To prevent numerical instabilities and unphysical waves propagating from the interior of the black hole, \textsc{Gkeyll} imposes excision boundary conditions at the (outer) horizon:

\begin{equation}
r = r_{+} = M + \sqrt{M^2 - a^2},
\end{equation}
and a region of zero flux in the interior ${r < r_{+}}$.

\begin{figure}[ht]
\centering
\includegraphics[trim={1cm, 0.5cm, 1cm, 0.5cm}, clip, width=0.45\textwidth]{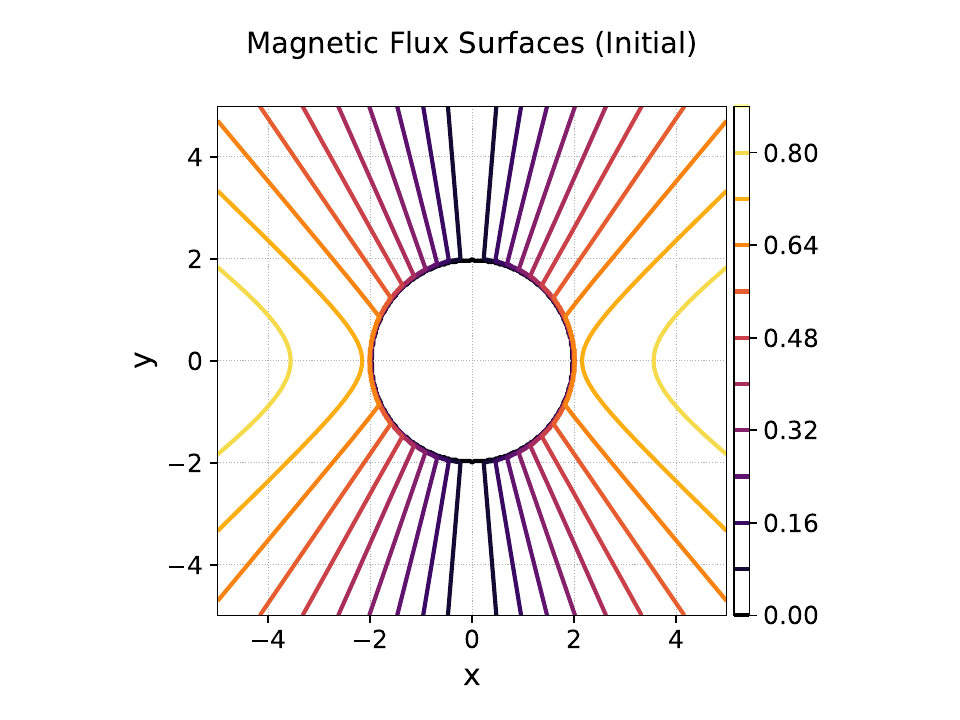}
\caption{The initial configuration of magnetic flux surfaces at time ${t = 0}$ for the Blandford-Znajek black hole magnetosphere problem with ${B_0 = 1}$.}
\label{fig:bz_monopole_slow_B_initial}
\end{figure}

\begin{figure*}[htp]
\centering
\subfigure{\includegraphics[trim={1cm, 0.5cm, 1cm, 0.5cm}, clip, width=0.45\textwidth]{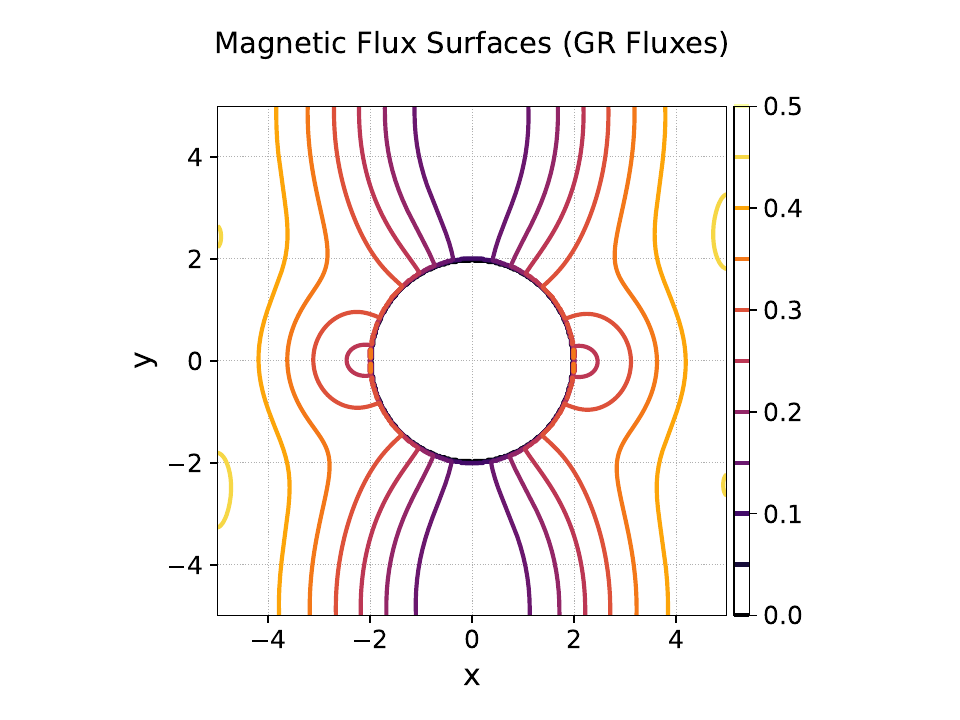}}
\subfigure{\includegraphics[trim={1cm, 0.5cm, 1cm, 0.5cm}, clip, width=0.45\textwidth]{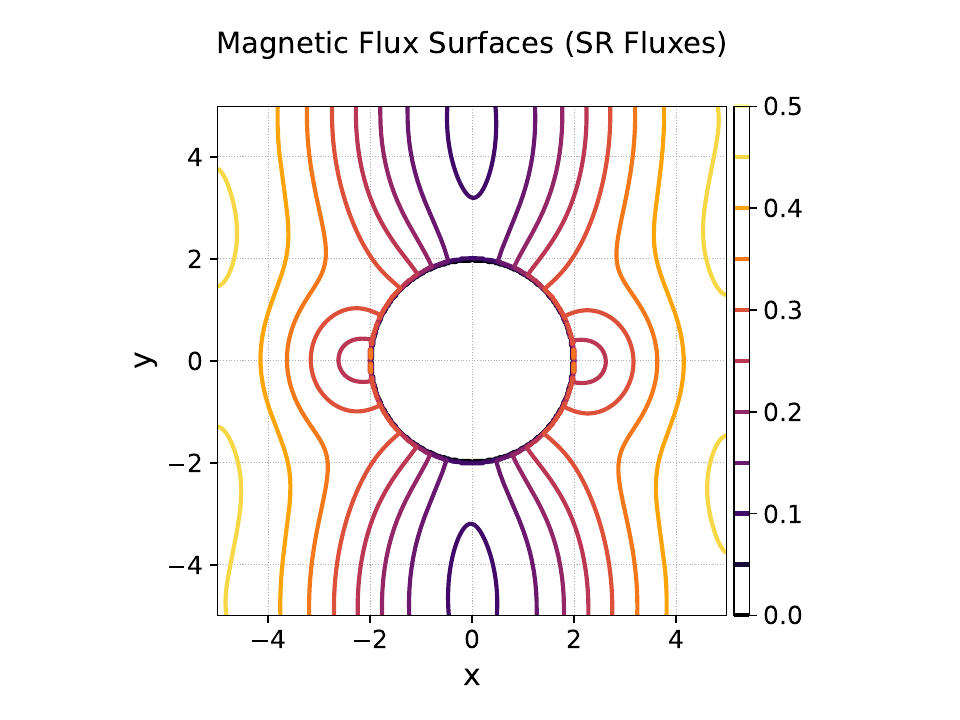}}
\caption{The configuration of magnetic flux surfaces at time ${t = 50}$ for the Blandford-Znajek magnetosphere problem for a slowly-rotating black hole with ${B_0 = 1}$ and ${a = 0.1}$, obtained using standard general relativistic fluxes (left) and local special relativistic fluxes in an orthonormal tetrad basis (right).}
\label{fig:bz_monopole_slow_B}
\end{figure*}

\begin{figure*}[htp]
\centering
\subfigure{\includegraphics[width=0.45\textwidth]{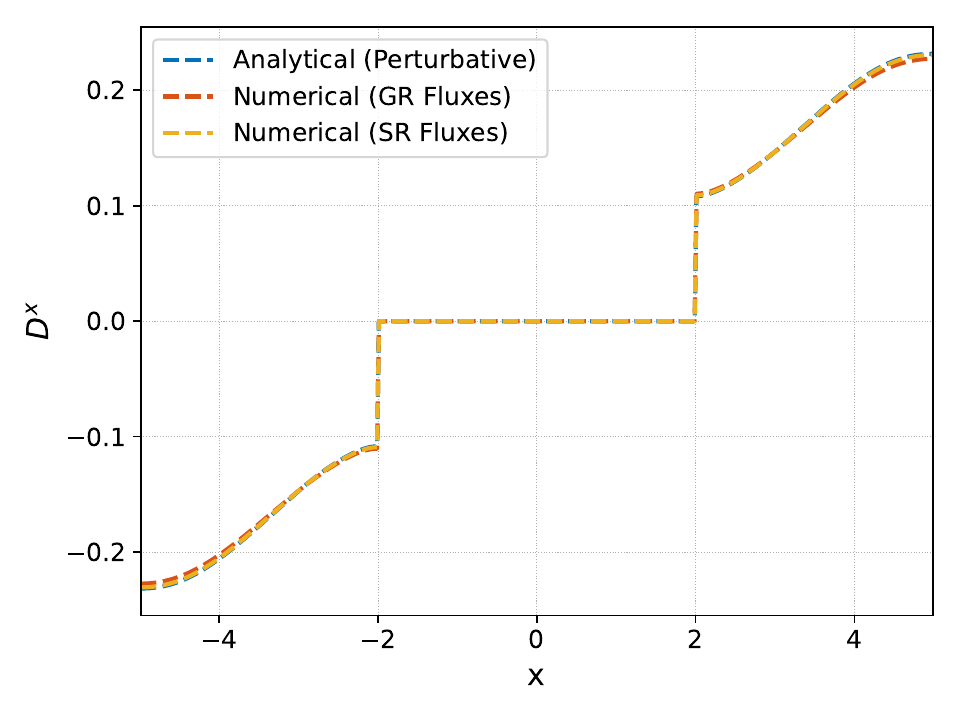}}
\subfigure{\includegraphics[width=0.45\textwidth]{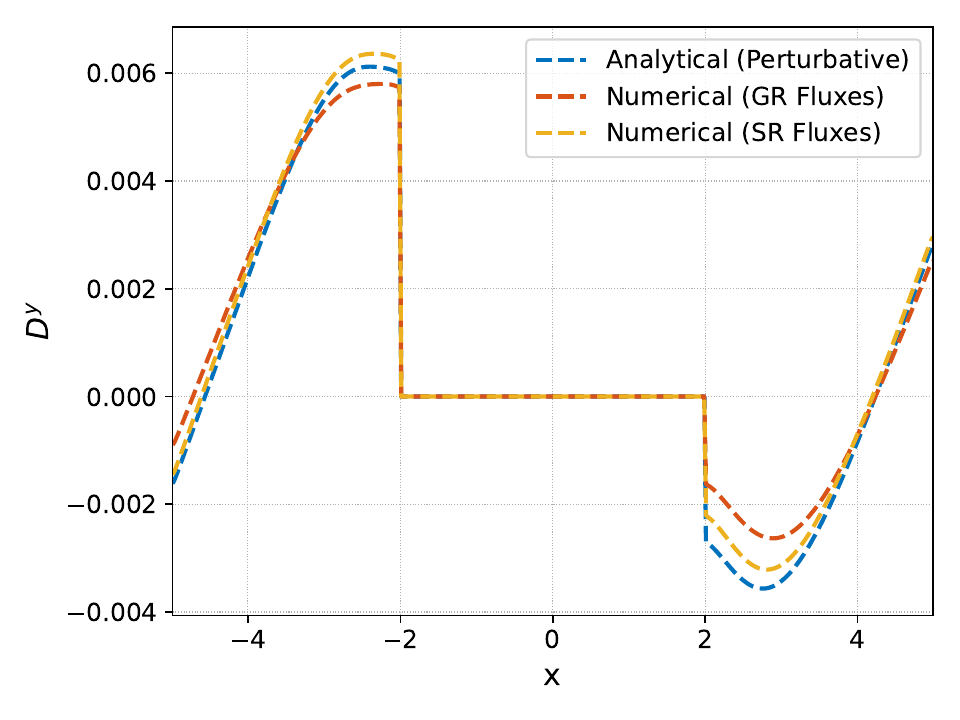}}
\caption{A cross-sectional profile of the $x$-component (left) and $y$-component (right) of the electric field ${\mathbf{D}}$ (as perceived by Eulerian observers) at time ${t = 50}$ for the Blandford-Znajek magnetosphere problem for a slowly-rotating black hole with ${B_0 = 1}$ and ${a = 0.1}$, validating both the special relativistic and general relativistic Riemann solver approaches against the analytical (perturbative) solution. Cross sections are taken at ${y = 0}$.}
\label{fig:bz_monopole_slow_D}
\end{figure*}

\begin{figure*}[htp]
\centering
\subfigure{\includegraphics[width=0.45\textwidth]{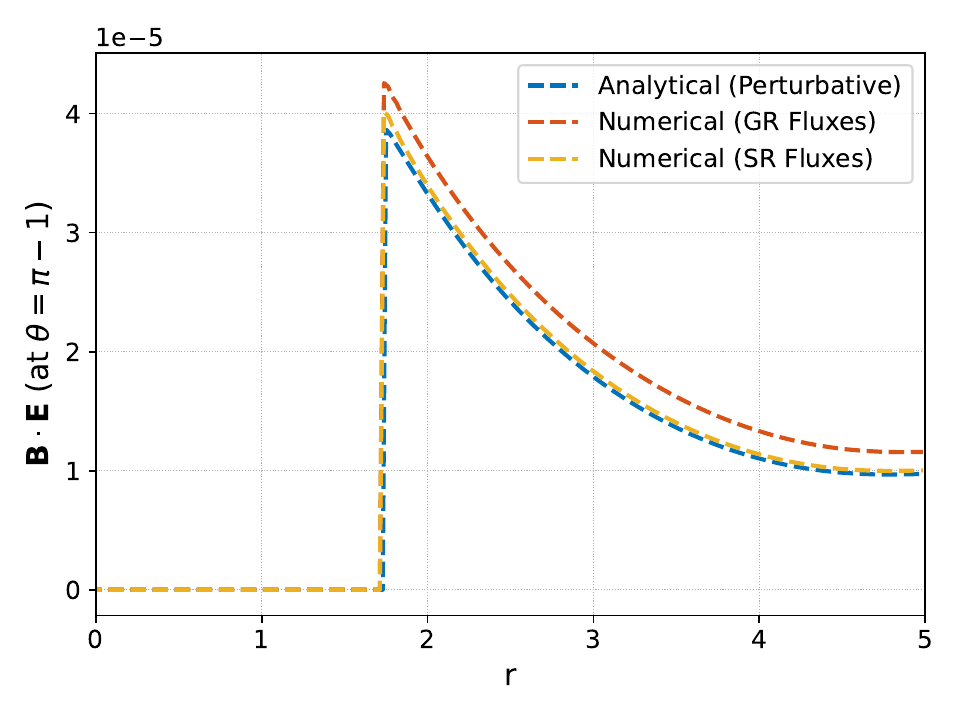}}
\subfigure{\includegraphics[width=0.45\textwidth]{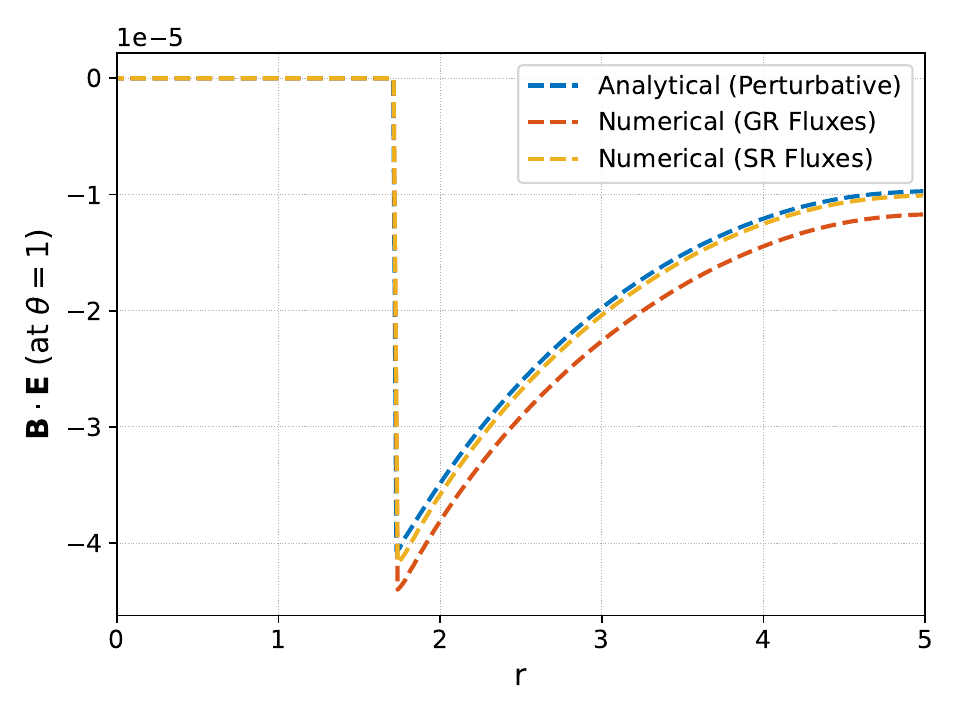}}
\caption{A radial profile of the Lorentz invariant scalar quantity ${\mathbf{E} \cdot \mathbf{B}}$ along ${\theta = \pi - 1}$ (left) and ${\theta = 1}$ (right) at time ${t = 50}$ for the Blandford-Znajek magnetosphere problem for a slowly-rotating black hole with ${B_0 = 1}$ and ${a = 0.1}$, comparing the special relativistic and general relativistic Riemann solver approaches against the analytical (perturbative) solution. The purpose of this figure is to facilitate direct comparison against Figure 2 (c) in \citet{komissarov2004}.}
\label{fig:bz_monopole_slow_EdotB}
\end{figure*}

For our magnetosphere simulations based upon the \citet{blandford_znajek1977} setup, we initialize a purely radial magnetic field,

\begin{equation}
B^r = \frac{1}{\sqrt{\gamma}} \left[ B_0 \sin \left( \theta \right) \right],
\end{equation}
adapted from the flat spacetime monopole solution of \citet{michel1973}, surrounding a slowly-rotating black hole of mass ${M = 1}$ and spin ${a = 0.1}$ (so-chosen because the numerical solution was found by \citet{komissarov2001} to be in very close agreement with the perturbative solution in this case). Agreement with the flat spacetime monopole solution of \citet{michel1973}, when combined with the requirement that the toroidal magnetic field ${B^{\phi}}$ remain finite across the event horizon (i.e., the \textit{Znajek condition} of \citet{znajek1977}), further necessitates that:

\begin{equation}
H_{\phi} = -\frac{1}{8} a B_0 \sin^2 \left( \theta \right).
\end{equation}
The initial configuration of magnetic flux surfaces, assuming reference magnetic field strength ${B_0 = 1}$ and a symmetric domain ${\left( x, y \right) \in \left[ -5, 5 \right] \times \left[ -5, 5 \right]}$, is shown in Figure \ref{fig:bz_monopole_slow_B_initial}. All simulations are run with a CFL coefficient of 0.95, using a spatial discretization of ${400 \times 400}$ cells, and up to a final time of ${t = 50}$, first using a standard general relativistic Riemann solver and then using a local special relativistic Riemann solver adapted to an orthonormal tetrad basis (as shown in Figure \ref{fig:bz_monopole_slow_B}). Visually, the results obtained using the local special relativistic Riemann solver appear sharper and better resolved, even at the relatively coarse ${400 \times 400}$ resolution of the simulation domain, than those obtained using the standard general relativistic Riemann solver. This result can be confirmed quantitatively by examining the cross-sectional profiles of certain field components through the ${y = 0}$ axis, such as the $x$- and $y$-components of the electric field vector ${\mathbf{D}}$ perceived by Eulerian observers, as shown in Figure \ref{fig:bz_monopole_slow_D}. The ${D^y}$ component in particular shows significantly better convergence to the perturbative Blandford-Znajek solution in the near-horizon region. This result is likely because the characteristic wave-speeds appearing in the special relativistic Riemann problem are -1 and 1, while in the general relativistic Riemann problem they are ${-\alpha - \beta^x}$ and ${\alpha - \beta^x}$. Therefore, in regions of high spacetime curvature in which ${\alpha \ll 1}$ (such as near a black hole horizon), a special relativistic Riemann solver is able to use more uniform wave-speed estimates across the entire simulation domain, resulting in faster and more uniform convergence overall. We also show a radial profile of the Lorentz invariant scalar quantity ${\mathbf{E} \cdot \mathbf{B}}$ along ${\theta = \pi - 1}$ and ${\theta = 1}$ in Figure \ref{fig:bz_monopole_slow_EdotB}, so as to facilitate direct comparison against Figure 2 (c) of \citet{komissarov2004}.

\begin{figure*}[htp]
\centering
\subfigure{\includegraphics[trim={1cm, 0.5cm, 1cm, 0.5cm}, clip, width=0.45\textwidth]{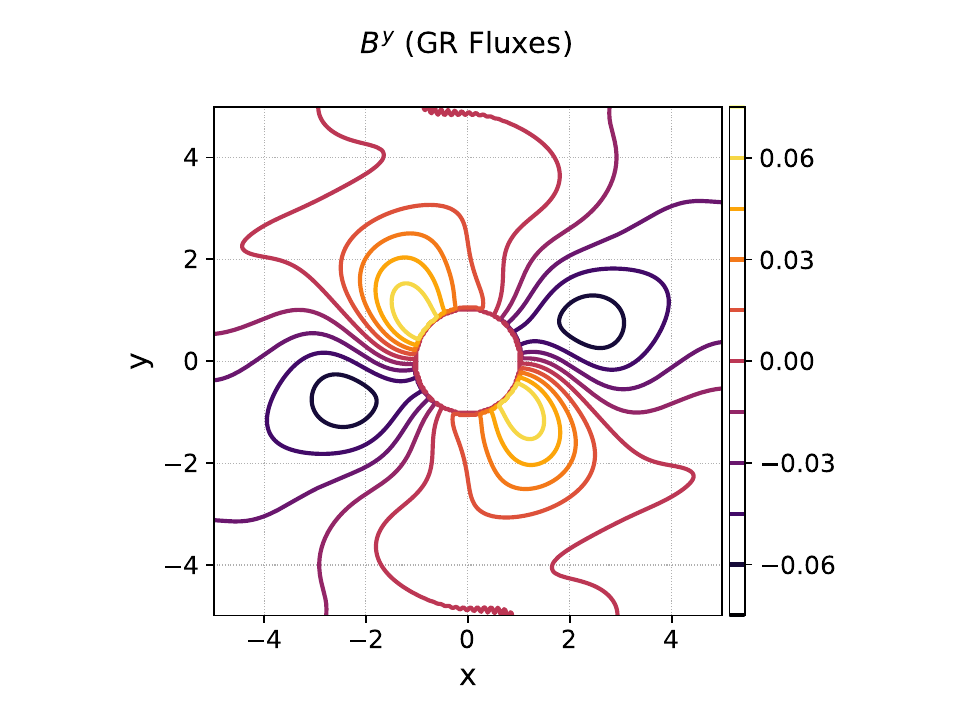}}
\subfigure{\includegraphics[trim={1cm, 0.5cm, 1cm, 0.5cm}, clip, width=0.45\textwidth]{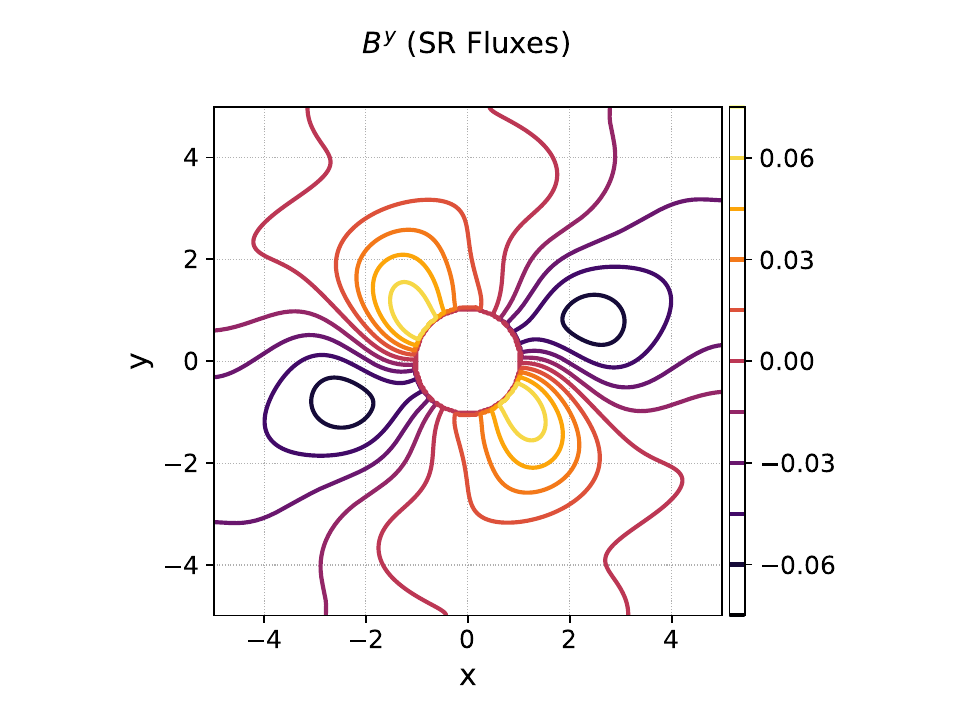}}
\caption{The $y$-component of the magnetic field at time ${t = 50}$ for the Blandford-Znajek magnetosphere problem for a rapidly-rotating black hole with ${B_0 = 1}$ and ${a = 0.999}$, obtained using standard general relativistic fluxes (left) and local special relativistic fluxes in an orthonormal tetrad basis (right).}
\label{fig:bz_monopole_fast_By}
\end{figure*}

Since the analytical solution of \citet{blandford_znajek1977} is derived perturbatively, their approximation breaks down in the limit of large black hole spin ${a \sim 1}$. Nevertheless, as an illustration of the relative robustness and generality of our tetrad-first approach, we also simulate a Blandford-Znajek-like magnetosphere solution for the case of a rapidly-spinning black hole with ${a = 0.999}$ (with the initial field configuration still identical to the slowly-rotating case, with ${B_0 = 1}$). The results at time ${t = 50}$, obtained using both general relativistic and special relativistic Riemann solvers, are shown in Figure \ref{fig:bz_monopole_fast_By}. Although there is no longer any analytical solution to compare against directly, we see that the general relativistic and special relativistic Riemann solver approaches yield near-identical results for this test. However, small numerical oscillations can be seen in the $y$-component of the magnetic field vector ${\mathbf{B}}$ perceived by Eulerian observers for the case of the general relativistic Riemann solver results (in the region near the ${y = -5}$ and ${y = 5}$ limits of the domain, due to the magnitude of frame-dragging effects being exerted on the magnetic field from the high black hole spin), yet these instabilities are absent in the corresponding special relativistic Riemann solver results obtained in the orthonormal tetrad basis, without any apparent loss of numerical accuracy. This result confirms that, as one would expect, solving a Riemann problem directly within a curved spacetime is sometimes less numerically robust than solving a locally-transformed version of the same Riemann problem within a flat spacetime, and vindicates the superiority of the tetrad-first approach in such scenarios.

\subsection{The Wald Black Hole Magnetosphere}

\begin{figure}[ht]
\centering
\includegraphics[trim={1cm, 0.5cm, 1cm, 0.5cm}, clip, width=0.45\textwidth]{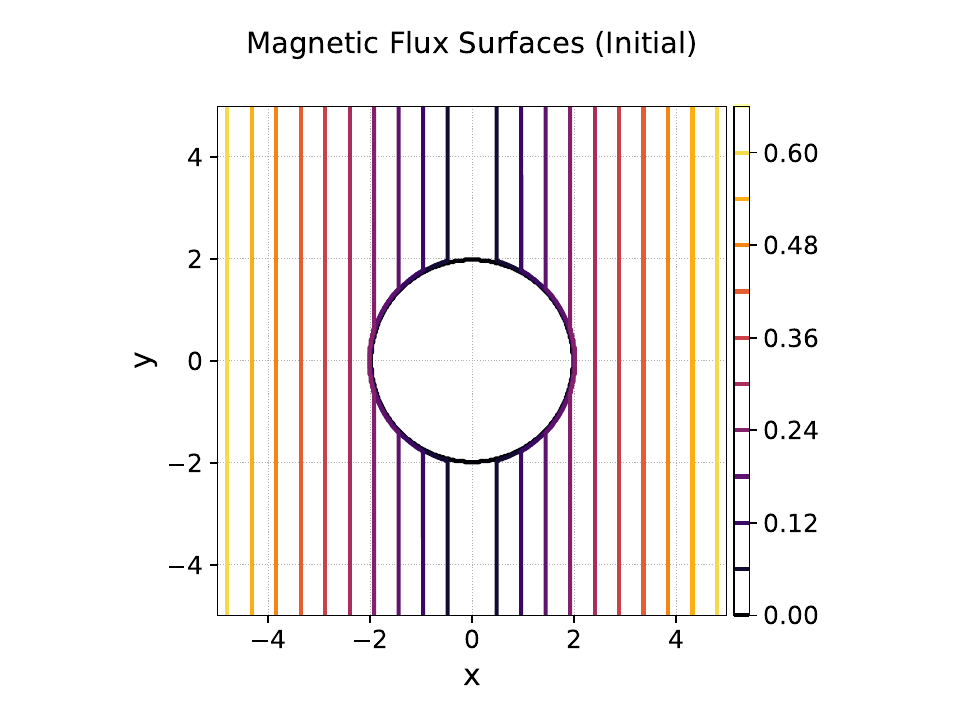}
\caption{The initial configuration of magnetic flux surfaces at time ${t = 0}$ for the Wald black hole magnetosphere problem with ${B_0 = 1}$.}
\label{fig:wald_magnetosphere_static_B_initial}
\end{figure}

\begin{figure*}[htp]
\centering
\subfigure{\includegraphics[trim={1cm, 0.5cm, 1cm, 0.5cm}, clip, width=0.45\textwidth]{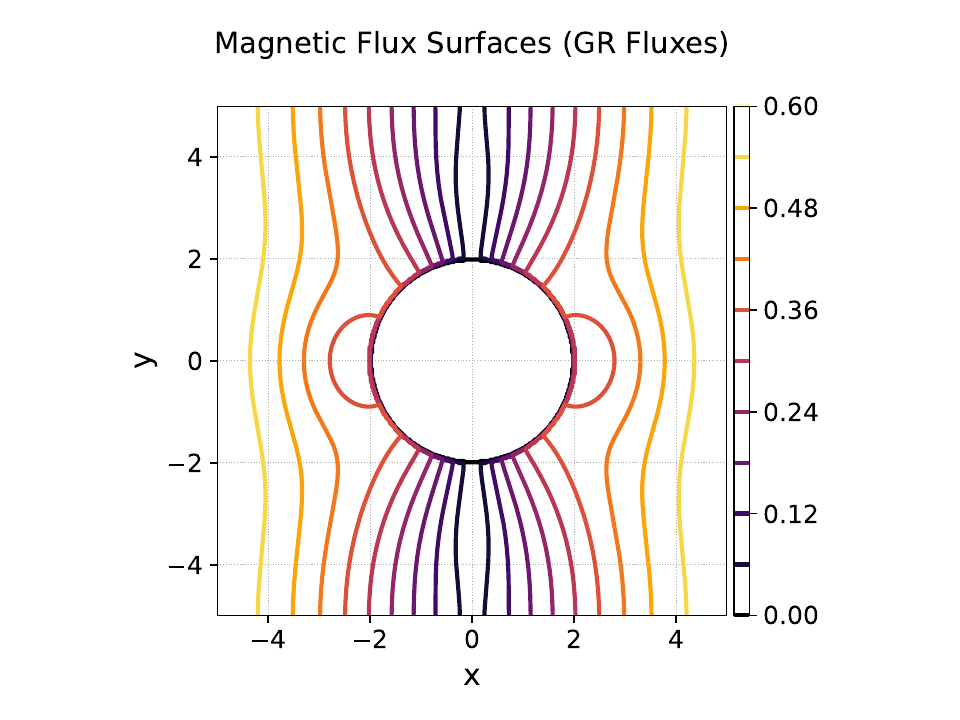}}
\subfigure{\includegraphics[trim={1cm, 0.5cm, 1cm, 0.5cm}, clip, width=0.45\textwidth]{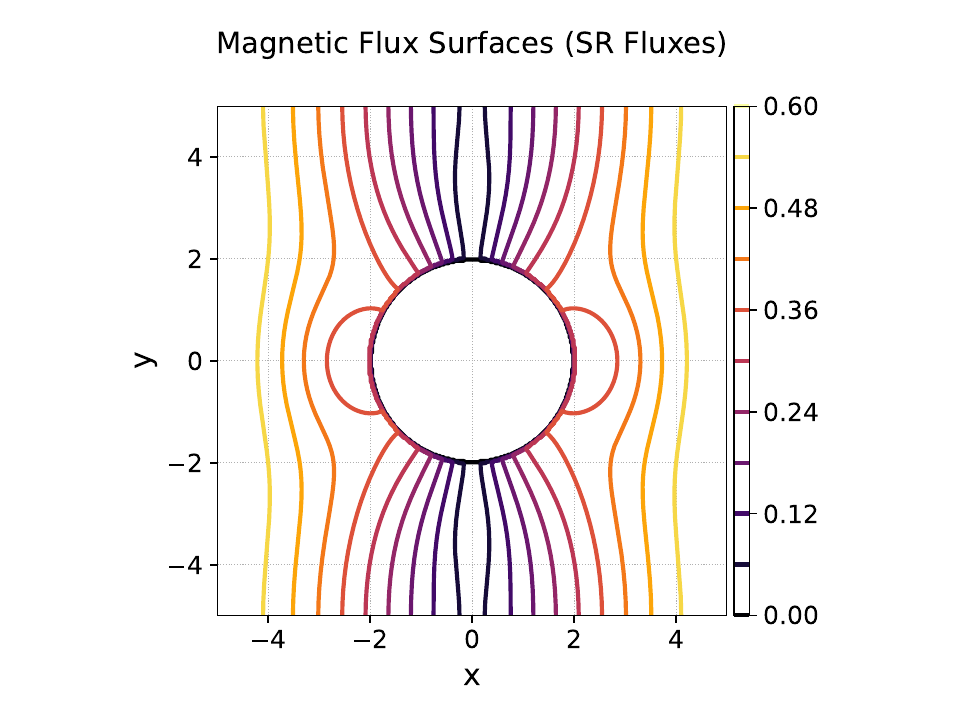}}
\caption{The configuration of magnetic flux surfaces at time ${t = 50}$ for the Wald magnetosphere problem for a static black hole with ${B_0 = 1}$ and ${a = 0}$, obtained using standard general relativistic fluxes (left) and local special relativistic fluxes in an orthonormal tetrad basis (right).}
\label{fig:wald_magnetosphere_static_B}
\end{figure*}

\begin{figure*}[htp]
\centering
\subfigure{\includegraphics[width=0.45\textwidth]{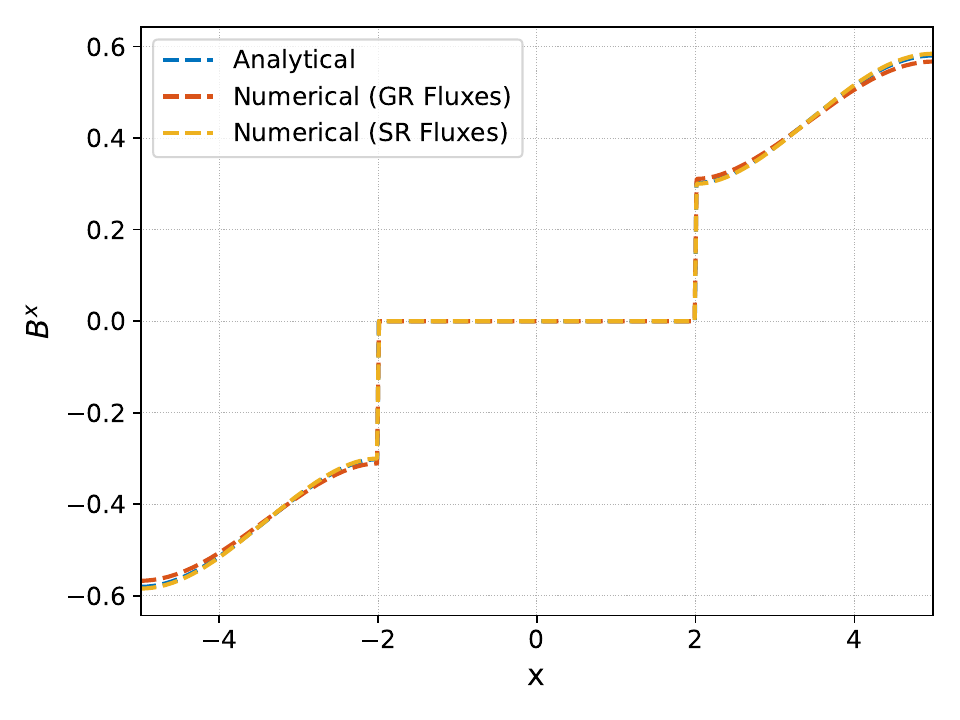}}
\subfigure{\includegraphics[width=0.45\textwidth]{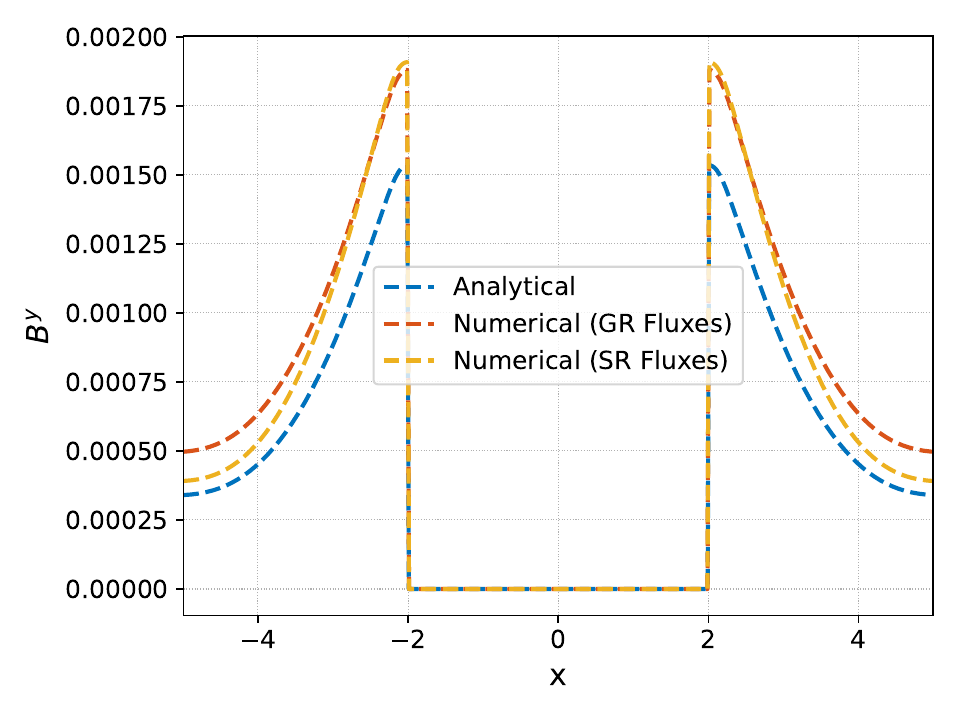}}
\caption{A cross-sectional profile of the $x$-component (left) and $y$-component (right) of the magnetic field ${\mathbf{B}}$ (as perceived by Eulerian observers) at time ${t = 50}$ for the Wald magnetosphere problem for a static black hole with ${B_0 = 1}$ and ${a = 0}$, validating both the special relativistic and general relativistic Riemann solver approaches against the analytical solution. Cross sections are taken at ${y = 0}$.}
\label{fig:wald_magnetosphere_static_B_cross}
\end{figure*}

\citet{wald1974} obtained an exact solution to the vacuum Einstein-Maxwell equations for a spinning black hole immersed in an initially-uniform magnetic field that is aligned with the axis of black hole spin. This problem is somewhat more artificial than the \citet{blandford_znajek1977} solution, with a greater proportion of the magnetic field lines penetrating the black hole horizon, but the resulting analytical solution for the electromagnetic field tensor ${F_{\mu \nu}}$ is significantly more elegant:

\begin{equation}
F_{\mu \nu} = B_0 \left( m_{\mu} - m_{\nu} + 2 a \left( k_{\mu} - k_{\nu} \right) \right),
\end{equation}
with ${\mathbf{k} = \partial_t}$ and ${\mathbf{m} = \partial_{\phi}}$ being the two Killing vector fields of the Kerr metric. The \citet{wald1974} solution also has the considerable advantage over the \citet{blandford_znajek1977} solution of being derived non-perturbatively, thereby remaining valid even in the limit of large black hole spin ${a \sim 1}$. For the case of a static (Schwarzschild) black hole with ${a = 0}$, the \citet{wald1974} solution reduces to:

\begin{equation}
B_{\theta} = - \frac{B_0}{2 \sqrt{\gamma}} \left( \partial_{\theta} \gamma_{\phi \phi} \right), \qquad B_{\phi} = \frac{B_0}{2 \sqrt{\gamma}} \left( \partial_{r} \gamma_{\phi \phi} \right),
\end{equation}
with ${B_r = 0}$ and ${\mathbf{E} = \mathbf{0}}$. The initial configuration of magnetic flux surfaces for our magnetosphere simulations based upon the \citet{wald1974} setup is shown in Figure \ref{fig:wald_magnetosphere_static_B_initial}, again assuming reference magnetic field strength ${B_0 = 1}$ and a symmetric domain ${\left( x, y \right) \in \left[ -5, 5 \right] \times \left[ -5, 5 \right]}$. Once again, all simulations are run with a CFL coefficient of 0.95, a spatial discretization of ${400 \times 400}$ cells, and a final time of ${t = 50}$. The results for a static (Schwarzschild) black hole with mass ${M = 1}$ and spin ${a = 0}$ are shown in Figure \ref{fig:wald_magnetosphere_static_B} using both general relativistic fluxes and special relativistic fluxes in an orthnormal tetrad basis. Just as in the case of the \citet{blandford_znajek1977} magnetosphere simulations, we can take cross-sectional profiles of various field components through the ${y = 0}$ axis in order to confirm superior convergence of the special relativistic Riemann solver results to the analytical solution; the $x$- and $y$-components of the magnetic field vector ${\mathbf{B}}$ perceived by Eulerian observers are shown in Figure \ref{fig:wald_magnetosphere_static_B_cross}. Unlike in previous results, however, we note that both the special and general relativistic Riemann solvers significantly overestimate the strength of the $y$-component of the magnetic field in the near-horizon region for a Schwarzschild black hole at coarse resolution, with the favorable convergence properties of the tetrad-first approach only becoming noticeable at greater distances from the black hole; we confirm that these discrepancies in the near-horizon region converge to zero with increased grid resolution.

\begin{figure*}[htp]
\centering
\subfigure{\includegraphics[trim={1cm, 0.5cm, 1cm, 0.5cm}, clip, width=0.45\textwidth]{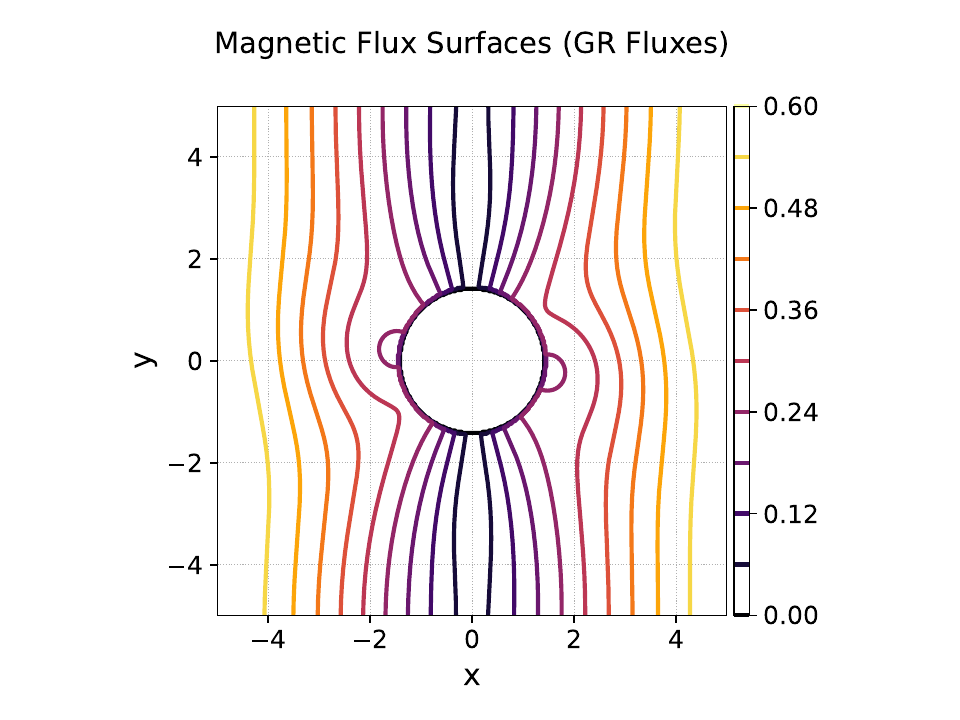}}
\subfigure{\includegraphics[trim={1cm, 0.5cm, 1cm, 0.5cm}, clip, width=0.45\textwidth]{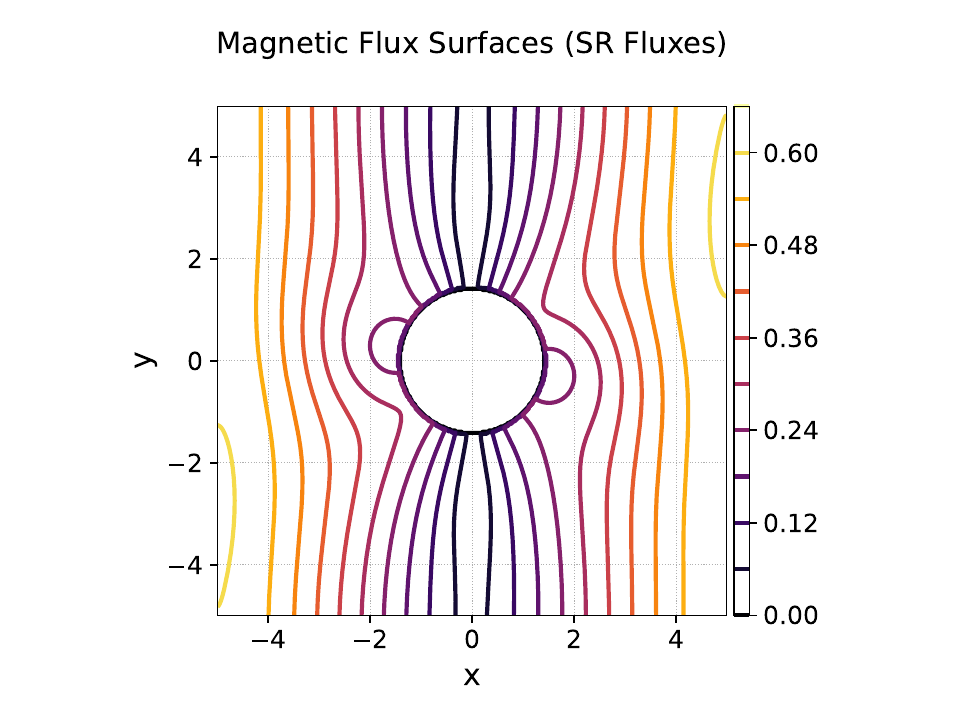}}
\caption{The configuration of magnetic flux surfaces at time ${t = 50}$ for the Wald magnetosphere problem for a spinning black hole with ${B_0 = 1}$ and ${a = 0.9}$, obtained using standard general relativistic fluxes (left) and local special relativistic fluxes in an orthonormal tetrad basis (right).}
\label{fig:wald_magnetosphere_spinning_B}
\end{figure*}

\begin{figure}[ht]
\centering
\includegraphics[width=0.45\textwidth]{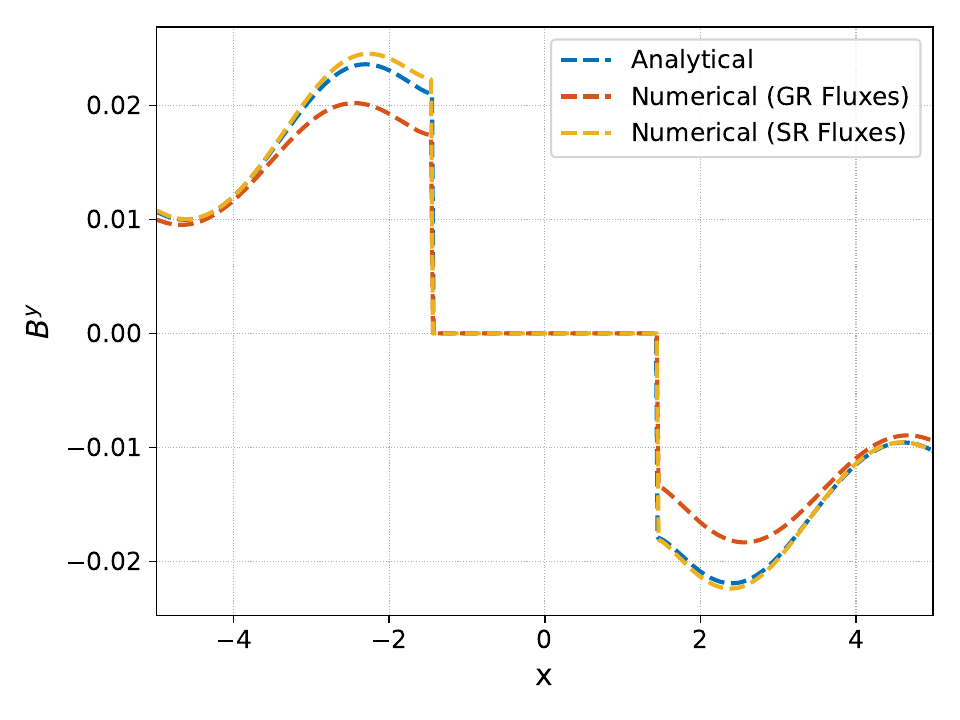}
\caption{A cross-sectional profile of the $y$-component of the magnetic field ${\mathbf{B}}$ (as perceived by Eulerian observers) at time ${t = 50}$ for the Wald magnetosphere problem for a spinning black hole with ${B_0 = 1}$ and ${a = 0.9}$, validating both the special relativistic and general relativistic Riemann solver approaches against the analytical solution. Cross sections are taken at ${y = 0}$.}
\label{fig:wald_magnetosphere_spinning_By}
\end{figure}

\begin{figure*}[htp]
\centering
\subfigure{\includegraphics[width=0.45\textwidth]{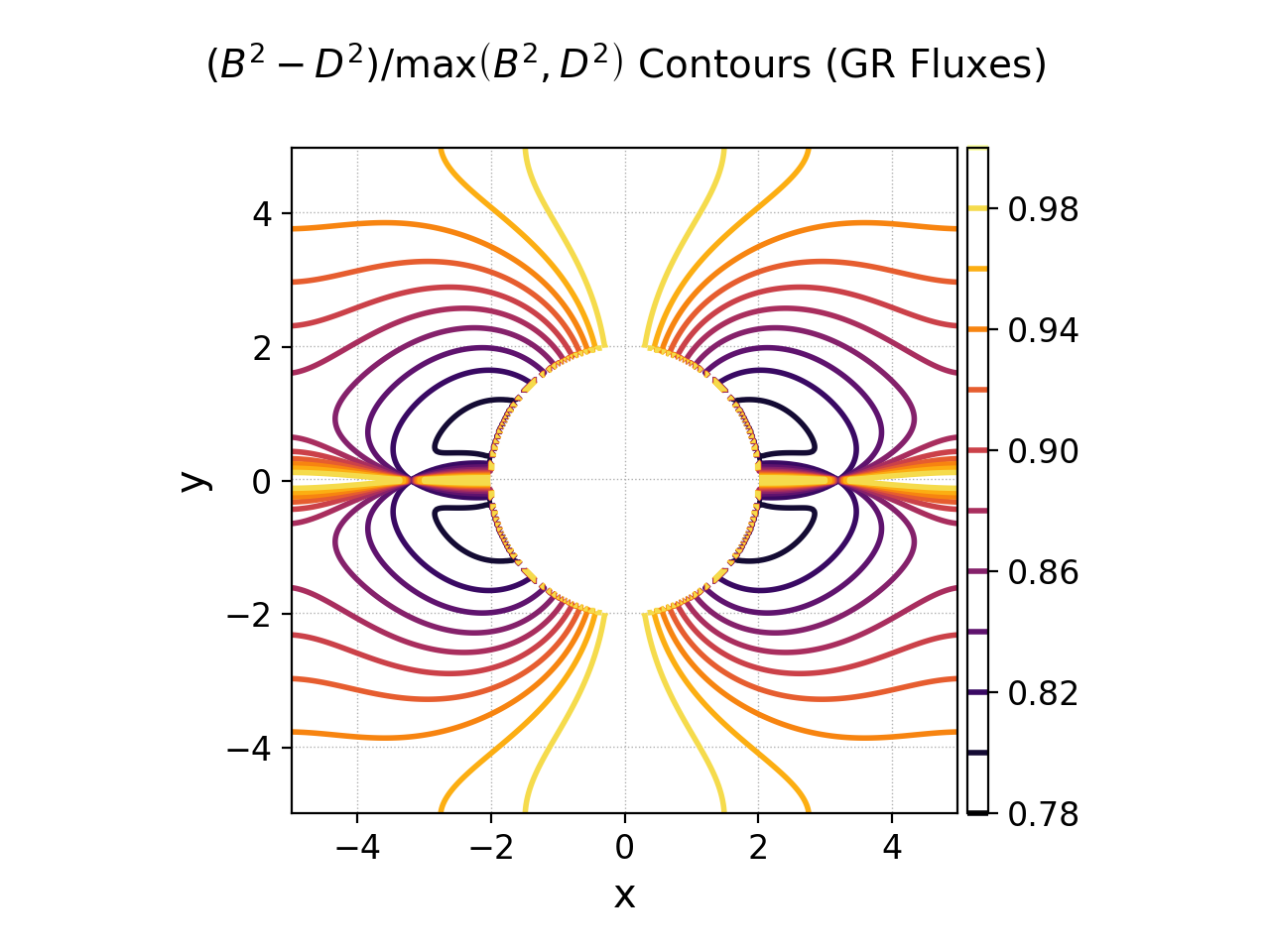}}
\subfigure{\includegraphics[width=0.45\textwidth]{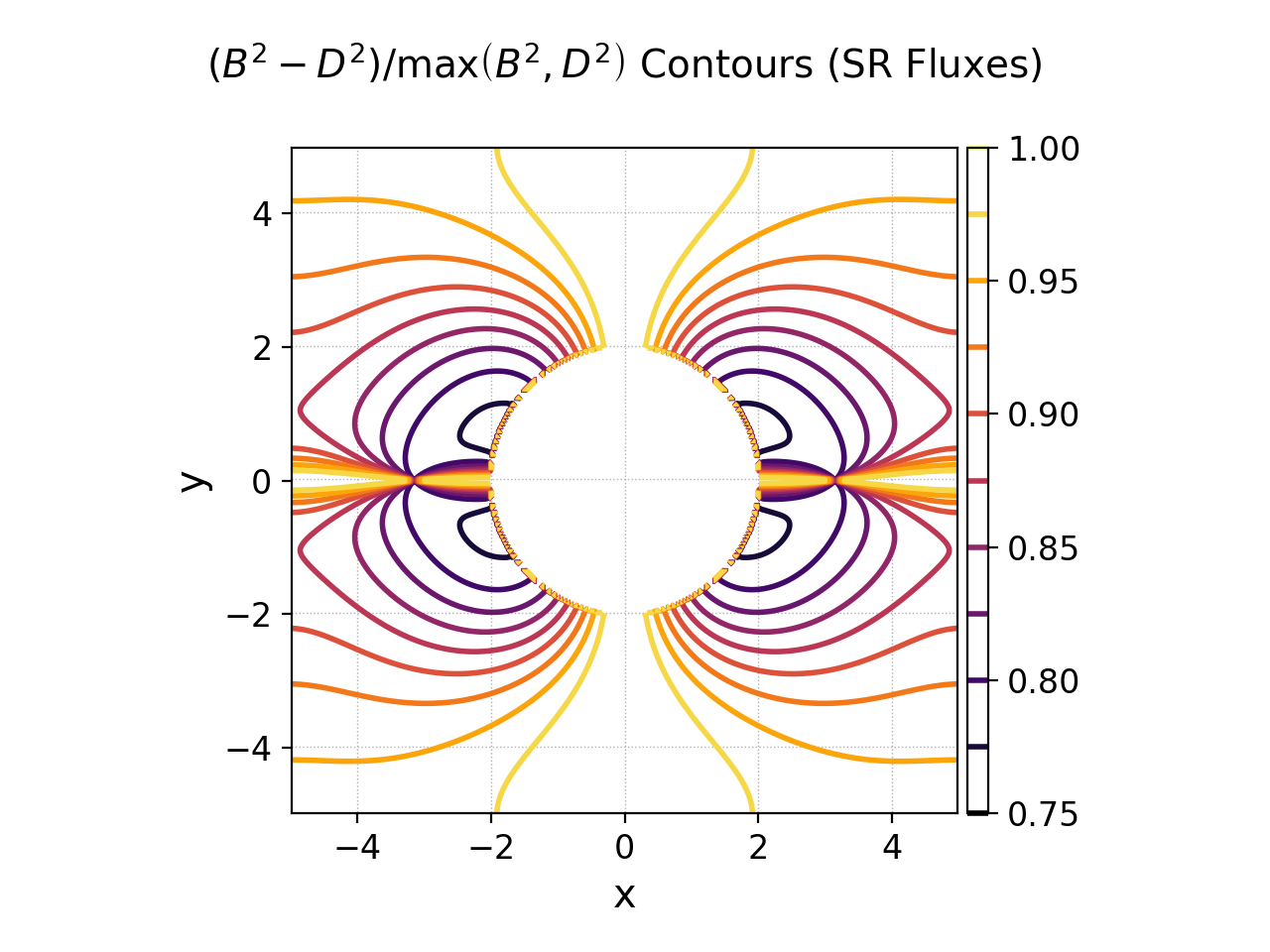}}
\caption{The contours of ${\frac{B^2 - D^2}{\max \left( B^2, D^2 \right)}}$ at time ${t = 50}$ for the Wald magnetosphere problem for a spinning black hole with ${B_0 = 1}$ and ${a = 0.9}$, obtained using standard general relativistic fluxes (left) and local special relativistic fluxes in an orthonormal tetrad basis (right), showing formation of the equatorial current sheet. The purpose of these figures is to facilitate direct comparison against plots like Figure 4 (right) in \citet{komissarov2004}.}
\label{fig:wald_magnetosphere_static_BD}
\end{figure*}

\begin{figure}[ht]
\centering
\includegraphics[width=0.45\textwidth]{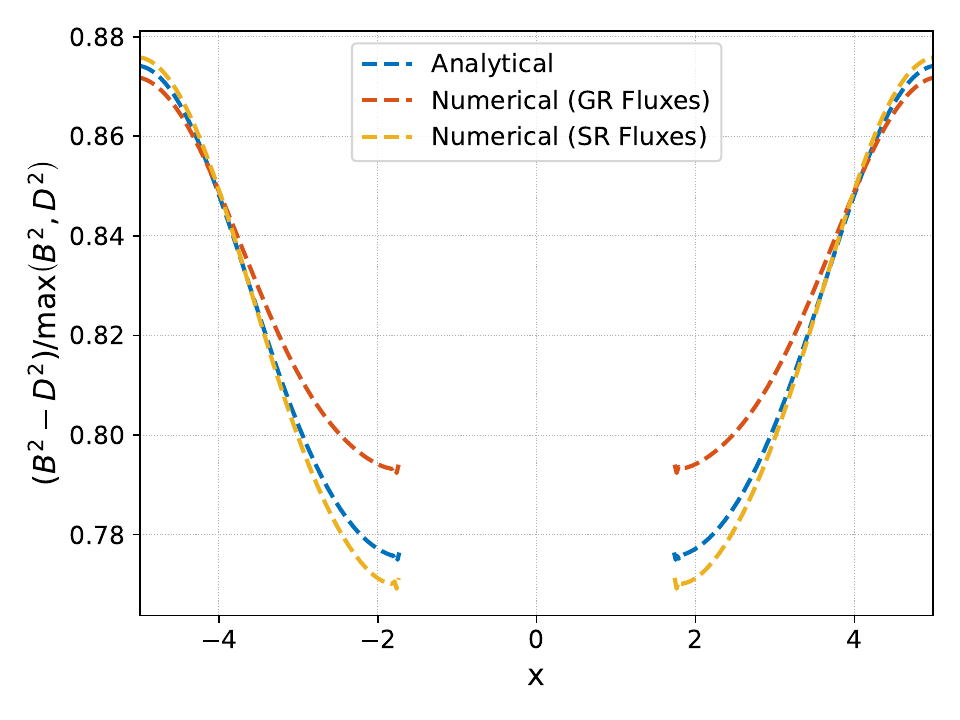}
\caption{A cross-sectional profile of ${\frac{B^2 - D^2}{\max \left( B^2, D^2 \right)}}$ at time ${t = 50}$ for the Wald magnetosphere problem for a spinning black hole with ${B_0 = 1}$ and ${a = 0.9}$, validating both the special relativistic and general relativistic Riemann solver approaches against the analytical solution. Cross sections are taken at ${y = 0}$.}
\label{fig:wald_magnetosphere_static_BD_profile}
\end{figure}

\begin{figure*}[htp]
\centering
\subfigure{\includegraphics[trim={1cm, 0.5cm, 1cm, 0.5cm}, clip, width=0.45\textwidth]{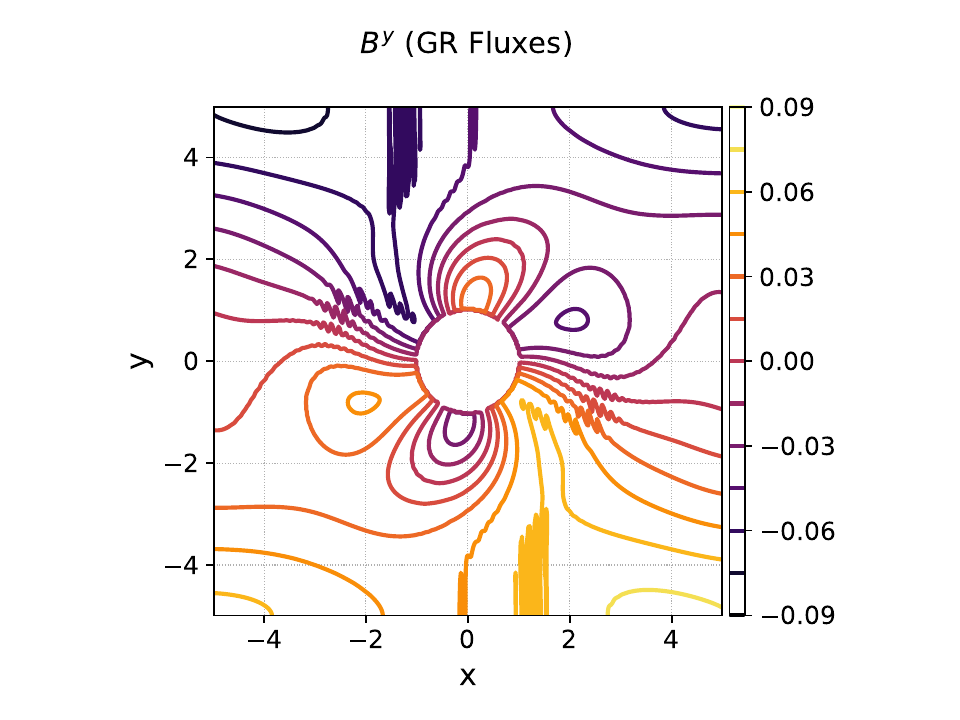}}
\subfigure{\includegraphics[trim={1cm, 0.5cm, 1cm, 0.5cm}, clip, width=0.45\textwidth]{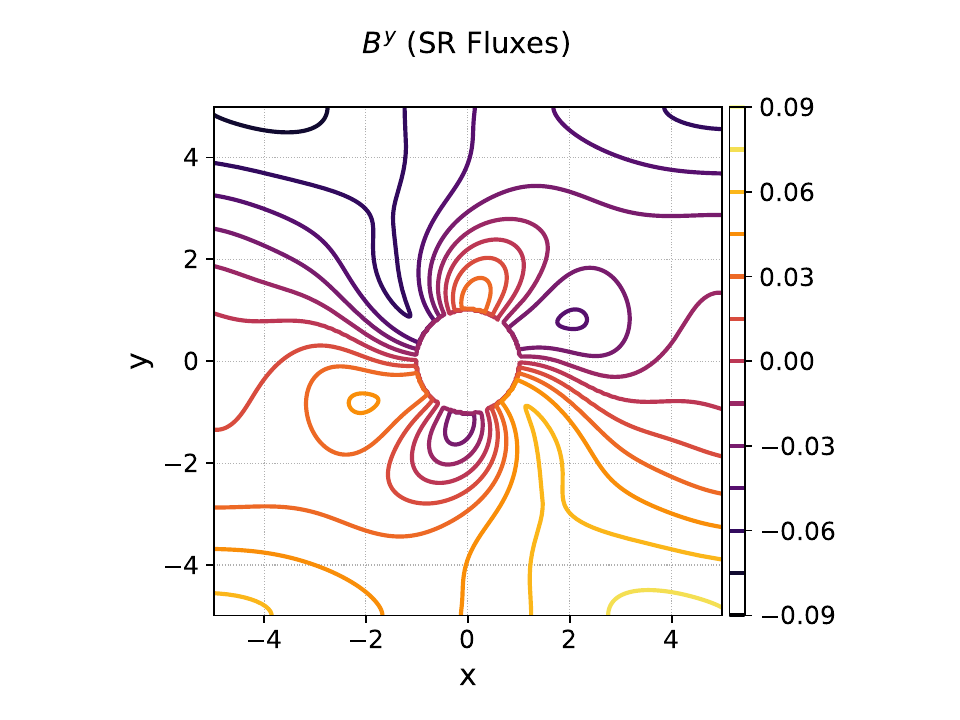}}
\caption{The configuration of magnetic flux surfaces at time ${t = 50}$ for the Wald magnetosphere problem for a rapidly-rotating black hole with ${B_0 = 1}$ and ${a = 0.9999}$, obtained using standard general relativistic fluxes (left) and local special relativistic fluxes in an orthonormal tetrad basis (right).}
\label{fig:wald_magnetosphere_fast_By}
\end{figure*}

On the other hand, the results for a spinning (Kerr) black hole with mass ${M = 1}$ and spin ${a = 0.9}$ are shown in Figure \ref{fig:wald_magnetosphere_spinning_B}. Unlike the static (Schwarzschild) case, we see from the cross-sectional profile of the ${B^y}$ field component through the ${y = 0}$ axis in Figure \ref{fig:wald_magnetosphere_spinning_By} that the tetrad-first approach yields favorable convergence to the analytical solution in \textit{both} the near-horizon region and at greater distances from the black hole. This suggests, perhaps unsurprisingly, that the relative advantages of the tetrad-first approach (as compared to using standard general relativistic Riemann solvers) only truly become manifest in highly distorted coordinate systems with ${\alpha \ll 1}$ and ${\left\lVert \boldsymbol\beta \right\rVert \gg 0}$, whilst for ``milder'' background spacetimes such as the Schwarzschild metric, the performance of the two approaches is comparable, and therefore the benefits of the tetrad-first approach are less apparent.  We also show a contour plot of the normalized quantity ${\left( B^2 - D^2 \right) / \max \left( B^2, D^2 \right)}$ in Figure \ref{fig:wald_magnetosphere_static_BD}, so as to facilitate direct comparison against Figure 4 (right) in \citet{komissarov2004}. Figure \ref{fig:wald_magnetosphere_static_BD_profile} shows a cross-sectional profile of the ${\left( B^2 - D^2 \right) / \max \left( B^2, D^2 \right)}$ scalar through the ${y = 0}$ axis, again demonstrating favorable convergence of the tetrad-first approach to the analytical solution. As in the case of the Blandford-Znajek simulations shown previously, as we begin to push the black hole to progressively more extreme values of the spin parameter, such as ${a = 0.9999}$, we start to see strong numerical instabilities appearing in the solutions obtained using standard general relativistic fluxes, while the corresponding solutions obtained using local special relativistic fluxes in an orthnormal tetrad basis remain relatively stable (as shown in Figure \ref{fig:wald_magnetosphere_fast_By}). All of these results serve to underscore the basic point that the true advantage of the tetrad-first approach lies in its superior convergence and stability properties in highly distorted spacetime coordinate systems with ${\alpha \ll 1}$ and/or ${\left\lVert \boldsymbol\beta \right\rVert \gg 0}$

\section{Case II: General Relativistic Hydrodynamics}
\label{sec:gr_hydrodynamics}

As our second illustrative example of the tetrad-first approach, we now turn to the case of curved spacetime hydrodynamics, whose governing equations may also be cast in the general relativistic conservation law form described in Section \ref{sec:general_algorithm}. A perfect relativistic fluid with density ${\rho}$, pressure $p$, and four-velocity ${u^{\mu}}$ (in the absence of heat conduction or viscous stresses), embedded within a background spacetime with inverse metric tensor ${g^{\mu \nu}}$, is described by a stress-energy tensor ${T^{\mu \nu}}$ of the general form,

\begin{equation}
T^{\mu \nu} = \rho h u^{\mu} u^{\nu} + p g^{\mu \nu},
\end{equation}
where $h$ denotes the specific enthalpy of the fluid, which is related to its specific internal energy ${\varepsilon \left( \rho, p \right)}$ by,

\begin{equation}
h = 1 + \varepsilon \left( \rho, p \right) + \frac{p}{\rho},
\end{equation}
and whose particular form is therefore dependent upon the equation of state. Following the so-called \textit{Valencia} formalism of \citet{banyuls_font_ibanez_marti_miralles1997} and taking timelike and spacelike projections of the general conservation law for energy-momentum,

\begin{equation}
\nabla_{\nu} T^{\mu \nu} = 0,
\end{equation}
we obtain individual conservation laws for the relativistic energy density ${\tau}$,

\begin{equation}
\begin{split}
\frac{1}{\sqrt{-g}} \left( \partial_t \left( \sqrt{\gamma} \tau \right) + \partial_i \left\lbrace \sqrt{-g} \left[ \tau \left( v^i - \frac{\beta^i}{\alpha} \right) + p v^i \right] \right\rbrace \right)\\
= \alpha \left( T^{\mu t} \partial_{\mu} \alpha - T^{\mu \nu} \Gamma_{\nu \mu}^{t} \right),
\end{split}
\end{equation}
and three-momentum density ${S_i}$,

\begin{equation}
\begin{split}
\frac{1}{\sqrt{-g}} \left( \partial_t \left( \sqrt{\gamma} S_j \right) + \partial_i \left\lbrace \sqrt{-g} \left[ S_j \left( v^i - \frac{\beta^i}{\alpha} \right) + p \delta_{j}^{i} \right] \right\rbrace \right)\\
= T^{\mu \nu} \left( \partial_{\mu} g_{\nu j} - \Gamma_{\nu \mu}^{\sigma} g_{\sigma j} \right),
\end{split}
\end{equation}
of a perfect fluid, respectively. In the above, the relativistic energy density ${\tau}$ and three-momentum density ${S_i}$ are defined in terms of the primitive fluid variables ${\rho}$, ${v^i}$, and $p$ as,

\begin{subequations}
\begin{align}
\tau &= \rho h W^2 - p - \rho W,\\
S_i &= \rho h W^2 v_i,
\end{align}
\end{subequations}
respectively, where the three-velocity ${v^i}$ of the fluid is related to its four-velocity ${u^{\mu}}$ by,

\begin{equation}
v^i = \frac{u^i}{\alpha u^t} + \frac{\beta^i}{\alpha},
\end{equation}
and where $W$ denotes the Lorentz factor of the fluid,

\begin{equation}
W = \alpha u^t = \frac{1}{\sqrt{1 - \gamma_{i j} v^i v^j}}.
\end{equation}
Furthermore, the conservation of (rest) mass current density ${J^{\mu}}$,

\begin{equation}
\nabla_{\mu} J^{\mu} = 0,
\end{equation}
yields, in the case of a perfect fluid with ${J^{\mu} = \rho u^{\mu}}$, an additional conservation law for baryon number density of the form:

\begin{equation}
\frac{1}{\sqrt{-g}} \left( \partial_t \left( \sqrt{\gamma} D \right) + \partial_i \left\lbrace \sqrt{-g} \left[ D \left( v^i - \frac{\beta^i}{\alpha} \right) \right] \right\rbrace \right) = 0,
\end{equation}
where $D$ is the relativistic mass density, ${D = \rho W}$.

Suppose now that our choice of spacetime foliation satisfies the ADM Hamiltonian and momentum constraint equations, namely,

\begin{equation}
\mathcal{H} = {}^{\left( 3 \right)} R + K^2 - K_{i j} K^{i j} - 2 \alpha^2 \, {}^{\left( 4 \right)} G^{t t} = 0,
\end{equation}
and,

\begin{equation}
\mathcal{M}^i = \nabla_j \left( K^{i j} - \gamma^{i j} K \right) - \alpha \, {}^{\left( 4 \right)} G^{t i} = 0,
\end{equation}
respectively (obtained by taking timelike and spacelike projections of the contracted Bianchi identities, respectively). In the above, ${K_{i j}}$ denotes the extrinsic curvature tensor on spacelike hypersurfaces, obtained as the Lie derivative of the spatial metric tensor ${\gamma_{i j}}$ with respect to the unit normal ${\mathbf{n}}$,

\begin{equation}
\begin{split}
K_{i j} &= - \frac{1}{2} \mathcal{L}_{\mathbf{n}} \gamma_{i j}\\
&= - \frac{1}{2 \alpha} \left( \partial_t \gamma_{i j} + \nabla_i \beta_j + \nabla_j \beta_i \right),
\end{split}
\end{equation}
$K$ denotes its trace,

\begin{equation}
K = \gamma^{i j} K_{i j},
\end{equation}
${{}^{\left( 3 \right)} R}$ denotes the Ricci scalar on spacelike hypersurfaces, and ${{}^{\left( 4 \right)} G_{\mu \nu}}$ denotes the Einstein tensor on spacetime. Subject to these constraints, following \citet{gorard2024a}, it becomes possible to rewrite the source terms appearing in the energy-momentum conservation laws purely in terms of the (first derivatives of the) ADM gauge variables ${K_{i j}}$, ${\beta^i}$, and ${\partial_i \alpha}$, as well as the perfect fluid stress-energy tensor ${T^{\mu \nu}}$, as,

\begin{equation}
\begin{split}
\alpha \left( T^{\mu t} \partial_{\mu} \alpha - T^{\mu \nu} \Gamma_{\nu \mu}^{t} \right) = T^{t t} \left( \beta^i \beta^j K_{i j} + \beta^i \partial_i \alpha \right)\\
+ T^{t i} \left( - \partial_i \alpha + 2 \beta^j K_{i j} \right) + T^{i j} K_{i j},
\end{split}
\end{equation}
and,

\begin{equation}
\begin{split}
T^{\mu \nu} \left( \partial_{\mu} g_{\nu j} - \Gamma_{\nu \mu}^{\sigma} g_{\sigma j} \right) = T^{t t} \left( \frac{1}{2} \beta^k \beta^l \partial_j \gamma_{k l} - \alpha \partial_j \alpha \right)\\
+ T^{t i} \beta^k \partial_j \gamma_{i k} + \frac{S_k}{\alpha} \partial_j \beta^k,
\end{split}
\end{equation}
respectively. We further exploit the fact that ${\sqrt{-g} = \alpha \sqrt{\gamma}}$ to write everything in terms of spatial variables only. Thus, the complete system of conservation equations that we actually solve numerically consists of:

\begin{equation}
\frac{1}{\alpha \sqrt{\gamma}} \left( \partial_t \left( \sqrt{\gamma} D \right) + \partial_i \left\lbrace \alpha \sqrt{\gamma} \left[ D \left( v^i - \frac{\beta^i}{\alpha} \right) \right] \right\rbrace \right) = 0,
\end{equation}
for the baryonic number density,

\begin{equation}
\begin{split}
\frac{1}{\alpha \sqrt{\gamma}} \left( \partial_t \left( \sqrt{\gamma} S_j \right) + \partial_i \left\lbrace \alpha \sqrt{\gamma} \left[ S_j \left( v^i - \frac{\beta^i}{\alpha} \right) + p \delta_{j}^{i} \right] \right\rbrace \right)\\
= T^{t t} \left( \frac{1}{2} \beta^k \beta^l \partial_j \gamma_{k l} - \alpha \partial_j \alpha \right) + T^{t i} \beta^k \partial_j \gamma_{i k} + \frac{S_k}{\alpha} \partial_j \beta^k,
\end{split}
\end{equation}
for the momentum density, and

\begin{equation}
\begin{split}
\frac{1}{\alpha \sqrt{\gamma}} \left( \partial_t \left( \sqrt{\gamma} \tau \right) + \partial_i \left\lbrace \alpha \sqrt{\gamma} \left[ \tau \left( v^i - \frac{\beta^i}{\alpha} \right) + p v^i \right] \right\rbrace \right)\\
= T^{t t} \left( \beta^i \beta^j K_{i j} + \beta^i \partial_i \alpha \right) + T^{t i} \left( - \partial_i \alpha + 2 \beta^j K_{i j} \right) + T^{i j} K_{i j},
\end{split}
\end{equation}
for the relativistic energy density.

Although the conserved variables $D$ (relativistic mass density), ${S_i}$ (three-momentum density) and ${\tau}$ (relativistic energy density) are the quantities that are actually evolved, the computation of the inter-cell fluxes depends upon the primitive fluid variables ${\rho}$ (density), ${v^i}$ (three-velocity) and $p$ (pressure). In order to reconstruct the primitive variables from the conserved ones, assuming a generic form of the equation of state, it is necessary to perform a non-linear root-finding operation (since, in contrast to Newtonian fluid mechanics, the presence of the Lorentz factor $W$ implies that the three-momenta ${S_i}$ are no longer algebraically independent of one another). For the case of an ideal gas equation of state, where the the specific enthalpy of the fluid is given by,

\begin{equation}
h = 1 + \frac{p}{\rho} \left( \frac{\Gamma}{\Gamma - 1} \right),
\end{equation}
with ${\Gamma}$ denoting the adiabatic index, we opt to follow the prescription of \citet{eulderink_mellema1995}, in which we apply a one-dimensional Newton-Raphson iteration in order to approximate the roots of the following quartic in ${\xi}$:

\begin{equation}
\alpha_4 \xi^3 \left( \xi - \eta \right) + \alpha_2 \xi^2 + \alpha_1 \xi + \alpha_0 = 0.
\end{equation}
In the above, the variable ${\xi}$ is defined as,

\begin{equation}
\xi = \frac{\sqrt{-g_{\mu \nu} T^{t \mu} T^{t \nu}}}{\rho h u^t} = \frac{\sqrt{\left( \tau + D \right)^2 - S_i S^i}}{\rho h W},
\end{equation}
the constant ${\eta}$ as,

\begin{equation}
\eta = \frac{2 \rho u^t \left( \Gamma - 1 \right)}{\left( \sqrt{-g_{\mu \nu} T^{t \mu} T^{t \nu}} \right) \Gamma} = \frac{2 D \left( \Gamma - 1 \right)}{\Gamma \sqrt{\left( \tau + D \right)^2 - S_i S^i}},
\end{equation}
and the coefficients ${\alpha_0}$, ${\alpha_1}$, ${\alpha_2}$, and ${\alpha_4}$ as,

\begin{equation}
\begin{split}
\alpha_0 &= - \frac{1}{\Gamma^2},\\
\alpha_1 &= - \frac{2 \rho u^t \left( \Gamma - 1 \right)}{\left( \sqrt{-g_{\mu \nu} T^{t \mu} T^{t \nu}} \right) \Gamma^2} = - \frac{2 D \left( \Gamma - 1 \right)}{\Gamma^2 \sqrt{ \left( \tau + D \right)^2 - S_i S^i}},\\
\alpha_2 &= \left( \frac{\Gamma - 2}{\Gamma} \right) \left( \frac{\left( T^{t t} \right)^2}{g^{t t} g_{\mu \nu} T^{t \mu} T^{t \nu}} - 1 \right) + 1\\
&+ \left( \frac{\left( \rho u^t \right)^2}{g_{\mu \nu} T^{t \mu} T^{t \nu}} \right) \left( \frac{\Gamma - 1}{\Gamma} \right)^2\\
&= \left( \frac{\Gamma - 2}{\Gamma} \right) \left( \frac{\left( \tau + D \right)^2}{\left( \tau + D \right)^2 - S_i S^i} - 1 \right) + 1\\
&- \frac{D^2 \left( \Gamma - 1 \right)^2}{\Gamma^2 \left( \left( \tau + D \right)^2 - S_i S^i \right)},\\
\alpha_4 &= \frac{\left( T^{t t} \right)^2}{g^{t t} g_{\mu \nu} T^{t \mu} T^{t \nu}} - 1 = \frac{\left( \tau + D \right)^2}{\left( \tau + D \right)^2 - S_i S^i} - 1,
\end{split}
\end{equation}
respectively. Once the solution ${\xi}$ has been approximated to the desired degree of accuracy, the primitive variables may be recovered by first calculating the Lorentz factor $W$ as,

\begin{equation}
\begin{split}
W &= \frac{1}{2} \left( \frac{\tau + D}{\sqrt{\left( \tau + D \right)^2 - S_i S^i}} \right) \xi\\
&\times \left( 1 + \sqrt{1 + 4 \left( \frac{\Gamma - 1}{\Gamma} \right) \left( \frac{1 - \left( \frac{D}{\sqrt{\left( \tau + D \right)^2 - S_i S^i}} \right) \xi}{\left( \frac{\left( \tau + D \right)^2}{\left( \tau + D \right)^2 - S_i S^i} \right) \xi^2} \right)} \right),
\end{split}
\end{equation}
followed by the density ${\rho}$ and enthalpy $h$ as,

\begin{subequations}
\begin{align}
\rho &= \frac{D}{W},\\
h &= \frac{1}{\left( \frac{D}{\sqrt{\left( \tau + D \right)^2 - S_i S^i}} \right) \xi},
\end{align}
\end{subequations}
from which the pressure $p$ can then be calculated directly, via the equation of state. Finally, the covariant components of the three-velocity ${v_i}$ are recovered straightforwardly as:

\begin{equation}
v_i = \frac{S_i}{\rho h W^2}.
\end{equation}
During the recovery procedure, if any of these quantities are found to have attained unphysical values (namely, if ${\rho \leq 10^{-8}}$, ${p \leq 10^{-8}}$ or ${v_i v^i \geq 1 - 10^{-8}}$, where we have selected ${10^{-8}}$ as our underlying numerical tolerance), then an appropriate numerical floor/ceiling is imposed, and all other quantities are recalculated as necessary.

In order to integrate the geometric source terms appearing on the right-hand side of the general relativistic hydrodynamics equations, we use the existing strong stability-preserving Runge-Kutta (SSP-RK) integration scheme described by \citet{gottlieb_shu_tadmor2001} and \citet{peterson_hammett2013}, which is already implemented and extensively tested within \textsc{Gkeyll}. Specifically, we use the four-stage, third-order SSP-RK3 scheme, in which the value of the integrated function $f$ at time ${t^{n + 1} = t^n + \Delta t}$ is computed from its value at time ${t^n}$ as follows:

\begin{equation}
\begin{split}
f^{\left( 1 \right)} &= \frac{1}{2} f^n + \frac{1}{2} \mathcal{F} \left[ f^n, t^n \right],\\
f^{\left( 2 \right)} &= \frac{1}{2} f^{\left( 1 \right)} + \frac{1}{2} \mathcal{F} \left[ f^{\left( 1 \right)}, t^n + \frac{\Delta t}{2} \right],\\
f^{\left( 3 \right)} &= \frac{2}{3} f^{n} + \frac{1}{6} f^{\left( 2 \right)} + \frac{1}{6} \mathcal{F} \left[ f^{\left( 2 \right)}, t^n + \Delta t \right],\\
f^{n + 1} &= \frac{1}{2} f^{\left( 3 \right)} + \frac{1}{2} \mathcal{F} \left[ f^{\left( 3 \right)}, t^n + \frac{\Delta t}{2} \right],
\end{split}
\end{equation}
where the operator ${\mathcal{F}}$ denotes the standard first-order forward-Euler integration step,

\begin{equation}
\mathcal{F} \left[ f, t \right] = f + \Delta t \, \mathrm{RHS} \left[ f, t \right].
\end{equation}
This scheme has been chosen due to its high levels of stability and flexible time-step restrictions (remaining stable even at twice the time-step permitted by the usual CFL criterion). The geometric source terms arise because, although a coordinate transformation has been applied at the inter-cell boundary, the fluid variables in the two neighboring cells are still generically represented in different spacetime coordinate systems. This raises the possibility that one may be able to eliminate the source terms altogether, and hence the need for any explicit integration algorithm, by instead performing \textit{two} coordinate transformations, one at each cell center, and then interpolating a Riemann problem across the inter-cell boundary. We intend to explore the feasibility of such an algorithm in future work.

All that remains now for the implementation of a Roe solver for the equations of both special and general relativistic hydrodynamics is knowledge of the complete eigensystem of the flux Jacobian (we use a small modification of the Roe-averaging procedure described by \citet{eulderink_mellema1995}, wherein the spatial metric components ${\gamma_{i j}}$ and gauge variables ${\alpha}$, ${\beta^i}$ are also averaged over between neighboring cells). For flat spacetime hydrodynamics, the flux Jacobian in the $i$-th spatial coordinate direction:

\begin{equation}
\mathcal{J}_{i}^{flat} = \frac{\partial \begin{bmatrix}
D v^i & S_j v^i + p \delta_{j}^{i} & \tau v^i + p v^i
\end{bmatrix}^{\intercal}}{\partial \begin{bmatrix}
D & S_j & \tau
\end{bmatrix}^{\intercal}},
\end{equation}
has eigenvalues ${\lambda_{-} = v^i - c_s}$, ${\lambda_{+} = v^i + c_s}$ (each with algebraic multiplicity 1, representing the two acoustic waves) and ${\lambda_0 = v^i}$ (with algebraic multiplicity 3, representing the three material waves). For the case of an ideal gas equation of state, the speed of sound ${c_s}$ is given by:

\begin{equation}
c_s = \sqrt{\frac{\Gamma p}{\rho \left( 1 + \left( \frac{p}{\rho} \right) \left( \frac{\Gamma}{\Gamma - 1} \right) \right)}}.
\end{equation}
In the $x$-direction, for instance, the corresponding right eigenvectors are given by:

\begin{equation}
\begin{split}
\mathbf{r}_{-} &= \begin{bmatrix}
1\\
h W \left( v^x - \frac{v^x - \lambda_{-}}{1 - v^x \lambda_{-}} \right)\\
h W v^y\\
h W v^z\\
\frac{h W \left( 1 - v^x v^x \right)}{1 - v^x \lambda_{-}} - 1
\end{bmatrix}, \qquad \mathbf{r}_{+} = \begin{bmatrix}
1\\
h W \left( v^x - \frac{v^x - \lambda_{+}}{1 - v^x \lambda_{+}} \right)\\
h W v^y\\
h W v^z\\
\frac{h W \left( 1 - v^x v^x \right)}{1 - v^x \lambda_{+}} - 1
\end{bmatrix},\\
\mathbf{r}_{0}^{1} &= \begin{bmatrix}
\frac{\frac{1}{\rho} \left. \left( \frac{\partial p}{\partial \varepsilon} \right) \right\vert_{p}}{h W \left( \frac{1}{\rho} \left. \left( \frac{\partial p}{\partial \varepsilon} \right) \right\vert_{p} - c_{s}^{2} \right)}\\
v^x\\
v^y\\
v^z\\
1 - \frac{\frac{1}{\rho} \left. \left( \frac{\partial p}{\partial \varepsilon} \right) \right\vert_{p}}{h W \left( \frac{1}{\rho} \left. \left( \frac{\partial p}{\partial \varepsilon} \right) \right\vert_{p} - c_{s}^{2} \right)}
\end{bmatrix}, \qquad \mathbf{r}_{0}^{2} = \begin{bmatrix}
W v^y\\
2 h W^2 v^x v^y\\
h \left( 1 + 2 W^2 v^y v^y \right)\\
2 h W^2 v^z v^y\\
W v^y \left( 2 h W - 1 \right)
\end{bmatrix},\\
\mathbf{r}_{0}^{3} &= \begin{bmatrix}
W v^z\\
2 h W^2 v^x v^z\\
2 h W^2 v^y v^z\\
h \left( 1 + 2 W^2 v^z v^z \right)\\
W v^z \left( 2 h W - 1 \right)
\end{bmatrix}.
\end{split}
\end{equation}
On the other hand, for curved spacetime hydrodynamics, the flux Jacobian in the $i$-th spatial coordinate direction,

\begin{equation}
\mathcal{J}_{i}^{curved} = \frac{\partial \begin{bmatrix}
D \left( v^i - \frac{\beta^i}{\alpha} \right)\\
S_j \left( v^i - \frac{\beta^i}{\alpha} \right) + p \delta_{j}^{i}\\
\tau \left( v^i - \frac{\beta^i}{\alpha} \right) + p v^i
\end{bmatrix}}{\partial \begin{bmatrix}
D & S_j & \tau
\end{bmatrix}^{\intercal}},
\end{equation}
has eigenvalues,

\begin{equation}
\begin{split}
\lambda_{\pm} &= \frac{\alpha}{1 - v^2 c_{s}^{2}} \left[ v^i \left( 1 - c_{s}^{2} \right) \right.\\
&\left. \pm c_s \sqrt{\left( 1 - v^2 \right) \left[ \gamma^{i i} \left( 1 - v^2 c_{s}^{2} \right) - v^i v^i \left( 1 - c_{s}^{2} \right) \right]} \right] - \beta^i,
\end{split}
\end{equation}
each with algebraic multiplicity 1 (representing the two acoustic waves), and ${\lambda_0 = \alpha v^i - \beta^i}$, with algebraic multiplicity 3 (representing the three material waves), with the right eigenvector ${\mathbf{r}_{0}^{1}}$ being identical to its flat spacetime counterpart, and with ${\mathbf{r}_{-}}$, ${\mathbf{r}_{+}}$, ${\mathbf{r}_{0}^{2}}$ and ${\mathbf{r}_{0}^{3}}$ now being given by:

\begin{equation}
\begin{split}
\mathbf{r}_{-} &= \begin{bmatrix}
1\\
h W \left( v_x - \frac{v^x - \left( \frac{\lambda_{-} + \beta^x}{\alpha} \right)}{\gamma^{x x} - v^x \left( \frac{\lambda_{-} + \beta^x}{\alpha} \right)} \right)\\
h W v_y\\
h W v_z\\
\frac{h W \left( \gamma^{x x} - v^x v^x \right)}{\gamma^{x x} - v^x \left( \frac{\lambda_{-} + \beta^x}{\alpha} \right)} - 1
\end{bmatrix},\\
\mathbf{r}_{+} &= \begin{bmatrix}
1\\
h W \left( v_x - \frac{v^x - \left( \frac{\lambda_{+} + \beta^x}{\alpha} \right)}{\gamma^{x x} - v^x \left( \frac{\lambda_{+} + \beta^x} {\alpha} \right)} \right)\\
h W v_y\\
h W v_z\\
\frac{h W \left( \gamma^{x x} - v^x v^x \right)}{\gamma^{x x} - v^x \left( \frac{\lambda_{+} + \beta^x}{\alpha} \right)} - 1
\end{bmatrix},\\
\mathbf{r}_{0}^{2} &= \begin{bmatrix}
W v_y\\
h \left( \gamma_{x y} + 2 W^2 v_x v_y \right)\\
h \left( \gamma_{y y} + 2 W^2 v_y v_y \right)\\
h \left( \gamma_{z y} + 2 W^2 v_z v_y \right)\\
W v_y \left( 2 h W - 1 \right)
\end{bmatrix}, \qquad \mathbf{r}_{0}^{3} = \begin{bmatrix}
W v_z\\
h \left( \gamma_{x z} + 2 W^2 v_x v_z \right)\\
h \left( \gamma_{y z} + 2 W^2 v_y v_z \right)\\
h \left( \gamma_{z z} + 2 W^2 v_z v_z \right)\\
W v_z \left( 2 h W - 1 \right)
\end{bmatrix}.
\end{split}
\end{equation}

\subsection{Special Relativistic Riemann Problems}

We begin by validating the special relativistic Riemann solver implementations in \textsc{Gkeyll} against a collection of standard Riemann problems for flat spacetime hydrodynamics. As shown by \citet{pons_marti_muller2000}, such Riemann problems may be solved analytically (and for arbitrary equations of state) by exploiting a certain self-similarity condition across the rarefaction waves, together with the standard Rankine-Hugoniot conditions across the shock waves, making them ideal test cases for validating our numerical solutions against. Assuming that the initial discontinuity lies in the $x$-direction, the rarefaction waves correspond to self-similar solutions of the special relativistic hydrodynamics equations parameterized by ${\xi = x / t}$, thus reducing the flat spacetime baryonic number density conservation equation,

\begin{equation}
\partial_t \left( \rho W \right) + \partial_x \left( \rho W v^x \right) = 0,
\end{equation}
to the ordinary differential equation,

\begin{equation}
\begin{split}
&\left( v^x - \xi \right) \frac{d \rho}{d \xi} + \left[ \rho W^2 v^x \left( v^x - \xi \right) + \rho \right] \frac{d v^x}{d \xi}\\
&+ \rho W^2 v^y \left( v^x - \xi \right) \frac{d v^y}{d \xi} + \rho W^2 v^z \left( v^x - \xi \right) \frac{d v^z}{d \xi} = 0;
\end{split}
\end{equation}
the momentum density conservation equations,

\begin{equation}
\partial_t \left( \rho h W^2 v^j \right) + \partial_x \left( \rho h W^2 v^j v^x + p \delta^{j x} \right) = 0,
\end{equation}
to the system of ordinary differential equations,

\begin{equation}
\rho h W^2 \left( v^x - \xi \right) \frac{d v^j}{d \xi} + \left( \delta^{x j} - v^j \xi \right) \frac{d p}{d \xi} = 0;
\end{equation}
and the energy density conservation equation,

\begin{equation}
\partial_t \left( \rho h W^2 - p - \rho W \right) + \partial_x \left( \left( \rho h W^2 - \rho W \right) v^x \right) = 0,
\end{equation}
to a final ordinary differential equation (representing, rather, the conservation of \textit{entropy} across the rarefaction wave),

\begin{equation}
\frac{d p}{d \xi} = h c_{s}^{2} \frac{d \rho}{d \xi} = \rho \frac{d h}{d \xi}.
\end{equation}
These equations may then be integrated directly. Likewise, the shock waves correspond to jump discontinuities, satisfying the relativistic Rankine-Hugoniot conditions first derived by \citet{taub1948},

\begin{equation}
\left( \rho_{pre} u_{pre}^{\mu} - \rho_{post} u_{post}^{\mu} \right) n_{\mu} = 0,
\end{equation}
for rest mass current, and,

\begin{equation}
\left( T_{pre}^{\mu \nu} - T_{post}^{\mu \nu} \right) n_{\nu} = 0,
\end{equation}
for energy-momentum, where ${n^{\mu}}$ denote the components of the unit normal vector to the hypersurface ${\Sigma}$ across which the fluid variables are discontinuous (for our purposes, ${\mathbf{n}}$ is taken to be a spatial unit vector in the $x$-direction), and where the ``pre'' and ``post'' subscripts designate the values on the two sides of ${\Sigma}$. By introducing the Lorentz factor ${W_s}$ of the shock wave,

\begin{equation}
W_s = \frac{1}{\sqrt{1 - v_{s}^{2}}},
\end{equation}
with ${v_s}$ being the shock velocity, and the invariant mass flux $j$ across the shock wave,

\begin{equation}
j = W_s D_{pre} \left( v_s - v_{pre}^{x} \right) = W_s D_{post} \left( v_s - v_{post}^{x} \right),
\end{equation}
we are able to represent the complete system of Rankine-Hugoniot conditions purely in terms of the conserved fluid variables as:

\begin{equation}
v_{pre}^{x} - v_{post}^{x} = - \frac{j}{W_s} \left( \frac{1}{D_{pre}} - \frac{1}{D_{post}} \right),
\end{equation}
\begin{equation}
p_{pre} - p_{post} = \frac{j}{W_s} \left( \frac{S_{pre}^{x}}{D_{pre}} - \frac{S_{post}^{x}}{D_{post}} \right),
\end{equation}
\begin{equation}
\left( \frac{S_{pre}^{y}}{D_{pre}} - \frac{S_{post}^{y}}{D_{post}} \right) = 0, \qquad \left( \frac{S_{pre}^{z}}{D_{pre}} - \frac{S_{post}^{z}}{D_{post}} \right) = 0,
\end{equation}
\begin{equation}
\left( v_{pre}^{x} p_{pre} - v_{post}^{x} p_{post} \right) = \frac{j}{W_s} \left( \frac{\tau_{pre}}{D_{pre}} - \frac{\tau_{post}}{D_{post}} \right).
\end{equation}
Since the only non-contact waves appearing within the solution to the special relativistic Riemann problem are rarefactions and shocks, these relations, when combined with the wave-speeds arising from the eigenvalues ${\lambda_{\pm}}$ and ${\lambda_0}$ of the flux Jacobian ${\mathcal{J}}$, allow us to determine an exact solution to the special relativistic Riemann problem for an arbitrary equation of state.

\begin{figure}[ht]
\centering
\includegraphics[width=0.45\textwidth]{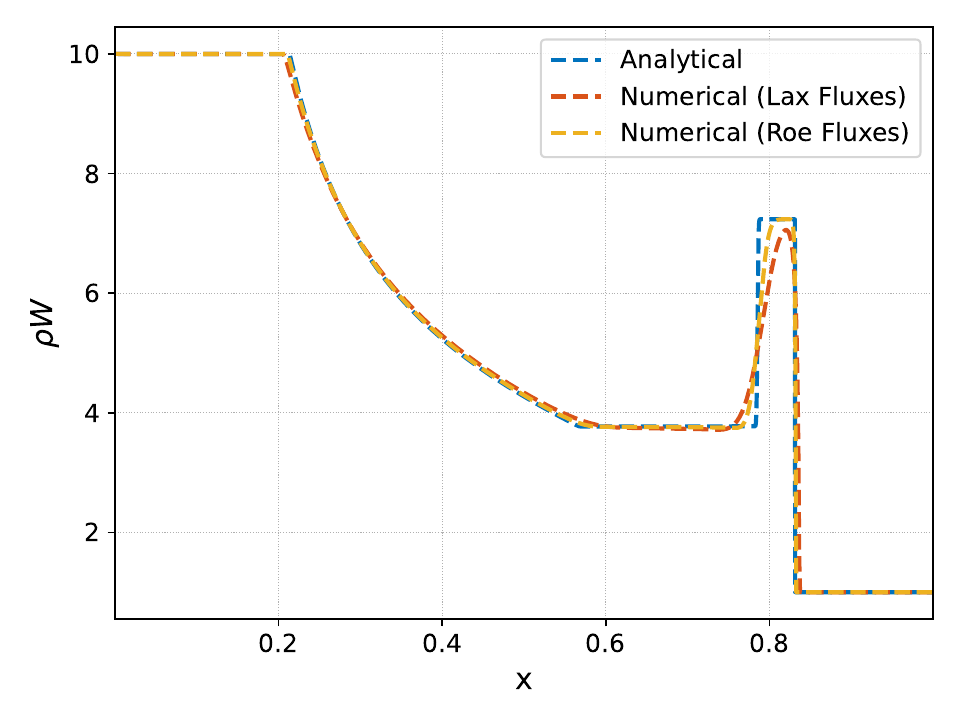}
\caption{The relativistic mass density ${D = \rho W}$ at time ${t = 0.4}$ for the mildly relativistic blast wave Riemann problem, validating the approximate Riemann solvers against the analytical solution.}
\label{fig:mild_shock}
\end{figure}

\begin{figure}[ht]
\centering
\includegraphics[width=0.45\textwidth]{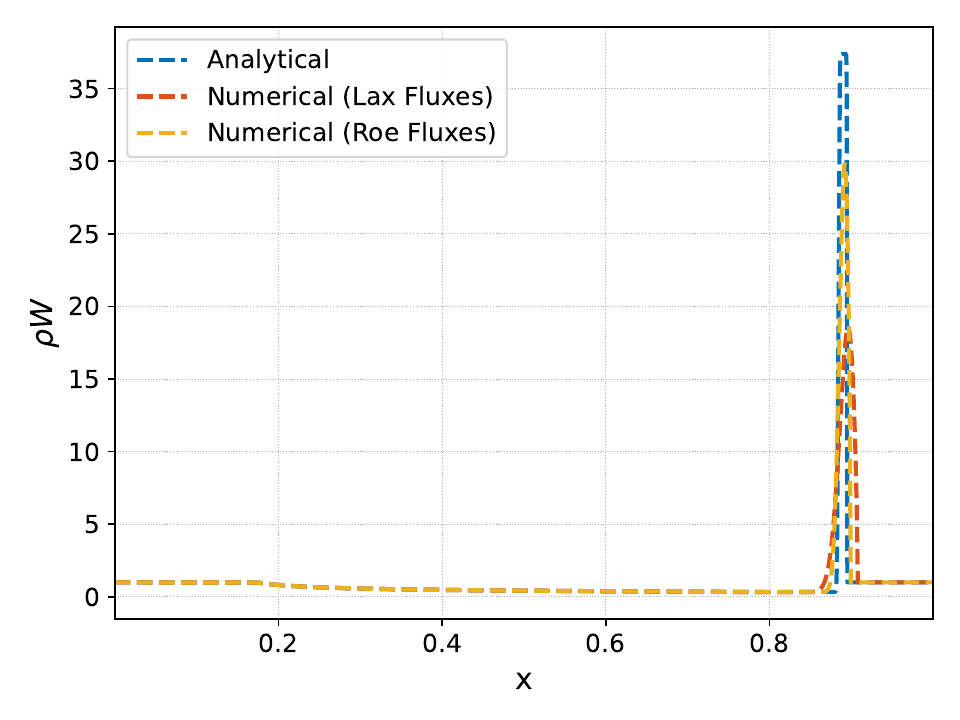}
\caption{The relativistic mass density ${D = \rho W}$ at time ${t = 0.4}$ for the strongly relativistic blast wave Riemann problem, validating the approximate Riemann solvers against the analytical solution.}
\label{fig:strong_blast}
\end{figure}

As our initial validation examples, we consider the two relativistic blast wave Riemann problems presented by \citet{delzanna_bucciantini2002}, based upon similar problems considered previously by \citet{donat_font_ibanez_marquina1998}. The \textit{mildly relativistic} blast wave Riemann problem uses initial conditions:

\begin{equation}
\left( \rho, p \right) = \begin{cases}
\left( 10, 13.3 \right), \qquad &\text{ for } \qquad x < \frac{1}{2},\\
\left( 1, 10^{-6} \right), \qquad &\text{ for } \qquad x > \frac{1}{2};
\end{cases}
\end{equation}
while the \textit{strongly relativistic} blast wave Riemann problem uses initial conditions:

\begin{equation}
\left( \rho, p \right) = \begin{cases}
\left( 1, 1000 \right), \qquad &\text{ for } \qquad x < \frac{1}{2},\\
\left( 1, 0.01 \right), \qquad &\text{ for } \qquad x > \frac{1}{2}.
\end{cases}
\end{equation}
In both cases, ${v^x = v^y = v^z = 0}$, and we use an ideal gas equation of state with an adiabatic index of ${\Gamma = \frac{5}{3}}$; the pressure is set to ${10^{-6}}$ on the right-hand side of the mildly relativistic blast wave problem due to the potential for numerical stability issues associated with setting pressures exactly to zero. Both approximate Riemann solvers (Lax-Friedrichs and Roe) were used in each case, with a CFL coefficient of 0.95 and using a computational domain ${x \in \left[ 0, 1 \right]}$, with a spatial discretizaton of 400 cells. Figures \ref{fig:mild_shock} and \ref{fig:strong_blast} show the relativistic mass densities ${D = \rho W}$ at time ${t = 0.4}$ in both cases. As expected, the Lax-Friedrichs fluxes exhibit higher levels of numerical diffusion than the Roe fluxes, yet still in both cases we see strong convergence to the respective analytical solutions; in particular, the peak density value obtained for the strongly relativistic blast wave problem is around 70\% that of the analytical value, which is comparable to the results obtained by \citet{lucasserrano_font_ibanez_marti2004} at equivalent resolution using a third-order reconstruction scheme in both space and time. These tests confirm not only the accuracy of our Riemann solvers, but also the robustness of our conservative-to-primitive variable reconstruction scheme in the presence of low densities, low pressures, large pressure gradients, and (at least in the case of the strongly relativistic blast wave test) fluid velocities approaching the speed of light, corresponding to a maximum Lorentz factor of ${W \approx 3.59}$.

\begin{figure}[ht]
\centering
\includegraphics[width=0.45\textwidth]{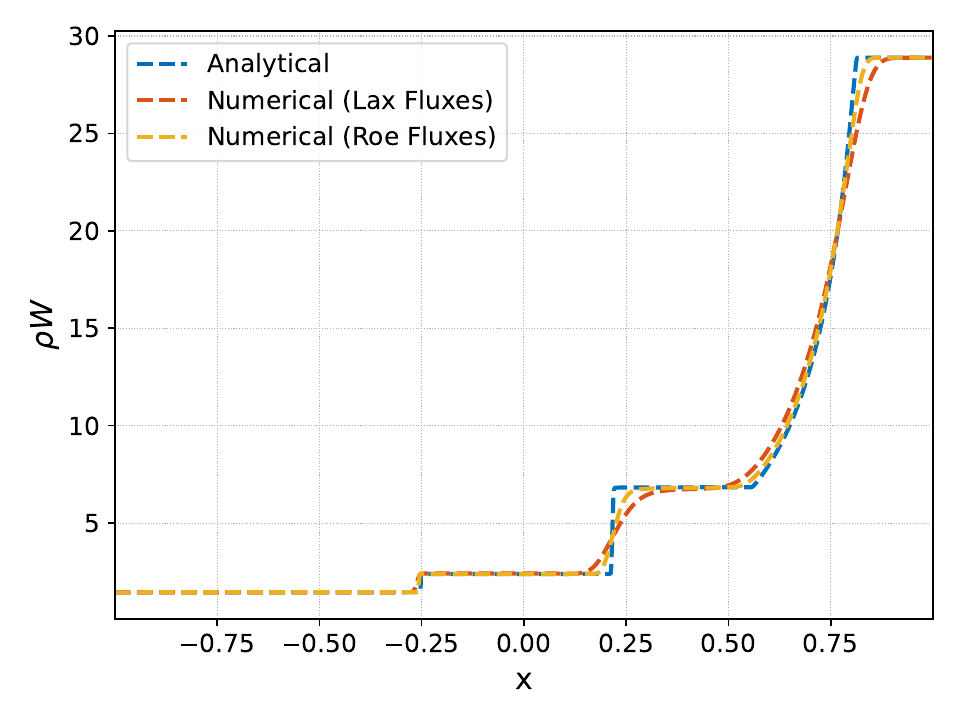}
\caption{The relativistic mass density ${D = \rho W}$ at time ${t = 1}$ for the ultra-relativistic shock Riemann problem, validating the approximate Riemann solvers against the analytical solution.}
\label{fig:ultra_rel_shock}
\end{figure}

The analytical black hole accretion problems that we aim to compare against are based on fluids obeying an \textit{ultra-relativistic} equation of state with adiabatic index ${\Gamma}$, in which the fluid pressure and density are no longer evolved independently as in an ideal gas, but are instead directly related by:

\begin{equation}
p = \left( \Gamma - 1 \right) \rho.
\end{equation}
Note therefore that, for ultra-relativistic fluids, we should no longer evolve the baryonic number density equation (since this would result in the system being overdetermined), and the relativistic energy density ${\tau}$ and three-momentum density ${S_i}$ now take on the simplified forms:

\begin{subequations}
\begin{align}
\tau &= \left( \rho + p \right) W^2 - p,\\
S_i &= \left( \rho + p \right) W^2 v_i.
\end{align}
\end{subequations}
Note moreover that, following the prescription of \citet{neilsen_choptuik2000}, the conservative-to-primitive variable reconstruction algorithm in the case of ultra-relativistic fluids reduces to a trivial algebraic procedure, whereby the fluid pressure $p$ may first be reconstructed as,

\begin{equation}
p = -2 \beta \tau + \sqrt{4 \beta^2 \tau^2 + \left( \Gamma - 1 \right) \left( \tau^2 - S_i S^i \right)},
\end{equation}
followed by the covector components of the fluid velocity ${v_i}$ as,

\begin{equation}
v_i = \frac{S_i}{\tau + p},
\end{equation}
where we have introduced the non-negative equation of state constant,

\begin{equation}
\beta = \frac{1}{4} \left( 2 - \Gamma \right).
\end{equation}
We have validated the specialized ultra-relativistic Riemann solver implementation within \textsc{Gkeyll} against the Riemann problems with non-vanishing tangential velocities proposed by \citet{mach_pietka2010}. For example, in order to reproduce their first Riemann problem solution, we use the initial conditions:

\begin{equation}
\left( \rho, v_x, v_y \right) = \begin{cases}
\left( 1, \frac{1}{2}, \frac{1}{3} \right), \qquad &\text{ for } \qquad x < 0,\\
\left( 20, \frac{1}{2}, \frac{1}{2} \right), \qquad &\text{ for } \qquad x > 0,
\end{cases}
\end{equation}
with ${v_z = 0}$ and an adiabatic index of ${\Gamma = \frac{4}{3}}$. Once again, we use both Lax-Friedrichs and Roe fluxes with a CFL coefficient of ${0.95}$, now with a symmetric domain of ${x \in \left[ -1, 1 \right]}$, yet still the same spatial discretization of 400 cells. Figure \ref{fig:ultra_rel_shock} shows the relativistic mass density ${D = \rho W}$ at time ${t = 1}$, demonstrating that one achieves comparable convergence to analytical Riemann problem solutions irrespective of whether one uses ideal gas or ultra-relativistic equations of state.

\subsection{Ultra-Relativistic Black Hole Accretion}

\begin{figure*}[htp]
\centering
\subfigure{\includegraphics[trim={1cm, 0.5cm, 1cm, 0.5cm}, clip, width=0.45\textwidth]{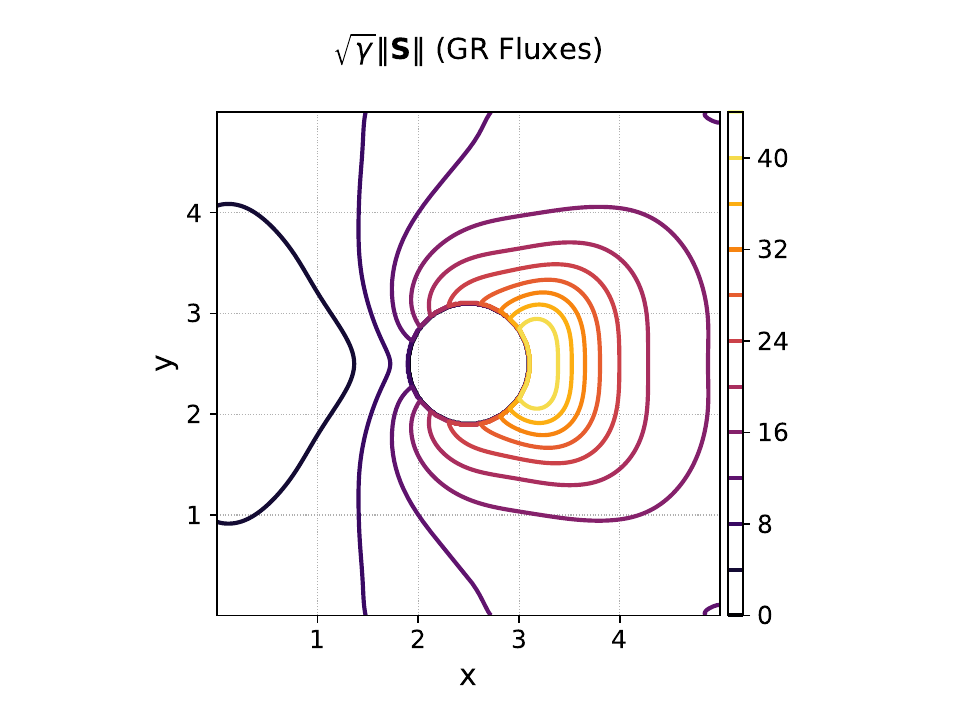}}
\subfigure{\includegraphics[trim={1cm, 0.5cm, 1cm, 0.5cm}, clip, width=0.45\textwidth]{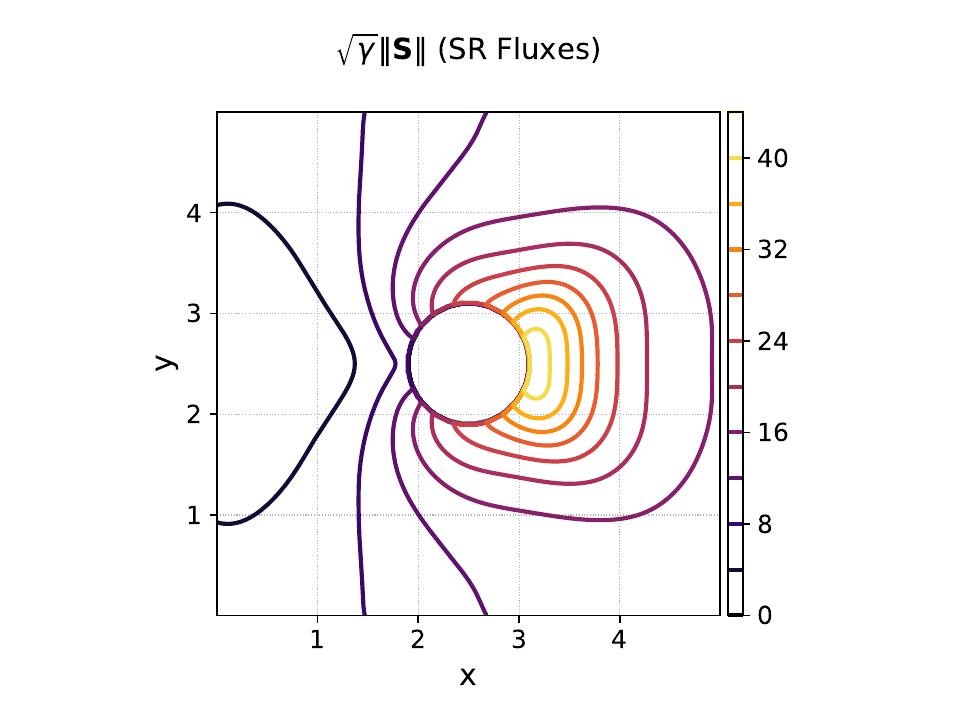}}
\caption{The configuration of fluid momentum contours at time ${t = 50}$ for the Petrich-Shapiro-Teukolsky stiff accretion problem onto a static black hole with ${v_{\infty} = 0.3}$ and ${a = 0}$, obtained using standard general relativistic fluxes (left) and local special relativistic fluxes in an orthonormal tetrad basis (right).}
\label{fig:bhl_static_S}
\end{figure*}

\begin{figure*}[htp]
\centering
\subfigure{\includegraphics[width=0.45\textwidth]{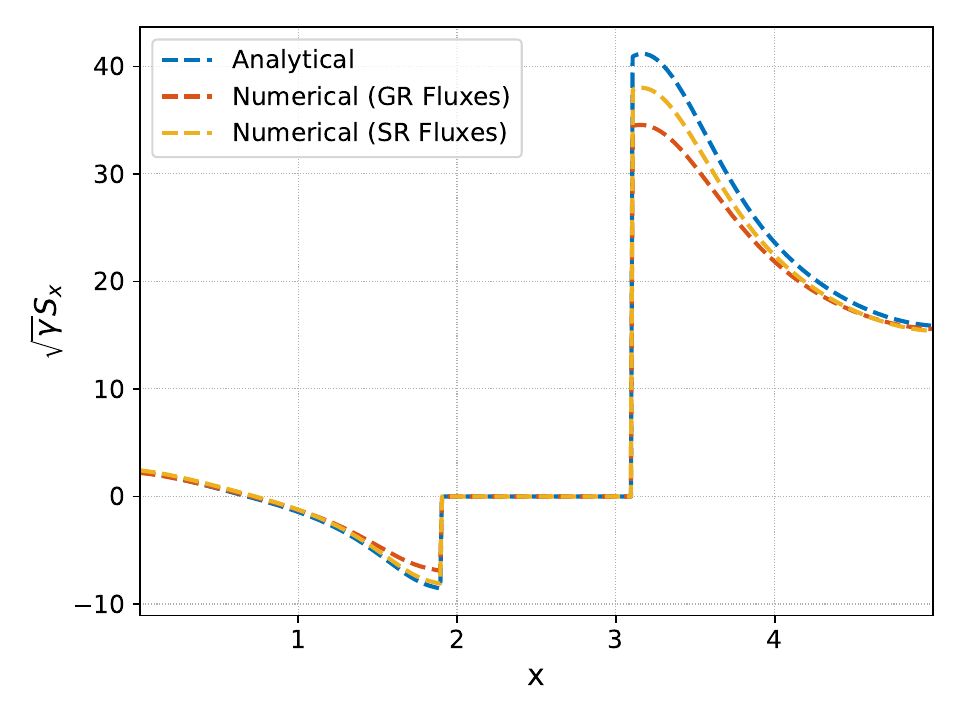}}
\subfigure{\includegraphics[width=0.45\textwidth]{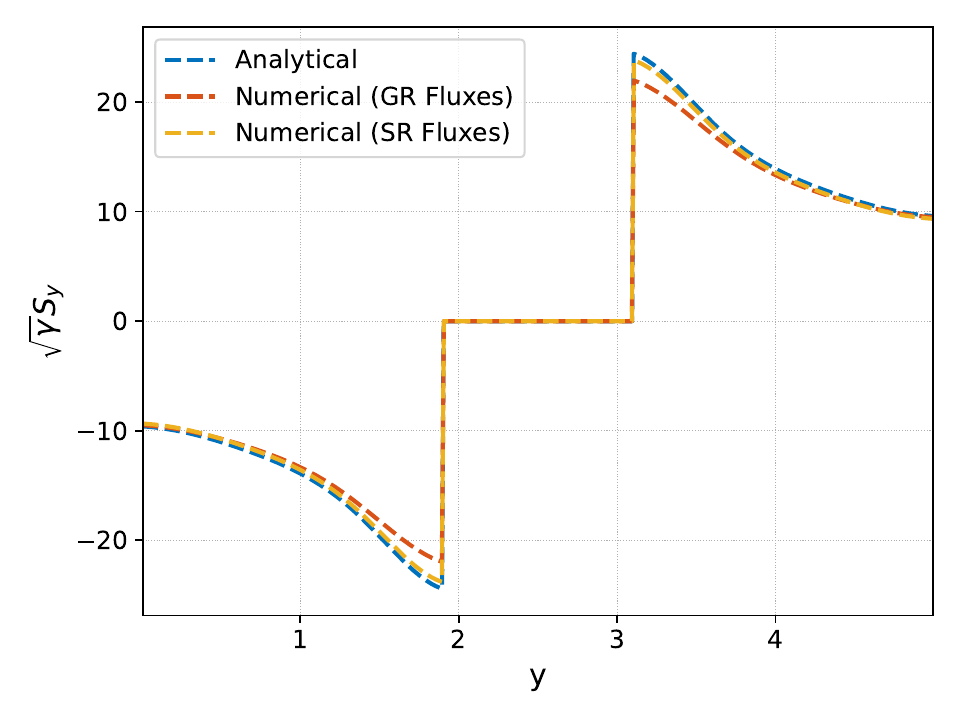}}
\caption{A cross-sectional profile of the $x$-component (left) and $y$-component (right) of the fluid momentum density ${\mathbf{S}}$ (as perceived by Eulerian observers) at time ${t = 50}$ for the Petrich-Shapiro-Teukolsky stiff accretion problem for a static black hole with ${v_{\infty} = 0.3}$ and ${a = 0}$, validating both the special and general relativistic Riemann solver approaches against the analytical solution. Cross sections are taken at ${y = 2.5}$ (left) and ${x = 2.5}$ (right).}
\label{fig:bhl_static_S_cross}
\end{figure*}

\citet{petrich_shapiro_teukolsky1988} derived a remarkable analytical solution to the equations of general relativistic hydrodynamics describing the subsonic accretion of an ultra-relativistic fluid onto a (potentially spinning) black hole. In particular, they assumed a \textit{stiff} ultra-relativistic equation of state, in which ${\Gamma = 2}$, and therefore ${\rho = p}$ and ${c_s = 1}$, thus guaranteeing that the fluid flow always remains strictly subsonic. Subject to these assumptions, the conservation of energy-momentum reduces to a covariant wave equation,

\begin{equation}
\nabla^{\mu} \nabla_{\mu} \psi = 0,
\end{equation}
in the \textit{stream function} ${\psi}$, from which the fluid density ${\rho}$ (and hence pressure $p$) and covariant components of the four-velocity ${u_{\mu}}$ may be reconstructed as:

\begin{equation}
\rho = p = \left( \partial_{\mu} \psi \right) \left( \partial^{\mu} \psi \right), \qquad u_{\mu} = \frac{1}{\sqrt{\rho}} \partial_{\mu} \psi,
\end{equation}
respectively. For the case of stiff fluid accretion onto a static (Schwarzschild) black hole of mass $M$ within the spherical coordinate system ${\left( t, r, \theta, \phi \right)}$, with the standard metric,

\begin{equation}
\begin{split}
d s^2 &= g_{\mu \nu} \, d x^{\mu} \, d x^{\nu}\\
&= - \left( 1 - \frac{2 M}{r} \right) d t^2 + \left( 1 - \frac{2 M}{r} \right)^{-1} d r^2\\
&+ r^2 \left( d \theta^2 + \sin^2 \left( \theta \right) d \phi^2 \right),
\end{split}
\end{equation}
they obtained the steady-state solution,

\begin{equation}
\psi = - u_{\infty}^{t} t - 2 M u_{\infty}^{t} \log \left( 1 - \frac{2 M}{r} \right) + u_{\infty} \left( r - M \right) \cos \left( \theta \right),
\end{equation}
yielding

\begin{equation}
\begin{split}
\rho = p &= \left( u_{\infty}^{t} \right)^2 \left[ 1 + \frac{2 M}{r} + \left( \frac{2 M}{r} \right)^2 + \left( \frac{2 M}{r} \right)^{3} \right]\\
&- \left( u_{\infty} \right)^2 \left[ 1 - \frac{2 M}{r} + \left( \frac{M}{r} \right)^2 \sin^2 \left( \theta \right) \right]\\
&+ 8 \left( \frac{M}{r} \right)^2 u_{\infty} u_{\infty}^{t} \cos \left( \theta \right)
\end{split}
\end{equation}
for the fluid density ${\rho}$ (and pressure $p$), and:

\begin{equation}
\begin{split}
\sqrt{\rho} \, u_t &= - u_{\infty}^{t},\\
\sqrt{\rho} \, u_r &= - \frac{4 M^2 u_{\infty}^{t}}{r \left( r - 2 M \right)} + u_{\infty} \cos \left( \theta \right),\\
\sqrt{\rho} \, u_{\theta} &= - u_{\infty} \left( r - M \right) \sin \left( \theta \right),\\
\sqrt{\rho} \, u_{\phi} &= 0,
\end{split}
\end{equation}
for the covariant components of the fluid four-velocity ${u_{\mu}}$. In the above, ${u_{\infty} = \mathbf{u}_{\infty} \cdot \mathbf{x}}$ designates the projection of the fluid four-velocity ${\mathbf{u}_{\infty}}$ at an infinite spatial distance from the black hole onto the wind direction ${\mathbf{x}}$ (with ${u_{\infty}^{t}}$ designating its timelike component). Assuming a square domain ${\left( x, y \right) \in \left[ 0, 5 \right] \times \left[ 0, 5 \right]}$ containing a static (Schwarzschild) black hole of mass ${M = 0.3}$ centered at ${\left( x, y \right) = \left( 2.5, 2.5 \right)}$, and a fluid velocity at spatial infinity of ${v_{\infty} = 0.3}$ (with the wind direction ${\mathbf{x}}$ oriented directly at the black hole, and a fluid density at spatial infinity of ${\rho_{\infty} = 10}$), we run all simulations using a spatial discretization of ${400 \times 400}$ cells, a CFL coefficient of 0.95, and up to a final time of ${t = 50}$. The relativistic momentum contours ${\sqrt{\gamma} \, S_i S^i}$ of the fluid, obtained using a general relativistic Riemann solver and a local special relativistic Riemann solver adapted to an orthonormal tetrad basis, are shown in Figure \ref{fig:bhl_static_S}. As in the case of the black hole magnetosphere simulations shown previously, we can see that there is (marginally) sharper resolution of the momentum contours when using the local special relativistic Riemann solver as compared to the general relativistic Riemann solver; this can be confirmed by taking cross-sectional profiles of the $x$- and $y$-components of the fluid momentum vector ${\sqrt{\gamma} \, \mathbf{S}}$ through the ${y = 2.5}$ and ${x = 2.5}$ axes, as shown in Figure \ref{fig:bhl_static_S_cross}. As in the general relativistic electromagnetic case, we observe faster convergence to the analytical solution of \citet{petrich_shapiro_teukolsky1988} for the special relativistic Riemann solver approach; once again, this is assumed to be because of our ability to use more uniform wave-speed estimates across the computational domain when applying a special relativistic Riemann solver than when applying a general relativistic one.

\begin{figure*}[htp]
\centering
\subfigure{\includegraphics[trim={1cm, 0.5cm, 1cm, 0.5cm}, clip, width=0.45\textwidth]{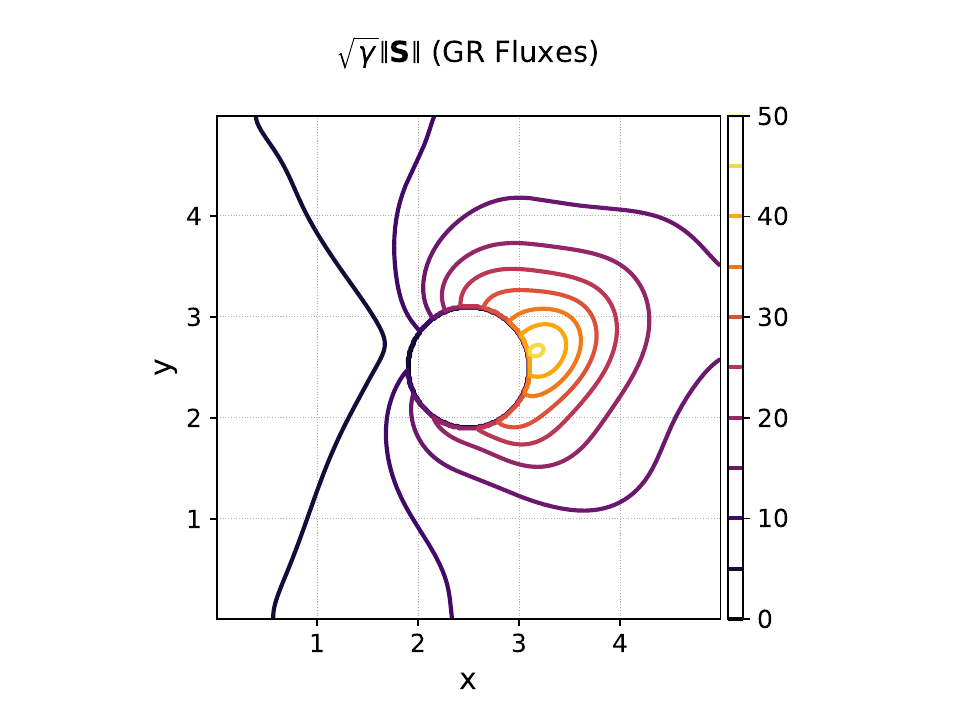}}
\subfigure{\includegraphics[trim={1cm, 0.5cm, 1cm, 0.5cm}, clip, width=0.45\textwidth]{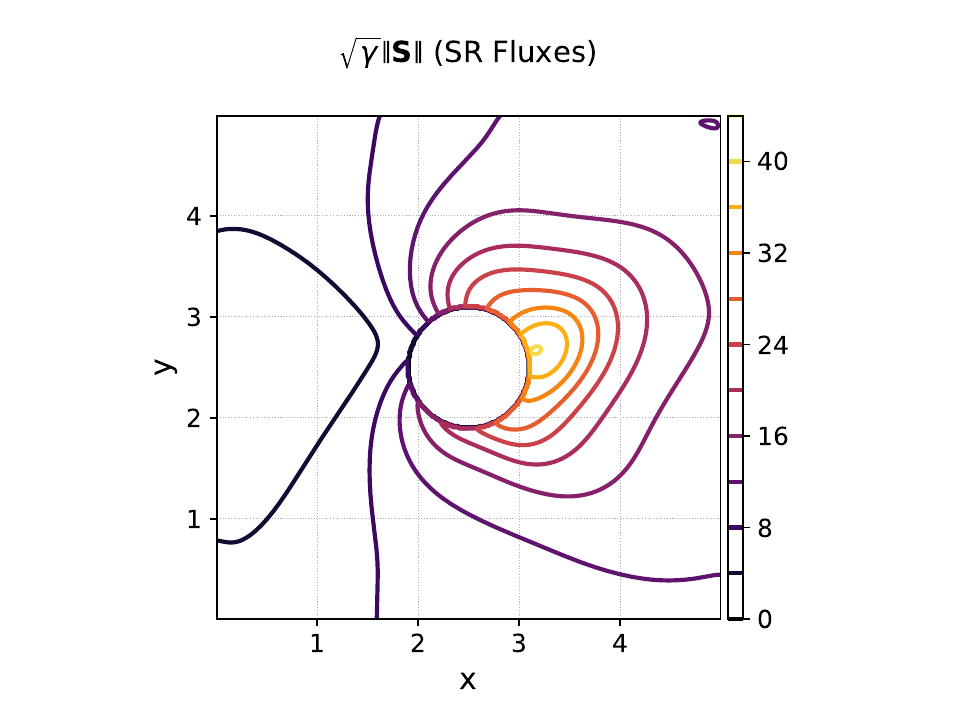}}
\caption{The configuration of momentum contours at time ${t = 9}$ for the Petrich-Shapiro-Teukolsky stiff fluid accretion problem onto a spinning black hole with ${v_{\infty} = 0.3}$ and ${a = 0.9}$, obtained using standard general relativistic fluxes (left) and local special relativistic fluxes in an orthonormal tetrad basis (right).}
\label{fig:bhl_spinning_S}
\end{figure*}

\begin{figure*}[htp]
\centering
\subfigure{\includegraphics[width=0.45\textwidth]{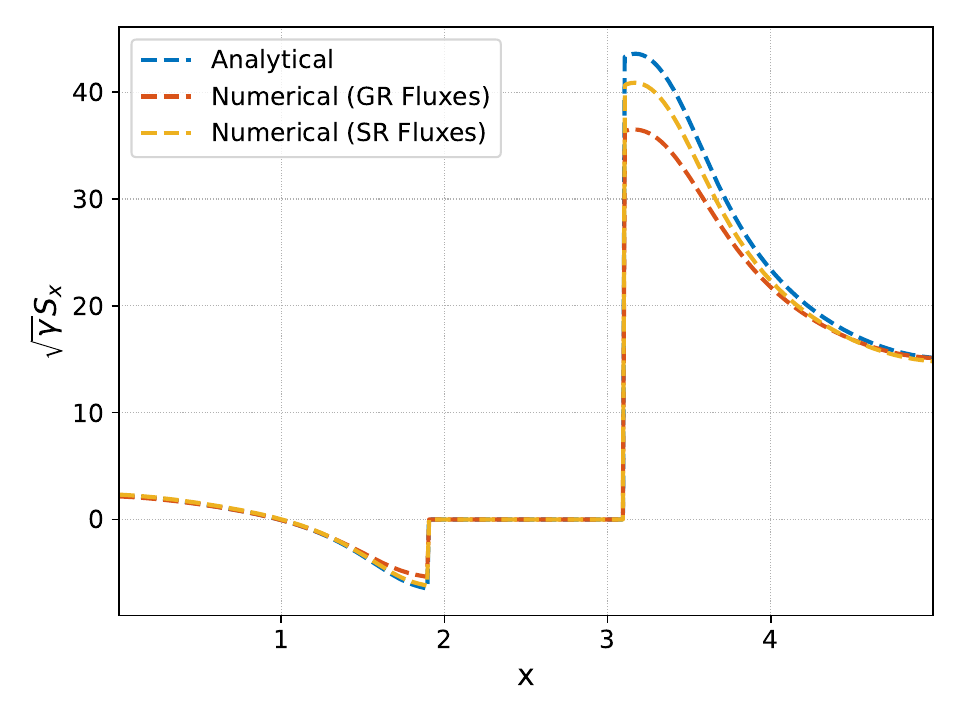}}
\subfigure{\includegraphics[width=0.45\textwidth]{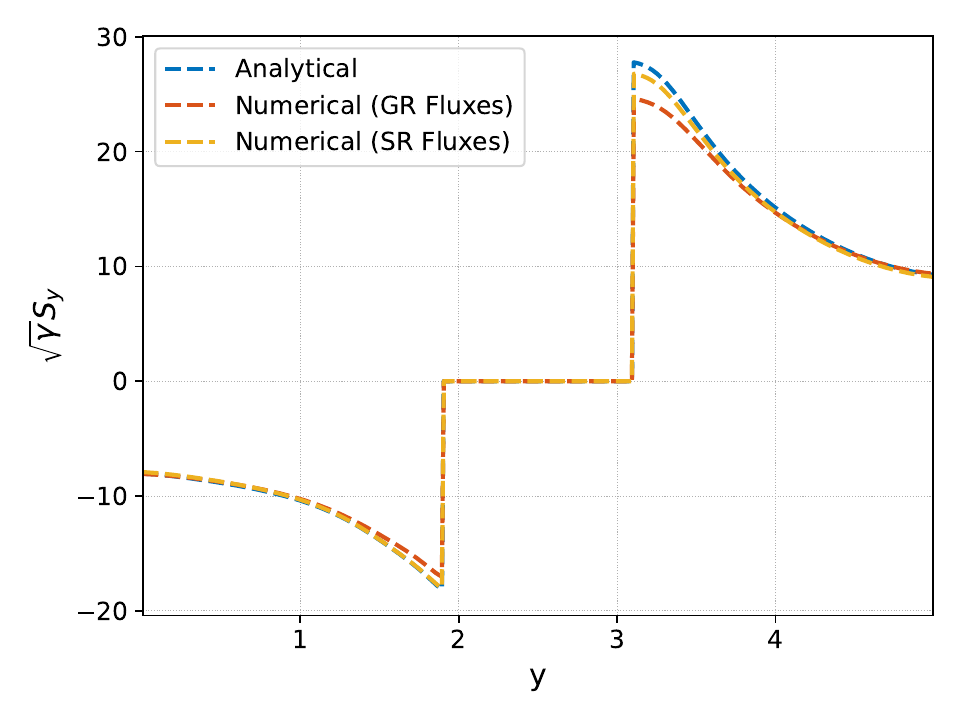}}
\caption{A cross-sectional profile of the $x$-component (left) and $y$-component (right) of the fluid momentum ${\mathbf{S}}$ (as perceived by Eulerian observers) at time ${t = 9}$ for the Petrich-Shapiro-Teukolsky stiff fluid accretion problem for a spinning black hole with ${v_{\infty} = 0.3}$ and ${a = 0.9}$, validating both the special and general relativistic Riemann solver approaches against the analytical solution. Cross sections are taken at ${y = 2.5}$ (left) and ${x = 2.5}$ (right).}
\label{fig:bhl_spinning_S_cross}
\end{figure*}

On the other hand, for the case of stiff fluid accretion onto a spinning (Kerr) black hole of mass $M$ and dimensionless spin ${a = J / M}$ within the Boyer-Lindquist oblate spheroidal coordinate system ${\left( t, r, \theta, \phi \right)}$, with the standard metric,

\begin{equation}
\begin{split}
d s^2 &= g_{\mu \nu} \, d x^{\mu} \, d x^{\nu}\\
&= - \left( 1 - \frac{2 M r}{\Sigma} \right) d t^2 + \frac{\Sigma}{\Delta} d r^2 + \Sigma d \theta^2\\
&+ \left( r^2 + a^2 + \frac{2 M r a^2}{\Sigma} \sin^2 \left( \theta \right) \right) \sin^2 \left( \theta \right) d \phi^2\\
&- \left( \frac{2 M r a \sin^2 \left( \theta \right)}{\Sigma} \right) dt \, d \phi,
\end{split}
\end{equation}
where ${\Sigma}$ and ${\Delta}$ have the equivalent definitions as in the spherical Kerr-Schild coordinate system,

\begin{subequations}
\begin{align}
\Sigma &= r^2 + a^2 \cos^2 \left( \theta \right),\\
\Delta &= \left( r - r_{+} \right) \left( r - r_{-} \right),
\end{align}
\end{subequations}
\citet{petrich_shapiro_teukolsky1988} obtained the steady-state solution:

\begin{equation}
\begin{split}
\psi &= - u_{\infty}^{t} + \left( \frac{\left( r_{+}^{2} + a^2 \right) u_{\infty}^{t}}{2 \sqrt{M^2 - a^2}} \right) \log \left( \frac{r - r_{-}}{r - r_{+}} \right)\\
&+ u_{\infty} \left( r - M \right) \cos \left( \theta \right) \cos \left( \theta_0 \right)\\
&+ u_{\infty} \mathrm{Re} \left[ \left( r - M + ia \right) \sin \left( \theta \right) \sin \left( \theta_0 \right) \exp \left( i \left( \phi - \phi_0 - \chi \right) \right) \right],
\end{split}
\end{equation}
where

\begin{equation}
\chi = \left( \frac{a}{2 \sqrt{M^2 - a^2}} \right) \log \left[ \left( r - r_{-} \right) \left( r - r_{+} \right) \right].
\end{equation}
From this, one may first reconstruct the covariant components of the fluid four-velocity ${u_{\mu}}$ as:

\begin{equation}
\begin{split}
\sqrt{\rho} \, u_t &= - u_{\infty}^{t},\\
\sqrt{\rho} \, u_r &= - \left( r_{+}^{2} + a^2 \right) \frac{u_{\infty}^{t}}{\Delta} + u_{\infty} \cos \left( \theta \right) \cos \left( \theta_0 \right)\\
&+ u_{\infty} \mathrm{Re} \left[ \left( 1 + ia \left( \frac{r - M + ia}{\Delta} \right) \right) \sin \left( \theta \right) \right.\\
&\left. \times \sin \left( \theta \right) \sin \left( \theta_0 \right) \exp \left( i \left( \phi - \phi_0 - \chi \right) \right) \right]\\
\sqrt{\rho} \, u_{\theta} &= u_{\infty} \left( r - M \right) \sin \left( \theta \right) \cos \left( \theta_0 \right)\\
&+ u_{\infty} \mathrm{Re} \left[ \left( r - M + ia \right) \cos \left( \theta \right) \sin \left( \theta_0 \right) \right.\\
&\left. \times \exp \left( i \left( \phi - \phi_0 - \chi \right) \right) \right],\\
\sqrt{\rho} \, u_{\phi} &= -u_{\infty} \mathrm{Im} \left[ \left( r - M + ia \right) \sin \left( \theta \right) \sin \left( \theta_0 \right) \right.\\
&\left. \times \exp \left( i \left( \phi - \phi_0 - \chi \right) \right) \right],
\end{split}
\end{equation}
followed by the fluid density ${\rho}$ (and fluid pressure $p$) as

\begin{equation}
\begin{split}
\rho = p &= \left( \Sigma \Delta \right)^{-1} \left[ \left( r^2 + a^2 \right) u_{\infty}^{t} - a \left( \sqrt{\rho} \, u_{\phi} \right) \right]^2\\
&- \left( \Sigma \sin^2 \left( \theta \right) \right)^{-1} \left[ \left( \sqrt{\rho} \, u_{\phi} \right) - a \sin^2 \left( \theta \right) \left( u_{\infty}^{t} \right) \right]^2\\
&- \left( \frac{\Delta}{\Sigma} \right) \left( \sqrt{\rho} \, u_r \right)^2 - \frac{\left( \sqrt{\rho} \, u_{\theta} \right)^2}{\Sigma}.
\end{split}
\end{equation}
Assuming now a spinning (Kerr) black hole of mass ${M = 0.3}$ and spin ${a = 0.9}$, but keeping all other simulation parameters the same as in the static (Schwarzschild) case, the relativistic momentum contours ${\sqrt{\gamma} \, S_i S^i}$ of the fluid, obtained using general relativistic and local special relativistic Riemann solvers, are shown in Figure \ref{fig:bhl_spinning_S}. In contrast to the static case, in the spinning case we now see a rather substantial difference in the sharpness of the two results, and the $x$- and $y$-components of the fluid momentum vector ${\sqrt{\gamma} \, \mathbf{S}}$ through the ${y = 2.5}$ and ${x = 2.5}$ axes, as shown in Figure \ref{fig:bhl_spinning_S_cross}, demonstrate an even greater discrepancy in the convergence rates to the analytical solution between the two Riemann solvers. This result is consistent with our previous findings for the curved spacetime Maxwell equations, wherein the relative advantages of the tetrad-first approach become progressively more manifest as the spacetime coordinate system becomes progressively more distorted (i.e., the advantage is most apparent when ${\alpha \ll 1}$ and ${\left\lVert \boldsymbol\beta \right\rVert \gg 0}$, as in the case of Kerr spacetimes with high values of the spin parameter $a$). Note that, in these plots, we have placed the excision boundary of the spinning black hole at the Schwarzschild radius ${r = 2M}$, rather than at the outer Kerr horizon ${r = r_{+} = M + \sqrt{M^2 - a^2}}$ as we did for the magnetospheric simulations; this placement is done to prevent numerical instabilities from appearing in the general relativistic Riemann solver results that would otherwise have prevented direct comparison with the analytical solutions.

In the above, in order to make the requisite comparisons between the analytical solutions of \citet{petrich_shapiro_teukolsky1988} in the non-horizon-adapted Boyer-Lindquist oblate spheroidal coordinate system, and the numerical solutions produced by \textsc{Gkeyll} in the horizon-adapted Cartesian Kerr-Schild coordinate system, we first transform the time and azimuthal angular coordinates ${t^{BL}}$ and ${\phi^{BL}}$ in the Boyer-Lindquist oblate spheroidal system into the corresponding time and azimuthal angular coordinates ${t^{KS}}$ and ${\phi^{KS}}$ in the \textit{spherical} Kerr-Schild system:

\begin{equation}
\begin{split}
dt^{KS} &= dt^{BL} + \left( \frac{2 M r}{r^2 - 2 M r + a^2} \right) dr,\\
d \phi^{KS} &= d \phi^{BL} + \left( \frac{a}{r^2 - 2 M r + a^2} \right) dr,
\end{split}
\end{equation}
and then apply the same transformation from spherical to Cartesian Kerr-Schild coordinates described previously in Section \ref{sec:gr_electromagnetism}.

\subsection{Ideal Black Hole Accretion}

\begin{figure*}[htp]
\centering
\subfigure{\includegraphics[trim={1cm, 0.5cm, 1cm, 0.5cm}, clip, width=0.45\textwidth]{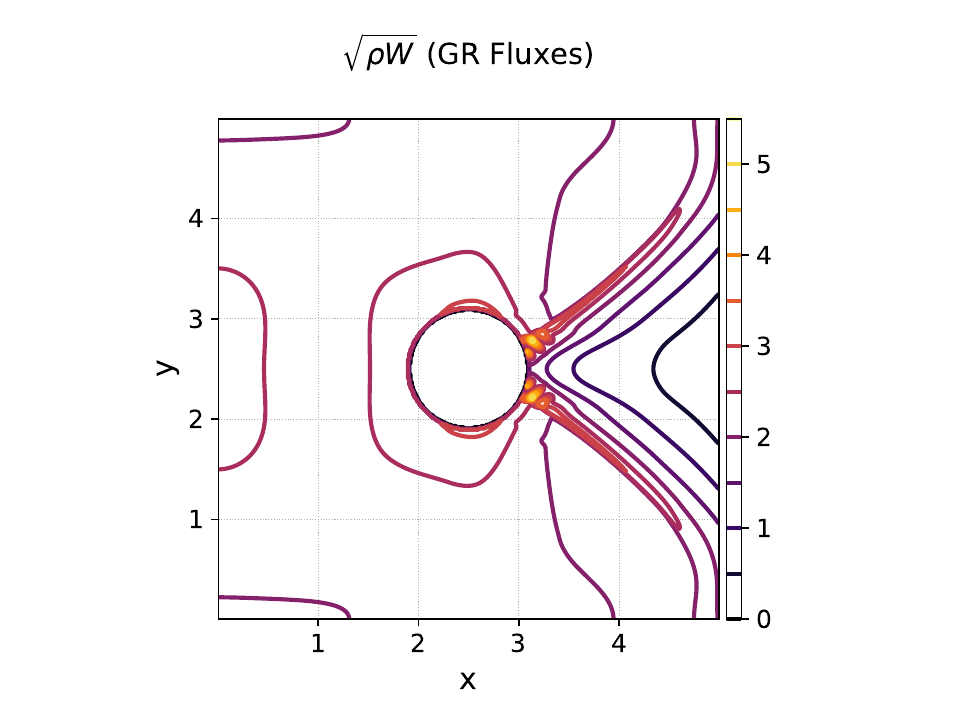}}
\subfigure{\includegraphics[trim={1cm, 0.5cm, 1cm, 0.5cm}, clip, width=0.45\textwidth]{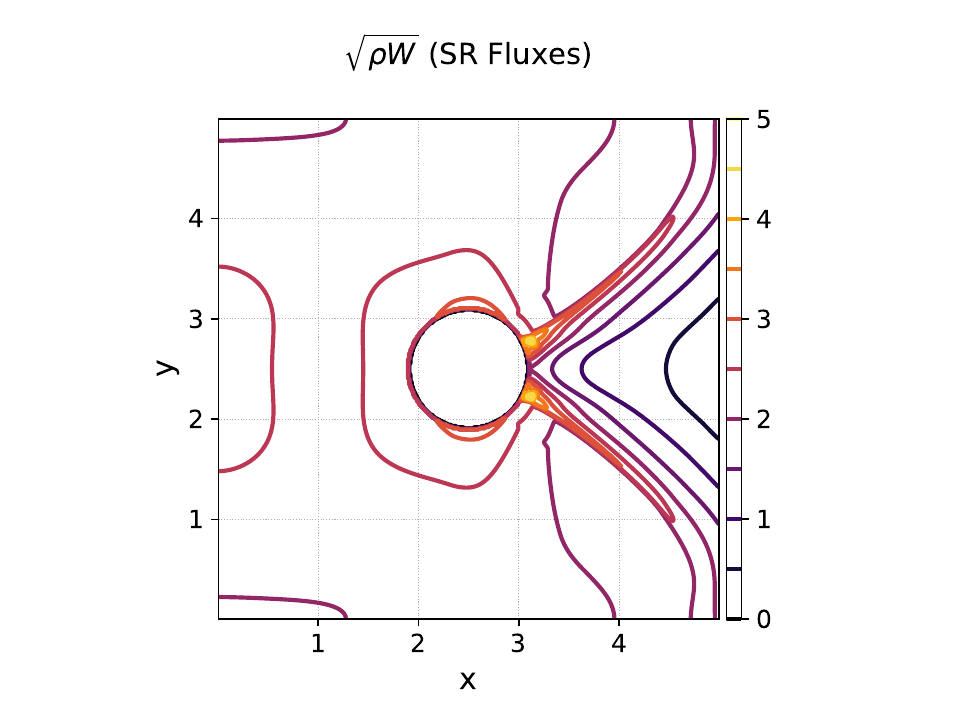}}
\caption{The configuration of mass density contours at time ${t = 15}$ for the ideal gas accretion problem onto a static black hole with ${v_{\infty} = 0.3}$ and ${a = 0}$, obtained using standard general relativistic fluxes (left) and local special relativistic fluxes in an orthonormal tetrad basis (right).}
\label{fig:bhl_ideal_static_D}
\end{figure*}

\begin{figure*}[htp]
\centering
\subfigure{\includegraphics[trim={1cm, 0.5cm, 1cm, 0.5cm}, clip, width=0.45\textwidth]{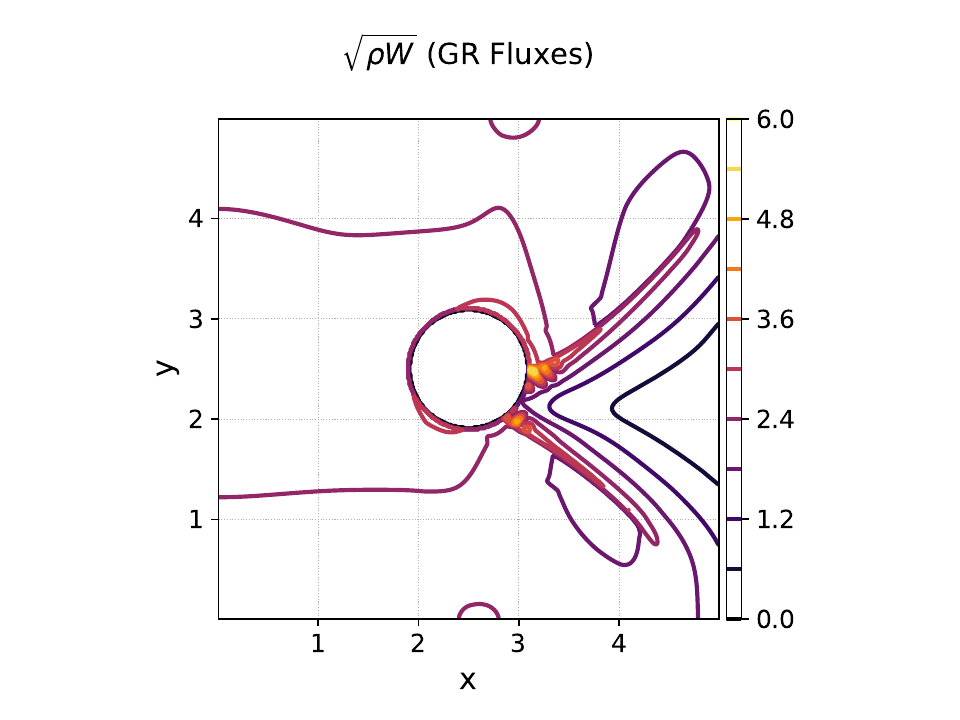}}
\subfigure{\includegraphics[trim={1cm, 0.5cm, 1cm, 0.5cm}, clip, width=0.45\textwidth]{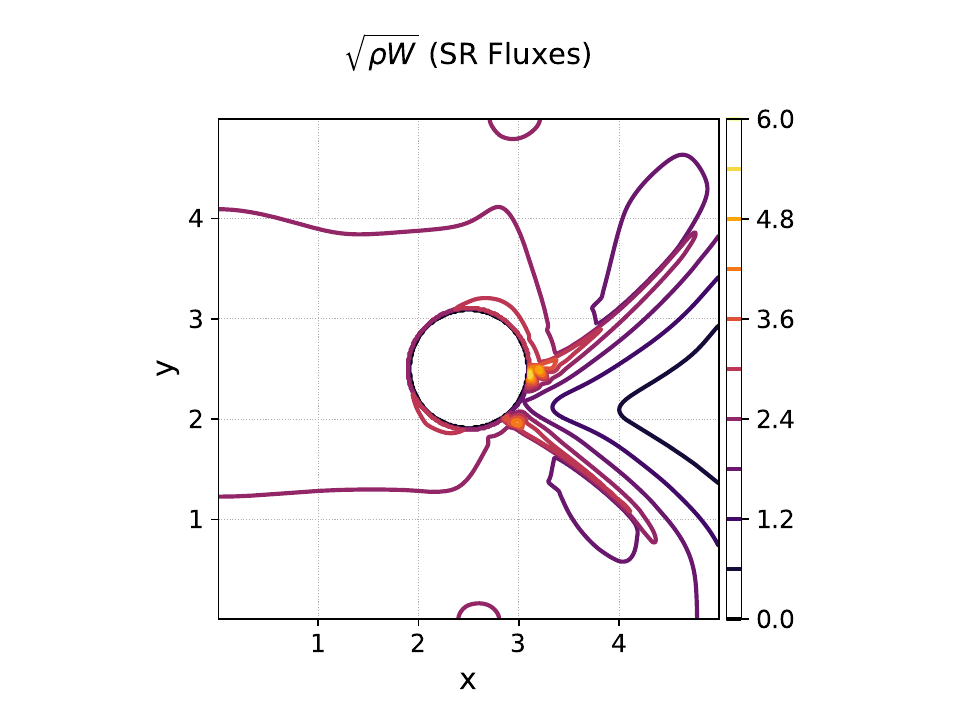}}
\caption{The configuration of mass density contours at time ${t = 15}$ for the ideal gas accretion problem onto a rapidly-spinning black hole with ${v_{\infty} = 0.3}$ and ${a = 0.999}$, obtained using standard general relativistic fluxes (left) and local special relativistic fluxes in an orthonormal tetrad basis (right).}
\label{fig:bhl_ideal_spinning_D}
\end{figure*}

As our final illustrative test problem in general relativistic hydrodynamics, we consider the supersonic accretion of a perfect fluid obeying an ideal gas equation of state onto a (potentially spinning) black hole; such problems have previously been considered by \citet{font_ibanez1998} in the Schwarzschild (static) case, and later by \citet{font_ibanez_papadopoulos1999} in the Kerr (spinning) case. There is no analytical solution against which to validate our algorithm for this problem, so we perform this test only to verify the qualitative plausibility of the results, and to make comparisons between the general relativistic and local special relativistic Riemann solver approaches for this more astrophysically realistic accretion scenario. We use the same basic wind accretion setup as in the ultra-relativistic accretion problem considered previously, with a square domain ${\left( x, y \right) \in \left[ 0, 5 \right] \times \left[ 0, 5 \right]}$ containing a static (Schwarzschild) black hole of mass ${M = 0.3}$ centered at ${\left( x, y \right) = \left( 2.5, 2.5 \right)}$, and a fluid velocity at spatial infinity of ${v_{\infty} = 0.3}$, with the wind direction oriented directly at the black hole. At spatial infinity, we set the fluid density and pressure to ${\rho_{\infty} = 3}$ and ${p_{\infty} = 0.05}$, respectively, while in the remainder of the spatial domain, the density and pressure are both set to ${\rho_0 = p_ 0 = 0.01}$. We have chosen an intentionally low fluid pressure so as to suppress some of the Rayleigh-Taylor instabilities that would otherwise appear downstream of the black hole. The adiabatic index is set to ${\Gamma = \frac{5}{3}}$, corresponding to accretion of a monatomic gas. Once again, we run all simulations using a spatial discretization of ${400 \times 400}$ cells, a CFL coefficient of 0.95, and now up to a final time of ${t = 15}$. The relativistic mass density contours ${\sqrt{D} = \sqrt{\rho W}}$ of the fluid, obtained using a general relativistic Riemann solver and a local special relativistic Riemann solver adapted to an orthonormal tetrad basis, are shown in Figure \ref{fig:bhl_ideal_static_D}. For the case of a rapidly-spinning Kerr black hole of mass ${M = 0.3}$ and spin ${a = 0.999}$, the corresponding general relativistic and local special relativistic mass density contours are shown in Figure \ref{fig:bhl_ideal_spinning_D}.

\begin{figure*}[htp]
\centering
\subfigure{\includegraphics[width=0.45\textwidth]{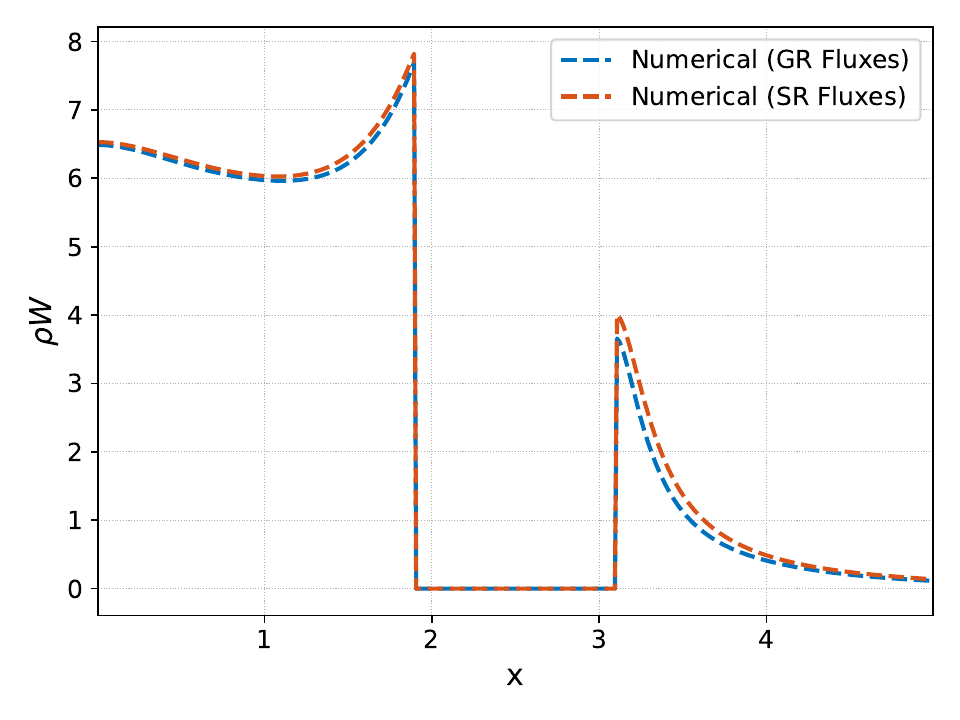}}
\subfigure{\includegraphics[width=0.45\textwidth]{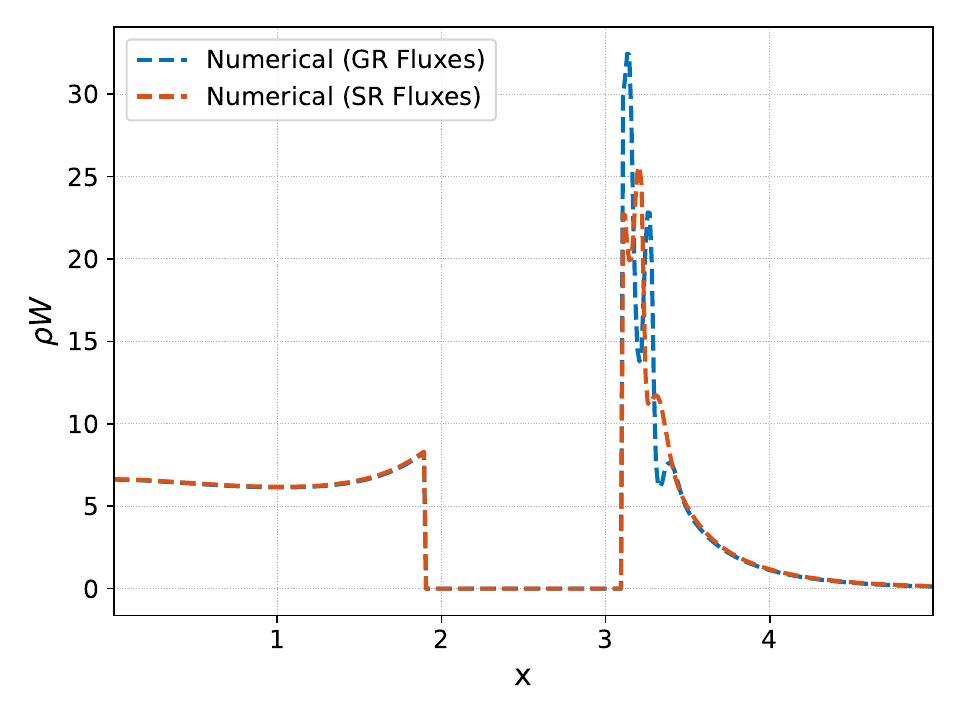}}
\caption{A cross-sectional profile of the fluid mass density ${D = \rho W}$ at time ${t = 15}$ for the ideal gas accretion problem onto a static black hole (left) and a rapidly-spinning black hole (right), with ${v_{\infty} = 0.3}$ and ${a = 0}$ (left) and ${a = 0.999}$ (right), comparing the performance of the special and general relativistic Riemann solver approaches. Cross sections are taken at ${y = 2.5}$.}
\label{fig:bhl_ideal_cross}
\end{figure*}

As in the case of ultra-relativistic accretion presented above, we have placed the excision boundary of the rapidly-spinning black hole at the Schwarzschild radius ${r = 2M}$, in order to prevent numerical instabilities from overwhelming the general relativistic Riemann solver. The resulting ideal gas accretion profiles are qualitatively similar to those obtained by \citet{font_ibanez1998} and \citet{font_ibanez_papadopoulos1999}. In both the static and rapidly-spinning cases, we note that significantly larger oscillations are observed in the fluid density profiles downstream of the black hole for the general relativistic Riemann solver than for the local special relativistic one; these oscillations may be confirmed to be numerical artifacts (as opposed to, for instance, the result of a true Rayleigh-Taylor instability) by the fact that they do not converge to a fixed profile, and instead only amplify, with increased grid resolution. In the rapidly-spinning case, these oscillations are easy to discern in the cross-sectional profiles of the fluid mass density ${D = \rho W}$ through the ${y = 2.5}$ axis, as shown in Figure \ref{fig:bhl_ideal_cross}. For higher values of the black hole spin (i.e., values beyond ${a = 0.999}$), these numerical oscillations become sufficient to destabilize the general relativistic Riemann solver altogether. The fact that these oscillations are significantly attenuated by the use of local special relativistic Riemann solvers, resulting in higher overall stability without any other apparent loss of numerical accuracy, again confirms the superiority of the tetrad-first approach in the case of highly distorted spacetime coordinate systems.

\section{Concluding Remarks}
\label{sec:concluding_remarks}

In this paper, we have proposed a new and general approach for solving systems of conservation law equations in curved spacetime using special relativistic Riemann solvers, based upon an earlier proposal of \citet{pons_font_ibanez_marti_miralles1998}, by transforming all primitive and conservative variables into a locally flat \textit{tetrad basis} at each inter-cell boundary. We have also presented an implementation of this approach within the \textsc{Gkeyll} simulation framework, with a focus upon two model equation systems in particular: general relativistic electromagnetism and general relativistic hydrodynamics. This implementation has been validated against the analytical solutions of \citet{wald1974} and \citet{blandford_znajek1977} for black hole magnetospheres, as well as the analytical solution of \citet{petrich_shapiro_teukolsky1988} for ultra-relativistic black hole accretion. Across all test cases, we find that the convergence and stability properties of local special relativistic Riemann solvers match or exceed those of standard general relativistic Riemann solvers. More specifically, we find that local special relativistic Riemann solvers typically exhibit faster and more uniform convergence to the analytical solutions than general relativistic ones, which we presume to be due to their more uniform wave-speed estimates, since the characteristic wave-speeds appearing in the special relativistic Riemann problem are not dependent upon the spacetime gauge variables ${\alpha}$ and ${\beta^i}$, and so are able to be kept more consistent across the computational domain. Moreover, we find that Riemann solvers which depend upon knowledge of the eigensystem of the flux Jacobian, such as the approximate solver of \citet{roe1981} (adapted for relativistic hydrodynamics by \citet{eulderink_mellema1995}), appear to be significantly more numerically robust in the flat spacetime case, where the eigensystem is substantially simpler.

Overall, we therefore find that, in regions of high spacetime coordinate distortion (i.e., regions within which the lapse function ${\alpha}$ becomes small, and/or the magnitude of the shift vector ${\boldsymbol\beta}$ becomes large), such as in the vicinity of rapidly-spinning black holes, our \textit{tetrad-first} approach is capable of achieving both higher numerical accuracy and greater numerical stability than a standard general relativistic Riemann solver. We further verify this observation by considering certain extreme test cases in the limit of very high (e.g., ${a = 0.999}$ or above) black hole spin, for which analytical solutions do not exist, such as the \citet{blandford_znajek1977} black hole magnetosphere problem beyond the perturbative limit, and the \citet{font_ibanez_papadopoulos1999} ideal gas accretion problem. In these cases, we find that the results generated by local special relativistic Riemann solvers remain qualitatively plausible, without any apparent loss of numerical accuracy, well into the regimes of black hole spin at which standard general relativistic Riemann solvers become prohibitively unstable. Thus, we conclude that the tetrad-first approach to the solution of general relativistic conservation laws does indeed constitute a robust numerical foundation upon which to develop future coupled multi-fluid simulations in curved spacetime. Although all of the test cases considered here have assumed a stationary background spacetime, we note that the same finite-volume numerical methods may also be applied to the solution of Einstein's equations, for instance by using the strongly hyperbolic, first-order conservation law form due to \citet{bona_masso_seidel_stela1997}. This extension of the tetrad-first approach to fully dynamical spacetimes remains a topic for future investigation.

The next major milestone in the development of a fully coupled multi-fluid solver in curved spacetime will be the implementation of a locally implicit integration scheme (following the same model as \citet{wang_hakim_ng_dong_germaschewski2020} in the case of non-relativistic plasmas) for coupling the momentum and energy equations of general relativistic hydrodynamics with the field equations of general relativistic electromagnetism. Such an implementation could then be verified against standard test problems for GRMHD codes in the appropriate limits, just as the non-relativistic multi-fluid formalism was validated by \citet{hakim_loverich_shumlak2006} against standard MHD Riemann problems, and also validated against GRPIC calculations to determine how well this multi-fluid approach approximates the kinetic response of the plasma, just as the non-relativistic multi-fluid formalism has been validated against fully kinetic simulations by \citet{wang_hakim_bhattacharjee_germaschewski2015}.

We emphasize at this point that not only do we expect the multi-fluid approach to provide advantages compared to GRMHD calculations in terms of the physics content of the underlying equations, allowing for the inclusion of e.g., Hall effects, electron inertia, and finite charge separation, but we also anticipate that this approach will provide a natural means of adding radiation reaction and quantum electrodynamic processes such as pair production. 
Close to the compact object, pair production is likely to be efficient in regions of low fluid density (see, e.g., \citet{parfrey_philippov_cerutti2019} and \citet{crinquand_cerutti_philippov_parfrey_dubus2020}), so not only does this approach allow for self-consistent fueling of the plasma in these regions, following similar prescriptions as GRPIC codes based on the local parallel electric field setting a density source, but we can even avoid a typical numerical difficulty in the single-fluid GRMHD approach of setting a density floor to avoid numerical instability. We note that single-fluid approaches which include both electron and ion dynamics, as in \citet{most_noronha_philippov2022}, may approximate the pair production within these regions of the compact objects, but treating accreting and pair-produced species as separate fluids not only simplifies the modeling, but also potentially incorporates the differential streaming between these different plasma populations: a source of free energy for launching instabilities. In fact, our tetrad-first approach, permitting the use of special relativistic Riemann solvers, may streamline the implementation of extended fluid models such as those discussed in \citet{most_noronha_philippov2022}, which include the effects of the self-consistent pressure tensor: the relativistic analogue of the ten-moment model pioneered in \citet{hakim2008} and \citet{wang_hakim_bhattacharjee_germaschewski2015}. 

Beyond the opportunity to add new physics to large-scale simulations of compact objects, there are potential numerical advantages from the multi-fluid modeling that our tetrad-first approach will permit. Since the hydrodynamic and electromagnetic equations are solved independently in the multi-fluid approach, we are able to circumvent the additional complexity of the coupled conservative-to-primitive variable reconstruction operation inherent to GRMHD, which, as outlined by \citet{noble_gammie_mckinney_delzanna2006}, generically involves solving a highly non-linear root-finding problem in higher dimensions. Further, not only is this non-linear root-finding problem more computationally expensive in GRMHD than in curved spacetime hydrodynamics, but the fact that the electromagnetic fields are solved for separately and the coupling handled through the source terms means that the maximum magnetization, ${\sigma = B^2/nmc^2}$, which can be safely handled numerically is avoided, thus allowing for the simulation of both lower densities and stronger magnetic fields. When combined with the increased numerical robustness of the tetrad-first approach, we expect that a curved spacetime multi-fluid code will therefore be capable of stably simulating plasmas with lower densities, lower pressures, higher pressure gradients, larger magnetizations, and higher Lorentz factors (in addition to including more physics) than a typical GRMHD code, thus providing a complementary approach to the simulation of a  variety of high-energy astrophysical systems.

\section*{Acknowledgments}

The authors would like to thank Kyle Parfrey for enlightening discussions regarding related approaches, Sasha Philippov for useful suggestions on the manuscript, and Kiran Eiden for earlier work on prototyping an alternative approach to the curved spacetime Maxwell solver in \textsc{Gkeyll}. J.G. was partially funded by the Princeton University Research Computing group. J.G., A.H., \& J.J. were partially funded by the U.S. Department of Energy under Contract No. DE-AC02-09CH1146 via an LDRD grant. J.M.T. was supported by the Simons Foundation and NSF Award 2206756. The development of \textsc{Gkeyll} was partially funded by the NSF-CSSI program, Award Number 2209471.

\end{document}

\maketitle
\section*{Preface}

\texttt{lineno.sty} is a macro package made by
Stephan~I.~Böttcher for attaching line numbers to
\LaTeX\ documents. Some people have used it for revising
submittings in collaboration with referees or co-authors.
Documentations are nowadays preferred to be in
Adobe's \texttt{PDF}---so \texttt{lineno.sty}'s
documentation is \lcurl[lineno/]{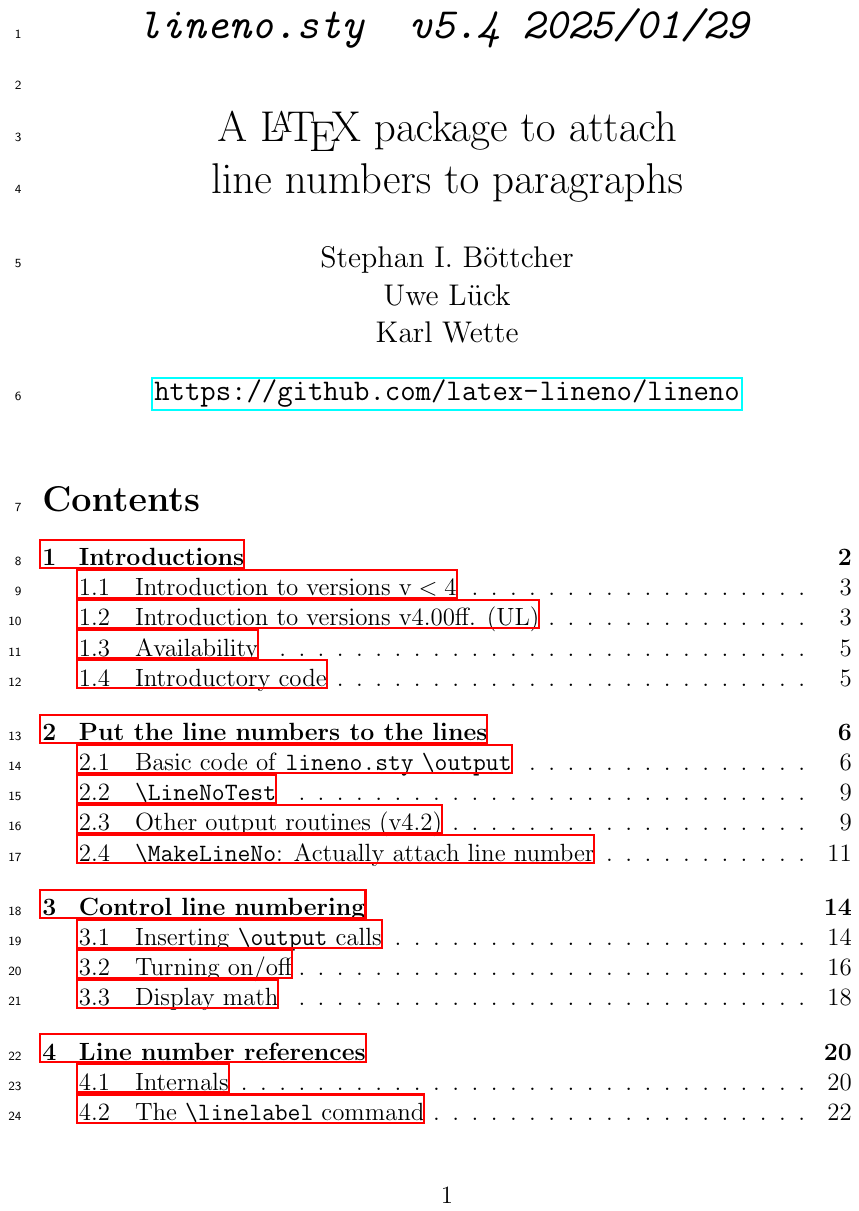}.

\texttt{ednotes.sty} uses \texttt{lineno.sty} for critical
editions, combining it with Alexander~I.~Rozhenko's
\texttt{manyfoot.sty}---this was Christian Tapp's idea,
who then hired me for adding the \TeX nical details.
In doing this, I had to change some internals of
\texttt{lineno.sty}, so Stephan transferred maintenance
to me; then some of my macro files that I originally had
made for \texttt{ednotes.sty} wandered into the
\texttt{lineno} directory of CTAN---because they turned
out not to need \texttt{ednotes.sty},
just to work as extensions of \texttt{lineno.sty}\,.

Now, I haven't had the time for making \texttt{.dtx} versions
of the \texttt{.sty} files for \texttt{ednotes}.
Therefore, ordinary \texttt{.pdf} documentation for
the remaining \texttt{.sty} files of \texttt{lineno}
is missing.
What you see here is nothing but a somewhat structured listing
of the additional \texttt{.txt} and \texttt{.sty} files in
\texttt{PDF}, deriving from the \texttt{verbatim} package and
its \cs{verbatiminput} command. I hope the high quality
(scalable) output is worth it.

By contrast, the new package \texttt{fnlineno.sty} added in 2011 for
footnote line numbers is documented in \lcurl[lineno/]{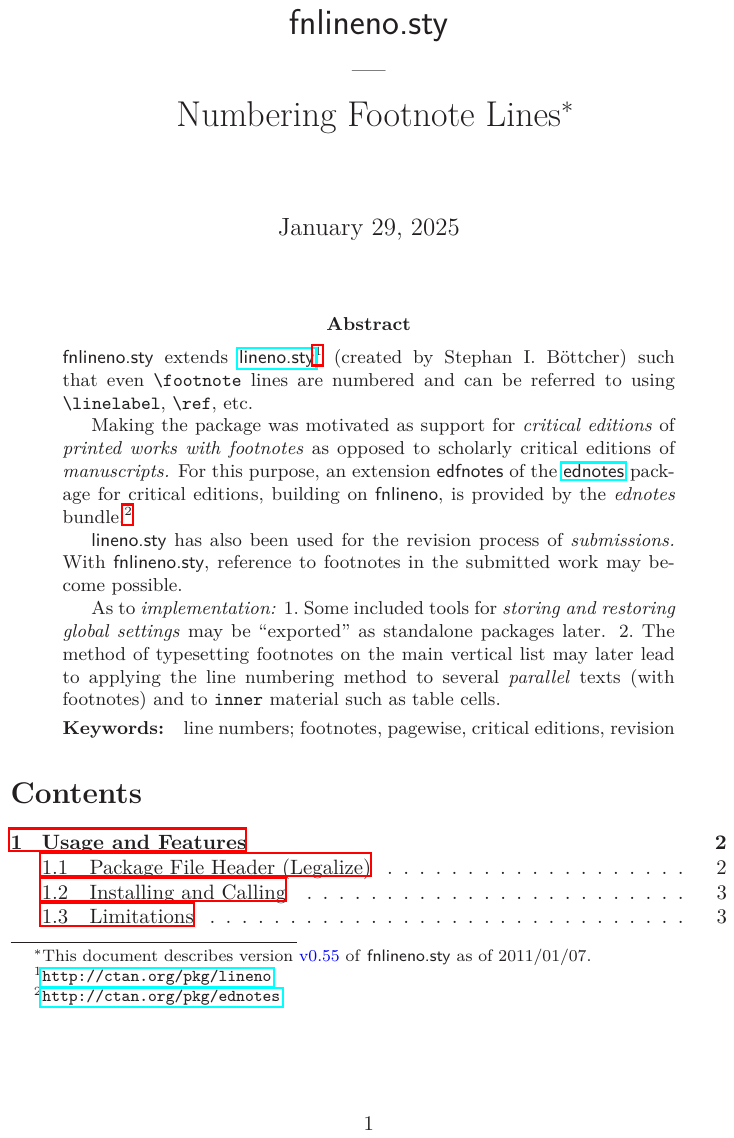}
in high quality, using the \lcurl{nicetext} bundle.

\leavevmode\hfill \textit{U.\,L.}

\newpage
\tableofcontents

\section{The \texttt{.txt} files}
\subsection{Summary: \texttt{README.txt}}
\verbatiminput{README.txt}
\subsection{Licenses/Copyright: \texttt{COPYING.txt}}
\begin{verbatim}
% lineno.sty
%
% The files in this directory are
%
% Copyright 1995--2003 Stephan I. Böttcher
% Copyright 2002--2005 Uwe Lück for versions 4.x and code from former Ednotes
% Copyright 2011       Uwe Lück for fnlineno.sty/tex/pdf
% Copyright 2021--2023 Karl Wette for versions 5.x
%
% The files can be redistributed and/or modified under
% the terms of the LaTeX Project Public License; either
% version 1.3a of the License, or any later version.
% The latest version of this license is in
%     http://www.latex-project.org/lppl.txt
% We did our best to help you, but there is NO WARRANTY.
%
% ============================================================================
<- (UL) End of official declaration.

History:
Update for version 5.x 2022/12/05 kw
add. fnlineno          2011/02/16 ul
This file -> The files 2004/10/12 ul
$Id: COPYING,v 3.0.2.1 2004/09/13 20:15:47 stephan Exp $
LPPL v1.3a 2004/10/26
\end{verbatim}
\subsection{Update summaries: \texttt{CHANGEs.txt}}
\begin{verbatim}
CHANGES  for lineno pkg v5.4   2025/01/29:

1. Option to allow ~\linelabel~ with ~\nolinenumbers~.


CHANGES  for lineno pkg v5.3   2023/05/20:

1. Handle special value of \prevdepth=-1000pt;
   thanks to Yukai Chou and Frank Mittelbach.
2. Add package options `sep' and `width' for
   setting line number separation and width.
3. Add `twocolumn` support for `bframe` env;
   thanks to Yukai Chou.


CHANGES  for lineno pkg v5.2   2023/05/19:

1. Support `amsmath` option `\allowdisplaybreaks`.


CHANGES  for lineno pkg v5.1   2023/01/19:

1. Patch `amsmath` with \AddToHook if possible.


CHANGES  for lineno pkg v5.0   2022/07/30:

1. Merge in `linenoamsmath' patches.


CHANGES  for lineno pkg        2011/02/16:

1. new fnlineno.sty for numbering footnote lines + \linelabel,
described in fnlineno.pdf

2. new overview SRCFILEs.txt

(From README.txt -- KW) numline obsolete, mirror.ctan.org, fnlineno,
amsmath compatibility, CHANGEs.txt, lineno.tds.zip


CHANGES  for lineno.sty v4.41  2005/11/02:

1. Loadable after amsmath.

2. Removed some nonsense from documentation.


CHANGES  for lineno.sty v4.4   2005/10/27:
[failed to be uploaded to CTAN, but distributed by mail]:

1. Proper effective line depth at end of paragraphs.
   The spacing bug was quite obvious in two-column mode
   when a paragraph end was at a column bottom.

2. Another bug concerning two-column mode that had been
   introduced in v4.22 has been removed again.

3. Support for \addvspace introduced more and more bugs
   in versions of v4.32 and v4.33. The reasons seem to
   be clear now, and v4.4 should be stable.


CHANGES  for lineno.sty v4.32  2005/10/17:

1. Support for \addvspace
   (a math display or a list meets a heading -- or a sub-heading
   follows a heading -- or the like)

2. Clearly explained former option `displaymath' and its change
   to a default.


CHANGES  for lineno.sty v4.31  2005/10/01:

1. \modulolinenumbers* and a package option `modulo*' for
   printing first line number after interrupting editor's
   text, regardless of the modulo.

2. Improved explanation of \firstlinenumber and package
   options.


CHANGES  for lineno.sty v4.3   2005/05/16:

1. Option `displaymath' (proper numbering at paragraphs
   containing math displays) becomes default.

2. Compatibility with hyperref now indeed (at least much more).

3. Tidied up documentation: terrible confusion of \newcounter
   vs. \stepcounter; sections on the same matter written at
   different times; ...

4. Additional internal improvements that perhaps hardly are
   observable (no more spurious linenumbers in math displays
   from vertical mode; some compatibity with packages that use
   \holdinginserts; \linelabel in headings etc.).


CHANGES  for lineno.sty v4.22  2005/05/09:

1. Restored "global" version of numbering lines of a \parbox or
   minipage or ... (I had missed and disabled this in taking
   over lineno.sty), explained in documentation (lineno.pdf/dvi
   subsec. 7.2).

2. Enabled \flushbottom in two-column pagewise or switching
   mode.

3. Re-implemented modulo mode -- disabling certain users' tricks
   see lineno.pdf sec. 5.5, also for a still supported
   substituting trick.

4. Tidied up setting the next line number globally vs. locally
   (TeXbook p. 301).

5. Tidied up discussions (in documentation) of possible changes
   in implementation.


CHANGES  for lineno.sty v4.21  2005/04/28:

Removed serious flaws with math display, + something ...


CHANGES  for lineno.sty v4.2   2005/04/26:

1. Re-enabled package option `displaymath' (needed rearrangement
   after sec. 5).

2. New package option `addpageno' for adding page numbers to
   line number references -- see sec. 6.1 of lineno.pdf.

3. Improved support for \includeonly (and improved sec. 5.3,
   p. 27).

4. Improved compatibility with other packages that change \output
   (tameflts.sty, e.g., for saving footnotes against \marginpar
   and floats), added advice on this matter -- sec. 2.3,
   pp. 7, 14f.


CHANGES  for lineno.sty v4.11  2005/03/08:

1. The `edtable' option now supports math environments like
`array'. This requires updating edtable.sty to v1.3. (The claim
in previous versions of edtable on this support simply was
wrong, sorry.) For how to make use of this support, we urge you
to read the usage instructions in edtable.sty (v1.3). These have
been extended very much, structured more clearly, and supplied
with examples.

2. \linelabel now complains when appearing outside line
numbering mode. This counters Stephan Boettcher's original
intention, but I generalize from my own experience that it is
helpful to be told when you have forgotten to switch into line
numbering mode and (e.g.) wonder why all the line number
references are 1 ... ednotes.sty users profit as well (at least
I expect and hope that anybody profits, which, to be sure, does
not mean that I hope that anybody forgets to switch ...).

3. The subsection on `edtable' in lineno.sty/tex/pdf was not
quite correct or complete -- corrected, improved.

4. The final list of user commands in lineno.sty/tex/pdf has not
been complete. This has not changed, but it is briefly explained
what is missing and where it can be found.


CHANGE  2005/01/20:

We have devised macros for indexing with line numbers,
yet we don't take the time to release them officially.
If you are interested, please ask via http://contact-ednotes.sty.de.vu


CHANGES  for lineno.sty v4.1   2004/10/19:

Extension packages from the Ednotes bundle for enabling
\linelabel in math mode and tabular environments are now
handled through new package options `mathrefs', `edtable',
`longtable', and `nolongtablepatch'. Two of these extension
packages moved from the ednotes folder to the lineno folder.
lineno.tex/pdf has been updated accordingly.
(From README.txt -- KW) minor fixes (removed `supported/', e.g.);
bracketed text; lines with `(UL)'; redistribution with README only.

CHANGES  for lineno.sty v4.00  2004/09/03:

o incorporated earlier extension packages linenox0.sty,
  linenox1.sty, lnopatch.sty (which belonged to the
  Ednotes bundle before);

o adopted LaTeX Project Public License v1.3.


EARLIER CHANGES (From README.txt -- KW)

-  2004-10-19  UL: package options for tabular and math
-  2004-09-03  UL: merge Ednotes changes, taking over lineno.sty
-  2002 .. 2003 FMi, UL, SiB: bug fixes
-  2001-08-04 SiB: linenomath wrapping for \[ \]
-  2001-07-30 SiB: [hyperref] option obsolete.
-  2001-01-17 SiB: LaTeX class option [twocolumn] support
-  2001-01-04 SiB: LaTeX class option [fleqn] support
-  2000-12-18 SiB: longtable compatibility
-  2000-07-01 SiB: extra \newlabel items, [hyperref] option
-  2000-03-10 SiB: indirect call of \output, to work with multicol.
-  1999-08-28 SiB: fixed the footnote problem using \holdinginserts
-  1999-06-11 SiB: included the extensions into lineno.sty
-  1999-03-02 SIB: Added LPPL License
\end{verbatim}
%% rm. 2011/02/16:
% \subsection{Files and subdirectories: \texttt{FILEs.txt}}
% \verbatiminput{FILEs.txt}
\subsection{Source file infos: \texttt{SRCFILEs.txt}}
\begin{verbatim}
ednmath0.sty            2005/01/10 v0.2b math support for lineno/ednotes (ul)
edtable.sty             2005/10/03 v1.3c arrays with lineno + ednotes (ul)
fnlineno.sty            2011/01/07 v0.55 numbers to footnote lines (UL)
lineno.sty              2025/01/29 line numbers on paragraphs v5.4
vplref.sty              2005/04/25 v0.2a page-line cross-refs
fnlineno.tex            2011/02/14 documenting fnlineno.sty (UL)
lineno.tex              2025/01/29 line numbers on paragraphs v5.4
lnosuppl.tex            2011/02/16 documenting supplementary files
ulineno.tex             2001/08/03 lineno.sty users manual
linenoamsmathdemo.tex   2021/09/30 Make amsmath work with lineno
\end{verbatim}

\section{Tabular and array environments}
\texttt{lineno.sty}'s package options \texttt{edtable},
\texttt{longtable}, and \texttt{nolongtablepatch}
redefine \LaTeX\ tabular and array environments
such that \texttt{lineno} and \texttt{ednotes} commands
can be used inside. The code for these options resides
in separate files at present. We are listing them here.
\subsection{\texttt{edtable.sty}}
\begin{verbatim}
%% `edtable.sty'---Uwe Lück, direction Christian Tapp.
%% LaTeX package for tables with line numbers and
%% editorial notes.
%%
%% Copyright (C) 2003-2005 Uwe Lück
%%
\def\fileversion{1.3c} \def\filedate{2005/10/03}
%%
%% This file can be redistributed and/or modified under
%% the terms of the LaTeX Project Public License; either
%% version 1.3 of the License, or any later version.
%% The latest version of this license is in
%%   http://www.latex-project.org/lppl.txt
%% We did our best to help you, but there is NO WARRANTY.
%%
%% Please send your comments via
%%
%%   https://github.com/latex-lineno/lineno
%%
%% * USAGE: *
%
% *Requirements/overview:*
%
% The package is made for the background of the LaTeX2e macro
% package and enhances functionality of the following packages:
% 1.) Stephan I. Böttcher's `lineno.sty' for printing line
%     numbers in the margin and a \label version \linelabel
%     relating labels to line numbers;
% 2.) our `ednotes.sty' for indicating variant readings and
%     other editorial remarks in separate footnote apparatuses;
% 3.) David Carlisle's `longtable.sty' for multi-page tables.
%                                                 %% TODO: supertabular etc.!?
% 4.) our `ltabptch.sty' for a patch of `longtable.sty'
%     (i.e., optionally, see below).
%
% Actually, only the first package is necessary for using the
% present one. The present one needs your line numbering commands
% according to the first one and its documentation to which we
% refer here. We likewise refer to the remaining packages for the
% details of their functionality.
%
% `lineno.sty' version 4.1 and (in case it is used) `ednotes.sty'
% version 0.8 (onwards) are needed. For obtaining recent versions
% of required packages, see the CTAN folder
%     /macros/latex/contrib/ednotes.
%
% *Install/load:*
% 1.) To be used, the present file must be put into a folder
%     that (La)TeX searches. You should have obtained a file
%     containing more detailed hints about this.
% 2.) We recommend loading this file *not* by
%     \usepackage[<options>]{edtable} but by loading `lineno.sty'
%     or `ednotes.sty' with package option `edtable'.
% 3.) To use the package options `longtable' and
%     `nolongtablepatch' that are described below, enter them
%     as options for `lineno.sty' or `ednotes.sty'.
%
% *User Commands:*
%
% 1. The package defines an environment `edtable' -- for its
% syntax we consider two cases:
% (a) Let <stdtable> be a tabular environment "like `tabular'".
% "Like" here means: (i) we have tested it with `tabular', but
% it should work with many more (e.g., from the `array' and
% `tabularx' packages) -- which share certain properties of
% implementation and requirements. Sorry, you must try, or we
% hope you will (and tell us). (For wizards: <stdtable> may
% probably be anything using a single \halign and \tabskip=0pt.)
% Definitively:
% (ii) LaTeX's standard `array' and other environments are *not*
% "like" `tabular'. Namely, environments that work only in math
% mode are not meant here. They are considered below in `b.'.
% (iii) `longtable' from the `longtable' package of the `tools'
% bundle neither is meant. An option of the present package
% deals with it as described below. Neither `supertabular' is
% meant.
% (iv) In general, usage with tabular environments that can
% break across pages as in (iii) is not recommended, at least
% when working with `ednotes.sty'. Even if some worked here, a
% shortcoming with `ednotes.sty' would be that the footnotes can
% appear on bad pages.
% -- Now the syntax is:
%    \begin{edtable}{<stdtable>}<args><entries>\end{edtable}
% -- where <args> and <entries> are usual arguments and entries
% for <stdtable>. So an example is:
%    \begin{edtable}{tabular}{cc}
%      left upper & right upper \\ left lower & right lower
%    \end{edtable}
% This produces just what <stdtable> does in an extra line,
% only adding line numbers in the margin and processing `ednotes'
% commands in the entries appropriately.
% (b) Let <stdarray> be an "array environment" like -- standard
% LaTeX's `array'! "Like" here means (i) ... analogously to
% <stdtable> above. (ii) "Text" ... oh, sorry, this is not clear
% to me at present, and I am in a hurry.                 %% TODO
% (iii) Nothing from the `amsmath' package seems to work here,
% sorry, I have tried a lot ... (iv) For standard LaTeX's
% `eqnarray', it is better to try the `linenomath' environment
% of `lineno.sty' [with package option `mathlines']. Its
% modifications in `amsmath' produce (at best) a spurious line
% number with the `linenomath' environment (up to now).
% -- Now the syntax is:
%    \begin{edtable}[<$$$>]{<stdarray>}
%      <args><entries>\end{edtable}
% -- where <args> and <entries> are usual arguments and entries
% for <stdarray>. <$$$> can be one of `$' or `$$'. Indeed, the
% <stdtable> syntax above (without [<$$$>]) *does not work*
% inside $...$ or $$...$$. -- An example:
%    \begin{edtable}[$$]{array}{cc}
%      a_{11} & a_{12} \\ a_{21} & a_{22}
%    \end{edtable}
% Now, this produces just what <stdarray> does in an extra line
% etc. as with <stdtable>. However, \linelabel and `ednotes.sty'
% footnotes may need the `mathrefs' option of `lineno.sty'. For
% the difference between `[$]' and `[$$]', see below.
%
% 2. When working with `ednotes.sty', in
%   \Anote{L1\<L2\>L3}{NOTE}
% L2 may contain &'s and \\'s, but L1, L3 must not! Analogously
% for \Bnote, etc., \Anotelabel...\donote..., etc. -- Well this
% holds for <stdtable>, hardly for <stdarray> ...        %% TODO
%
% 3. On positioning: You may have wondered about `extra line'
% above. This means that `edtable' starts a new line at
% \begin{edtable} and at \end{edtable}.          %% TODO: sure!?
% (a) It should be placed within a `center', `flushleft', or
% `flushright' environment for proper (vertical) spacing, but
% also works without. However, with <stdarray> and optional
% parameter `$$' of `edtable' (i.e., \begin{edtable}[$$]...),
% you can obtain the usual math display spacing *within a
% paragraph*; especially, the vertical space is less when the
% previous line is short ... and so on.
% (b) Horizontal positioning of line numbers (usually) needs a
% second run (after changing a table)! %% TODO: automatic warning.
% (For wizards: It may fail if some of \leftskip, \rightskip,
% \linewidth, and \@totalleftmargin are used in an unusual way.)
% A <dimen> register \ETextraoffset is provided whose value is
% 0pt by default and which additionally moves line numbers
% to the left (right) if given a positive (negative) value.
%                                               %% TODO: any use?
%
% *Package options:*
%
% 1. Option `longtable' adjusts the `longtable' environment
% defined by David Carlisle's longtable.sty for use with
% `lineno.sty' and `ednotes.sty'. The option makes `longtable'
% environments appear with line numbers in the margin, according
% to `lineno.sty', and process `ednotes.sty' commands if line
% numbering is active according to `lineno.sty'.
% [We maintain an alternative package with just this function
% in a slightly different implementation.] %% -> ulnltab.sty.
% Lemmas may go across table entries as with `edtable' (see above).
% ---There might be options like `supertabular' for adjusting
% other tabular environments that cannot be handled by the
% `edtable' environment provided here---they have not been
% implemented yet!                                             %% TODO
%
% 2. Option `nolongtablepatch' avoids loading/asking for
% `ltabptch.sty'. I.e.: according to LaTeX bugs database,
% tools/3180 and tools/3485, there have been problems with
% vertical spacing around `longtable' environments. By default
% the present package loads our `ltabptch.sty' or asks for it.
% Option `nolongtablepatch' overrides this. This is useful when
% you don't want to have the patch, especially when you use an
% "emergency stop" installation of TeX.
%
% *Wizard interface:*
% Macros \@ET@makeLineNumber, \@ET@use@outerhook,
% \@ET@execute@outerhook, \@ET@ampnotes \@ET@startlinewith,
% \@ET@trivialize@linelabel are provided as tools for adjusting
% tabular environments for use with `lineno.sty' and
% `ednotes.sty'. \@ET@step@linenumber is \let \stepLineNumber
% (from `lineno.sty'), but could be used as a hook for
% something different. See Environment `edtable' and Option
% `longtable' below for examples of application.
%                                        %% TODO: specification here.
%
%
%% * Acknowledgements: *
%
% Stephan I. Böttcher told us how to do it in extensive discussion
% and by providing some first code lines. We changed these essentially
% in some respects, but kept his general ideas and some parts of his
% macros, the latter even without knowing what we are doing.
%
% v1.3 is due to a request by Martin Brandenburg.
%
%  * Now for internals: *
%
\NeedsTeXFormat{LaTeX2e}[1994/12/01]
% 1994/12/01: \newenvironment* etc. %% TODO: more recent needed?
\ProvidesPackage{edtable}[\filedate\space v\fileversion\space
  arrays with lineno + ednotes (ul)]
%
% Alternative ideas for implementation:
% 1. There is a German package `TABMAC' enhancing `EDMAC' (cf.
% documentation of `ednotes.sty'). We could have rewritten its
% macros so it would work with `lineno.sty' and `ednotes.sty'
% in place of `EDMAC'. [TODO: "tabmac.sty"]
% 2. We redefine `longtable' by Option `longtable' because
% there seems to be only one reasonable use of the `longtable'
% environment in the course of a critical edition. This may be
% different for environments like `array' and `tabular' which
% can be used within text lines. Therefore, these environments
% are not redefined.
% 3. One might use \everycr, however: (a) it is executed
% before the first row, (b) it is executed in fake lines that
% `longtable' uses.
% 4. As an alternative to \everycr, there is the approach of
% redefining \cr and \crcr for setting \noalign. This needs
% redefining \@tabularcr etc.---not nice. We use this approach
% here for `longtable' which uses a stretching \tabskip.
% For \tabskip=0pt, we attach line numbers by a template.
%
% \RequirePackage{lineno}[2004/10/06]
% Wanted to check lineno version--causes unknown option error.
% See LaTeX bug latex/3730.

% Options:

%% TODO: underful page with `longtable.sty'!?
\let\if@ET@longtable\iffalse % \newif\if@ET@longtable@
\DeclareOption{longtable}{\let\if@ET@longtable\iftrue}
\let\if@ET@ltabptch\iftrue
\DeclareOption{nolongtablepatch}{\let\if@ET@ltabptch\iffalse}
\ProcessOptions

% Redefine \longtable if option:
\if@ET@longtable
  % Stephan's direction for attaching a line number to each row
  % is using \noalign in the course of \\.
  % However, (1) the user should not worry about closing the table
  % with \\ or without, and (2) the line number thing should not
  % happen twice if \\ is the last token before \end{longtable}.
  % Our solution: attach something to the beginning of \endlongtable
  % which attaches line number unless an unsucceeded \\ has done it.
  % This thing is activated by each row start and turned off by \\.
  % Since the switch at row start may happen to be in a pbox,
  % let it act globally on \endlongtable. (The \crcr at the beginning
  % of \endlongtable must not be changed directly, since original
  % \crcr may be required in table entries using \oalign etc.)
  %
  \RequirePackage{longtable}%
%   \IfFileExists{longtable.sty}{\RequirePackage{longtable}}%
%                               {\RequirePackage{longtabl}}%
% For Atari problem, it suffices to rename `longtable.sty'
% into `longtabl.sty'.
%
% Patch for tools/3180 and tools/3485 of LaTeX bug database:
 \if@ET@ltabptch
  \IfFileExists{ltabptch.sty}{\RequirePackage{ltabptch}}{%
    \PackageError{edtable}{%
      ltabptch.sty (for improving spacing around\MessageBreak
      longtable) missing! Be sure to use it always\MessageBreak
      or never!%
    }{%
      To omit ltabptch.sty *and* escape this error,\MessageBreak
      use package option `nolongtablepatch'.%
    }%
 }
 \fi
  \let\@ET@@longtable\longtable
  \def\longtable{%
    \ifLineNumbers
      \expandafter\@ET@longtable
    \else
      \let\@ET@sw@cr\@ET@crcr % ... in \endlongtable.
      \expandafter\@ET@@longtable
    \fi
  }
  \def\@ET@longtable{%
  % Since we have made it anyway, we use the method of redefining
  % \halign for inserting the activating row start as well.
  % However, redefinition must be repeated before every \LT@bchunk.
  % \longtable and \LT@get@withs are good places for this.
    \@ET@startlinewith\@ET@sw@cr@on
    \@ET@trivialize@linelabel
    \@ET@use@outerhook
    \@ET@ampnotes
    \let\LT@tabularcr\@ET@LT@tabularcr
    \expandafter\def\expandafter\LT@get@widths\expandafter
      {\LT@get@widths\@ET@startlinewith\@ET@sw@cr@on}
    % Admittedly, we could make it less of a hack by using @{...}
    % ---we could then leave \LT@get@widths untouched.
    \@ET@@longtable
  }
  \def\@ET@sw@cr@on{\global\let\@ET@sw@cr\@ET@cr@attach}
  % We are hacking a longtable version offering
  %   \def\LT@tabularcr{%
  %     \relax\iffalse{\fi\ifnum0=`}\fi
  %     \@ifstar
  %       {\def\crcr{\LT@crcr\noalign{\nobreak}}\let\cr\crcr
  %        \LT@t@bularcr}%
  %       {\LT@t@bularcr}}
  % We need a redefinition of \cr which is not overridden by \@ifstar:
  \def\@ET@LT@tabularcr{%
    \relax\iffalse{\fi\ifnum0=`}\fi
    \def\cr{\@ET@sw@cr}%
    % So \crcr is affected by change of \@ET@sw@cr, cf. below.
    \let\crcr\cr
    \@ifstar
      {\gdef\@ET@sw@nobreak{\nobreak
       \global\let\@ET@sw@nobreak\relax}%
  %     {\def\cr{\@ET@sw@cr\noalign{\nobreak}}\let\cr\crcr
  % Why didn't this work? %% TODO: try shorter again.
       \LT@t@bularcr}%
      \LT@t@bularcr
  %     {\let\cr\@ET@sw@cr \let\crcr\cr
  %      \LT@t@bularcr}% This accompanied \def\cr... above.
  }
  % Attach \@ET@sw@cr to beginning of \endlongtable:
  \expandafter\def\expandafter\endlongtable\expandafter
    {\expandafter\@ET@sw@cr\endlongtable}
  \def\@ET@cr@attach{% Actually attaching line numbers.
    \@ET@crcr\noalign{% Basically Stephan's approach.
  % If we were not careful, following box containing line number
  % could fool interline glue after longtable, even if tools/3485
  % is repaired in some way. This box should have same depth as
  % the line composed previously.
      \nobreak %% \@ET@ex... might cause page break. 2003/10/30.
      \setbox\z@\@ET@makeLineNumber
      \ht\z@-\prevdepth \dp\z@\prevdepth \box\z@
      \@ET@step@linenumber
      \global\let\@ET@sw@cr\@ET@crcr % ... if called by \\.
  % This also resets the \crcr starting \LT@echunk following
  % \@xargarraycr or \@yargarraycr in \LT@argtabularcr.
  % \let...\relax seems to suffice at well, but in case ...
      \@ET@execute@outerhook %% 2003/10/30.
      \@ET@sw@nobreak
%       \@ET@execute@outerhook
    }%
  }
  \let\@ET@crcr\crcr
  \global\let\@ET@sw@nobreak\relax % Just to remind ...
\fi

\let\@ET@step@linenumber\stepLineNumber
% \def\@ET@step@linenumber{\global\advance\c@linenumber\@ne}
\def\@ET@makeLineNumber{\hb@xt@\z@{\makeLineNumber}}
%% TODO: export to lineno.sty!?
% Special \halign:
% For `array' etc., we insert line numbers by a leftmost template.
% @{...} in the last argument of `array' etc. is difficult for this
% since `tabular*' has an additional argument. So we redefine
% \halign to put something to the right of the next \bgroup.
% Now, as a macro, \halign would break in an \edef or \xdef.
% Horribly, this danger has become quite real in longtable.sty's
% definition of \LT@bchunk. Only its latest version 4.10 has
% preceded \halign by a \noexpand (for mathtext.sty which
% redefines \halign as well). We need not rely on such a
% provision if we let \halign expand to \the<token register>
% for some token register. However, using \toks@ or \@temptokena
% is neither very safe---might be filled by macro ahead with
% something new. So reserve an own token register.
\newtoks\@ET@toks
\@ET@toks{\@ET@specialhalign}
% v1.3: In 'AMSmath', there are \halign's followed by explicit
% left braces. Thus '\def\@ET@specialhalign#1\bgroup{%' broke.
%% TODO ...
\def\@ET@specialhalign{%
  \ifmeasuring@ \expandafter\@firstoftwo
% v1.3 2005/03/04, for `amsmath'.
  \else \expandafter\@secondoftwo
  \fi
  {\let\halign\@ET@@halign \halign}%
  {\@ifnextchar\bgroup\@ET@replace@arg\@ET@sphalign@to}}
%% TODO: change back, commenting out code above, report
%%       error with `alignat'.
%   \@ifnextchar\bgroup\@ET@replace@arg\@ET@sphalign@to}
% If `amsmath' not loaded (v1.3):
\AtBeginDocument{\@ifundefined{measuring@true}{%
  \let\ifmeasuring@\iffalse}\relax}
\def\@ET@replace@arg#1{%
  \def\@EN@tempa{#1}\def\@EN@tempb{\bgroup}% Corr. after v1.3b.
  \ifx\@EN@tempa\@EN@tempb
    \def\@EN@tempa{\@ET@sphalign@to\bgroup}%
  \else
    \def\@EN@tempa{\@ET@sphalign@to{#1}}%
% This may be wrong, #1 might be `t' from `to' ...
  \fi
  \@EN@tempa}
\def\@ET@sphalign@to#1\bgroup{%
  \let\halign\@ET@@halign % Reset for nested arrays.
  \halign#1\bgroup
%  \ifmeasuring@\else % For `amsmath', moved backwards.
    \@ET@startline % [Wizard interface, via \@ET@starlinewith.]
%  \fi
}
% Wizard interface:
\def\@ET@startlinewith{% Next token precedes preamble.
%   \let\@ET@@halign\halign % Or \AtBeginDocument?
% ... indeed: bad loop with long longtable (\LT@get@widths!?)
  \def\halign{\the\@ET@toks}%
  \let\@ET@startline
}
\AtBeginDocument{\let\@ET@@halign\halign}
% Or save it at start of environment only
% (don't repeat inside longtable).
%
% Outer hook for inserts: (wizard interface)
\def\@ET@execute@outerhook{% To be placed outside inner mode.
  \@ET@outerhook \global\let\@ET@outerhook\@empty
}
\global\let\@ET@outerhook\@empty % Just to remind: global!
% \@EN@hookfn (ednotes) sends to outer hook:
\def\@ET@use@outerhook{% Wizard interface.
  \def\@EN@hookfn{\g@addto@macro\@ET@outerhook}%
}%% TODO: move last line to `ednotes', \let\@ET@use...\@empty.
%%        But this requires re-installing both packages
%%        at the same time. See the `ampnotes' thing as well.
%
% Trivialize \linelabel: (wizard interface)
\def\@ET@trivialize@linelabel{\let\linelabel\@ET@linelabel}
\def\@ET@linelabel{%
  \protected@edef\@currentlabel{\theLineNumber}%
  \label
}
% Change ampersand at \Anote etc.: (wizard interface)
%% 2003/10/31.
% & is changed by \Anote etc. No \begin...\endgroup,
% next \\ or & after % note/donote command switches back.
% ---Global change seems to break table setup.
\begingroup
\gdef\@ET@hideamps{\catcode`\&\active}
\@ET@hideamps
\gdef\@ET@ampnotes{%
  \let&\@ET@hideamp
  %% & would be undefined after \\ in lemma.
  \let\@ET@EN@rnote\@EN@note
  \def\@EN@note{\@ET@hideamps\@ET@EN@rnote}%
  \let\@ET@EN@rnotelb\@EN@notelabel
  \def\@EN@notelabel{\@ET@hideamps\@ET@EN@rnotelb}%
}
%% TODO: & and \\ in \@EN@lemmaexpands!?
\endgroup
\def\@ET@hideamp{&}

\newenvironment*{edtable}[2][]{%
% #1 $ or $$, #2 standard environment name.
%% TODO: star version!? cf. lineno.sty's `numquote'.
  \def\@ET@mode{#1}%
  \def\@ET@currenvir{#2}%
  \ifhmode \ifnum\the\lastpenalty=-\@M\else
% \par must be executed for printing/stopping line numbers:
    \@centercr\relax \noindent
  \fi \fi
  \ifvmode \noindent \fi % Otherwise linenumbers are indented
% -- however, not recommended.
  \@ET@use@outerhook
  \@ET@trivialize@linelabel
  \@ET@ampnotes
  \@ET@startlinewith\@ET@normal@startline
  \global\advance\c@linenumber\m@ne % Stepping correction.
% (Must come after closing previous paragraph.)
% Calculate offset:
  \global\advance\c@ET@array\@ne
% Saving/calling width of table (`longtable's algorithm).
% Test if there is \@flushglue on the left (center or flushleft):
  \edef\@EN@tempa{\the\@flushglue}%
% ... v1.3: fool this for "$$".
  \in@{$$}{#1\@nil}% \in@{#1}{#2} searches #1 in #2#1 ...
  \edef\@EN@tempb{\the\ifin@\@flushglue\else\leftskip\fi}%
  \ifx\@EN@tempa\@EN@tempb
% Assume \@ET@offset is 0pt else here.
    \@ET@offset\linewidth
    \advance\@ET@offset-%
      \@ifundefined\@ET@arraywidth@csn % Shortened for v1.3.
        {\linewidth\G@refundefinedtrue
         \PackageWarningNoLine{edtab}{^^JLine numbers at
           \@ET@currenvir\space need \jobname.aux}}%
        {\csname\@ET@arraywidth@csn\endcsname}%
% Test if \@flushglue on the right (center):
% ... v1.3: fool it again. Repeat \in@ in case ...
    \in@{$$}{#1\@nil}%
    \edef\@EN@tempb{\the\ifin@\@flushglue\else\rightskip\fi}%
    \ifx\@EN@tempa\@EN@tempb \divide\@ET@offset\tw@ \fi
  \fi
  \advance\@ET@offset\@totalleftmargin
  \advance\@ET@offset\ETextraoffset % Offset ready.
  \setbox\@tempboxa\hbox\bgroup
% v1.3: `alignat' complains later when this is \vbox instead.
%    \vbox\bgroup % Tried in vain.
    \ifx\@ET@mode\@empty\else $\fi
    \csname#2\endcsname
}{%
    \csname end\@ET@currenvir \endcsname
    \ifx\@ET@mode\@empty\else $\fi
  \egroup
  \@ET@step@linenumber % Stepping correction.
  \if@filesw
    \immediate\write\@auxout{%
      \gdef\expandafter\noexpand
        \csname\@ET@arraywidth@csn\endcsname
          {\the\wd\@tempboxa}}%
%% TODO: warning if changed!?
  \fi
% No automatic line number at array line:
  \nolinenumbers
  \@ET@mode\ensuremath{\vcenter{\box\@tempboxa}}\@ET@mode % v1.3.
  \@ET@execute@outerhook
  \ifhmode \@centercr\relax \fi
% \par must be executed for proper restart of printing line numbers
% (think so).
  \@endpetrue % Outside \@flushglue environment, avoid indent ahead.
%% TODO: slightly change for math mode around? Or give an error
%%       in math mode.
}
%% TODO: make \edtable \outer!?
\def\@ET@normal@startline{%
  \@ET@step@linenumber
  \llap{\@ET@makeLineNumber \hskip\@ET@offset}%
}
\newdimen\@ET@offset
\newdimen\ETextraoffset
%% TODO: export to lineno.sty!?
\newcount\c@ET@array
\def\@ET@arraywidth@csn{ET@a@\romannumeral\c@ET@array}
%
%% TODO: .dtx
\endinput

%% VERSION HISTORY:
v0.22 2003/01/13  First version, named `edtab02.sty', sent around,
                  together with ednotes.sty.
v0.23 2003/01/22  Added version history.
v0.24 2003/01/27  Copyright notice.
%%% v0.3  2003/03/03  Requires linenox0.sty. %% RETREATED
%% TODO: reconsider,  see `lineno[x].sty'.
v0.25 2003/03/05  Added <ednotes.sty@web.de>.
v0.26 2003/03/23  Added \@ET@startlinewith to wizard command list;
                  moved redefinition \LT@get@widths from option code
                  into \@ET@longtable so this really happens in
                  line number mode only; moved \let\@ET@@halign\halign
                  from \@ET@startlinewith to \AtBeginDocument.
v0.27 2003/06/02  Added \ifvmode \noindent \fi to `edtable' definition;
      2003/06/10  added `lemma cannot run' to instructions.
v0.28 2003/10/31  \@ET@cr@attach: moved \@ET@execu..., added \nobreak.
                  Added \@ET@ampnotes and documentation: now working
                  across entries.
v1.0  2004/05/13  `ednotes.sty' etc., updated copyright. Renamed as
                  `edtable.sty' for TUGboat article.
                  \RequirePackage{lineno}, changed doc. accordingly.
                  Removed `may be released under different name'.
v1.10 2004/07/27  Looking for `longtabl.sty' for my Atari!
      2004/08/22  Added warning (screen/documentation) concerning
                  ltabptch.sty.
      2004/08/23  Added suggestion of option `nolongtablepatch';
                  LPPL v1.3.
v1.2  2004/08/31  Rearranged preamble concerning maintenance,
                  removed `preliminary release'.
      2004/10/06  \newif -> \let...; option `nolongtablepatch';
                  changed documentation accordingly and for
                  loading edtable.sty by lineno/ednotes option;
                  \let\@ET@step@linenumber\stepLineNumber; use
                  \ifLineNumbers.
      2004/10/07  Needed version numbers.
      2004/10/08  ltabptch warning -> error!
      2004/10/11  Disabled \RequirePackage{lineno}[...]; undid
                  the `longtabl.sty' thing from 2004/07/27.
      2004/10/12  latex/3730.
v1.2a 2004/11/07  LPPL v1.3a.
v1.2b 2005/01/10  Contact via http.
v1.3b 2005/03/04  Support for environments like `array'
                  environment. The description of `edtable' was
                  wrong in this respect!
v1.3  2005/03/07  Adaptations in documentation, and extended usage
                  instructions very much, correcting some
                  errors as well. Acknowledgement to M. B.
v1.3a 2005/03/09  Corrected numbering in `usage'.
v1.3c 2005/03/15  \@centercr\relax, TODO on math mode.
      2005/04/09  `editory' -> `editorial'.
\end{verbatim}
\subsection{\texttt{ltabptch.sty}}
\verbatiminput{ltabptch.sty}

\section{\cs{linelabel} and notes from \textit{math} mode:
         \notinaux{\\} \texttt{ednmath0.sty}}
\begin{verbatim}
%% Macro package `ednmath0.sty' for LaTeX2e,
%% copyright (C) 2004 Uwe Lück
%% math support for `lineno.sty' and `ednotes.sty'.
%%
\def\fileversion{v0.2b} \def\filedate{2005/01/10}
%% This program can be redistributed and/or modified under the
%% terms of the LaTeX Project Public License distributed from
%% CTAN archives in directory macros/latex/base/lppl.txt; either
%% version 1.3a of the License, or any later version.
%% The latest version of this license is in
%%   http://www.latex-project.org/lppl.txt
%% There is NO WARRANTY.
%% This code is very EXPERIMENTAL!
%%
%% Please report bugs, problems, and suggestions via
%%
%%   https://github.com/latex-lineno/lineno
%
%% * MAIN FEATURE *
%
% lineno.sty's \linelabel and ednotes.sty's commands are enabled
% to work in math mode if it's "entered in outer mode"
% (including `displaymath' and `equation' environments).
% (lineno.sty is the package by Stephan Boettcher.)
% They will even work in tabular environments that are adjusted
% to notes by package `edtable.sty'.
%
% CAVEATS:
% -- Does not work yet in environments like LaTeX's
% `eqnarray'. (This could probably repaired along the lines
% of Edtable.sty--we're short of time and will try later.)
% -- Useful error messages when (i) math mode is entered from
% inner mode or when (ii) a math display gets not line number
% are missing at present.
%
%% * USAGE: *
%
% * Most simple: *
% --If you are working with ednotes and want to use its
% commands in math mode, load ednotes.sty--version 0.8
% onwards--with its package option `mathnotes'.
% --If you don't work with ednotes, only with lineno, you
% get the main feature of making \linelabel work in math mode
% by loading lineno.sty--version 4.1 onwards--with its
% package option `mathrefs'.
%
% * Switch off and on: *
% To reduce danger resulting from missing error messages
% ("caveat" above), you may switch these new math facilities
% off by \NoNotesToMath where you don't expect to need them.
% You may switch them on again by \NotesToMath where you want
% to use them, being aware of the danger. Both commands work
% locally, so you can replace one of them by enclosing it in
% a group. E.g., even, after \NoNotesToMath you can use an
% environment as follows:
%   \begin{NotesToMath}
%     <text>
%   \end{NotesToMath}
% (I am not quite sure that this is useful.)
%
% * Customize ellipsis: *
% ednotes' \lemmaellipsis is changed to expand to
% \mathlemmaellipsis when entering math, and this is preset
% to be LaTeX's \mathellipsis. (This is three dots as
% \mathinner.) You can change this by redefining
% \mathlemmaellipsis, e.g.:
%   \renewcommand{\mathlemmaellipsis}{\cdots}
% If you need \cdots as the ellipsis at a single place only,
% you may, of course, use the `<...>' option of \<, e.g.:
%   $ x = \Anote{a\<<\cdots>bcd\>e}{Indeed?} - y $
%
% * Customize note mode: *
% For variant readings, you may want that the note is
% usually set in math mode--so you may want that you
% needn't type the dollar signs in the note text.
% Note that you can do this by customizing \notefmt,
% and you can do this by customizing \Anotefmt (e.g.)
% to have this feature for \Anote only.
%

\NeedsTeXFormat{LaTeX2e}
\ProvidesPackage{ednmath0}[\filedate\space\fileversion\space
  math support for lineno/ednotes (ul)]
%
%% User commands:
\def\NotesToMath{\let\@LN@mathhook\@LN@labelinmath
  \@bsphack \@esphack
% For \begin{NotesToMath}
}
\def\NoNotesToMath{\@bsphack
  \def\@LN@mathhook{\@parmoderr\@gobble}%
  \@esphack
}
\def\endNotesToMath{\@bsphack\@Esphack}
\let\endNoNotesToMath\endNotesToMath
%
%% Core code for lineno.sty:
\@ifundefined{@LN@postlabel}{%
  \PackageError{ednmath0}{%
    Bad lineno.sty version%
  }{%
    lineno.sty from 2004/08/16 or later
    must be loaded earlier.%
  }%
}{%
  \def\@LN@labelinmath#1{%
    \ifmmode
      \@LN@postlabel{#1}%
    \else
      \@parmoderr
    \fi
  }
}
%
%% Core code for ednotes.sty:
\@ifundefined{@EN@note}{%
% v0.01 sent a warning in this case. Considered superfluous now.
}{%
  \def\@EN@themathlemmatag{%
    \ifmmode
      \toks@\expandafter{\@EN@lemmatag}%
      \edef\@EN@lemmatag{%
        $%
          \def\noexpand\lemmaellipsis{%
            \noexpand\mathlemmaellipsis}%
          \the\toks@
        $%
      }%
%       \expandafter \def \expandafter \@EN@lemmatag
%         \expandafter {\expandafter $\expandafter
%         \def \expandafter \lemmaellipsis \expandafter {%
%           \expandafter \mathlemmaellipsis \expandafter }%
%         \@EN@lemmatag $}%
    \fi
  }
% To be sure, \lemmaellipsis doesn't need to be changed when
% ednotes `\<...\>' feature is not used. Though I prefer to
% use one hook only in ednotes for both situations, with and
% without `\<...\>'.
%
% The final \unskip in ednotes' \@EN@lemmatag would undo a final
% \quad. That's OK: outside math the same happens.
% In v0.01, \NoNotesToMath undid ednotes changes for math mode.
% However, re-appearence of \linelabel error messages suffices.
%
% Now add lemma switch to the left of \[No]NotesToMath:
  \toks@\expandafter{\NotesToMath}
  \edef\NotesToMath{%
    \let \noexpand\@EN@mathlemmatag \noexpand\@EN@themathlemmatag
    \the\toks@
  }
%  \typeout{\string\NotesToMath: \meaning\NotesToMath}
  \toks@\expandafter{\NoNotesToMath}
  \edef\NoNotesToMath{%
    \let \noexpand\@EN@mathlemmatag \relax
    \the\toks@
  }
%  \typeout{\string\NoNotesToMath: \meaning\NoNotesToMath}
}
% We need no extra device for a choice for users whether the *note*
% should be set in math mode or in horizontal mode by default
% (which might depend on the kind ["layer"] of notes).
% This can be done already by customization of ednotes' \notefmt.
% However, we might change ednotes' default \notefmt to default
%   \renewcommand*{\notefmt}[1]{$#1$}
%
\let\mathlemmaellipsis\mathellipsis
%% TODO: Since when has LaTeX provided \mathellipsis?
%% -> \Needs...
%
% Default:
\NotesToMath
%
\endinput

%% TODO: Without \linenumberdisplaymath, in displaymath,
%% an error should be shown. Use, e.g., that in a displaymath
%% \ifinner is false.
%% TODO: E.g., by changing \everymath, perhaps can be warned
%% that the math group is in a box already, so the vertical
%% items will get lost.
%% TODO: Adjust `eqnarray' (in Edtable?) as well.

%% VERSION HISTORY:
v0.01 2004/08/16  First version, sent to Christian.
v0.02 2004/08/16  Considerably simplified for ednotes.
      2004/08/19  Added ellipsis stuff, documentation, and
                  instructions. Uncapitalized package names.
                  Added \end[No]NotesToMath.
      2004/08/20  Added \@bsphack and \@esphack; corrected
                  ednotes extension (too much deleted, completely
                  wrong), introducing \@EN@themathlemmatag.
v0.02b .../08/31  Rearranged preamble concerning maintenance.
v0.1  2004/09/20  Removed mentions of `linenox0.sty'.
v0.2  2004/10/07  Removed another mention of `linenox0.sty';
                  Instructions: `lineno' or `ednotes.sty' option.
v0.2a 2004/11/07  LPPL v1.3a.
v0.2b 2005/01/10  Contact via http.
\end{verbatim}

\section{Extended line number references: \texttt{vplref.sty}}
\texttt{vplref.sty} is input through the \texttt{lineno}
package option \texttt{addpageno}. This adds page numbers
to line number references to distant sides---using the
\texttt{varioref} package from the \LaTeX\ distribution.
\begin{verbatim}
%% `vplref.sty'
%% -- extended line number referencing with lineno.sty.

\def\filedate{2005/04/25} \def\fileversion{0.2a}

%% Copyright (C) 2004, 2005 Uwe Lück
%% -- support of lineno.sty for varioref.sty.

%% This file can be redistributed and/or modified under
%% the terms of the LaTeX Project Public License; either
%% version 1.3 of the License, or any later version.
%% The latest version of this license is in
%%   http://www.latex-project.org/lppl.txt
%% We did our best to help you, but there is NO WARRANTY.

%% USAGE:
%
% \vpagelineref{<label>} expands to
%
% a) \ref{<label>}
%    -- if on same page as \linelabel{<label>}
%
% b) \LineWithPage{<label>} -- otherwise.
%
% \LineWithPage{<label>} expands -- by default -- to
%
%   \pageref{<label>}.\ref{<label>}
%
% This can be customized by editing
%
%   \renewcommand*{\LineWithPage}[1]{\pageref{#1}.\ref{#1}}
%
% in your document preamble, after vplref.sty has been loaded
% (which may have happened through lineno.sty with option
% `addpageno').

%% IMPLEMENTATION:

\NeedsTeXFormat{LaTeX2e}[1994/12/01] %% \Declare...*
\ProvidesPackage{vplref}[\filedate\space v\fileversion \space
                         page-line cross-refs] %% UL 2011/02/13

\AtBeginDocument{\RequirePackage{lineno,varioref}}

%% Anderer Ansatz: GPNo (\FirstOnPage)

\DeclareRobustCommand*\vpagelineref[1]{{%
%   \def\reftextcurrent{\lineref{#1}}%% First vpageref arg.
  \let\reftextfaraway\LineWithPage
  \def\reftextafter{\reftextfaraway{#1}}%
  \let\reftextbefore\reftextafter
  \let\reftextfaceafter\reftextafter
  \let\reftextfacebefore\reftextafter
%% <- Looks somewhat stupid, but varioref.sty has its merits
%%    as compared with the mechanism in ednotes.sty.
  \vpageref[\ref{#1}][]{#1}%% The robust alternative.
%% Here and with \LineWithPage, \lineref seems more appropriate
%% than \ref, but it produces errors when labels have not been
%% defined. This seems to be an incompatibility with lineno.sty.
}}

%% Customizable format for different page:
\newcommand*\LineWithPage[1]{\pageref{#1}.\ref{#1}}

\endinput

VERSION HISTORY:
v0.1   2004/10/19  First, sent to Sergei Mariev.
v0.11  2004/10/19  Fit to recent varioref version;
                   sent to Sergei.
v0.2   2005/04/25  \Require... \AtBeginDocument.
v0.2a  2011/02/13  add. caption to \ProvidesPackage
\end{verbatim}